\documentclass{emulateapj}

\newcommand{\alphafe}{[\alpha/{\rm Fe}]} 
\newcommand{\feh}{[{\rm Fe/H}]} 
\usepackage{amsmath}
\usepackage{bm}

\shorttitle{A Chemodynamic Study of Hydra~I}
\shortauthors{Hargis et al.}

\begin{document}

\title{Evidence That Hydra I is a Tidally Disrupting Milky Way Dwarf Galaxy}

\author{Jonathan R. Hargis\altaffilmark{1}
       Brian Kimmig\altaffilmark{1}, 
       Beth Willman\altaffilmark{1,2}, 
       Nelson Caldwell\altaffilmark{3},
       Matthew G. Walker\altaffilmark{4},
       Jay Strader\altaffilmark{5},
       David J. Sand\altaffilmark{6},
       Carl J. Grillmair\altaffilmark{7},
       Joo Heon Yoon\altaffilmark{8}
}
\altaffiltext{1}{Haverford College, Department of Astronomy, 370 Lancaster Avenue, Haverford, PA 19041, USA;}
\altaffiltext{2}{LSST and Steward Observatory, 933 North Cherry Avenue, Tucson, AZ 85721, USA;}
\altaffiltext{3}{Harvard-Smithsonian Center for Astrophysics, 60 Garden Street, Cambridge, MA 02138, USA}
\altaffiltext{4}{McWilliams Center for Cosmology, Department of Physics, Carnegie Mellon University, 5000 Forbes Avenue, Pittsburgh, PA 15213}
\altaffiltext{5}{Michigan State University, Department of Physics and Astronomy, East Lansing, MI 48824}
\altaffiltext{6}{Texas Tech University, Physics Department, Box 41051, Lubbock, TX 79409-1051, USA}
\altaffiltext{7}{Spitzer Science Center, 1200 E. California Blvd., Pasadena, CA 91125, USA}
\altaffiltext{8}{University of California Santa Barbara, Department of Physics, Santa Barbara, CA, 93106, USA}

\begin{abstract}
The Eastern Banded Structure (EBS) and Hydra~I halo overdensity are very nearby (d $\sim$ 10 kpc) objects discovered in SDSS data.  Previous studies of the region have shown that EBS and Hydra I are spatially coincident, cold structures at the same distance, suggesting that Hydra I may be the EBS's progenitor.  We combine new wide-field DECam imaging and MMT/Hectochelle spectroscopic observations of Hydra I with SDSS archival spectroscopic observations to quantify Hydra I's present-day chemodynamical properties, and to infer whether it originated as a star cluster or dwarf galaxy.  While previous work using shallow SDSS imaging assumed a standard old, metal-poor stellar population, our deeper DECam imaging reveals that Hydra~I has a thin, well-defined main sequence turnoff of intermediate age ($\sim 5-6$ Gyr) and metallicity ([Fe/H] = $-0.9$ dex).  We measure statistically significant spreads in both the iron and alpha-element abundances of $\sigma_{[Fe/H]} = 0.13 \pm 0.02$ dex and $\sigma_{[\alpha/{\rm Fe}]} = 0.09 \pm 0.03$ dex, respectively, and place upper limits on both the rotation and its proper motion.   Hydra~I's intermediate age and [Fe/H] -- as well as its low [$\alpha$/Fe], apparent [Fe/H] spread, and present-day low luminosity -- suggest that its progenitor was a dwarf galaxy, which has subsequently lost more than $99.99\%$ of its stellar mass.

\end{abstract}

\section{Introduction}

\begin{figure*}[hbt!]
\includegraphics[width=0.5\textwidth,trim={0cm 0cm 0cm 0cm},clip]{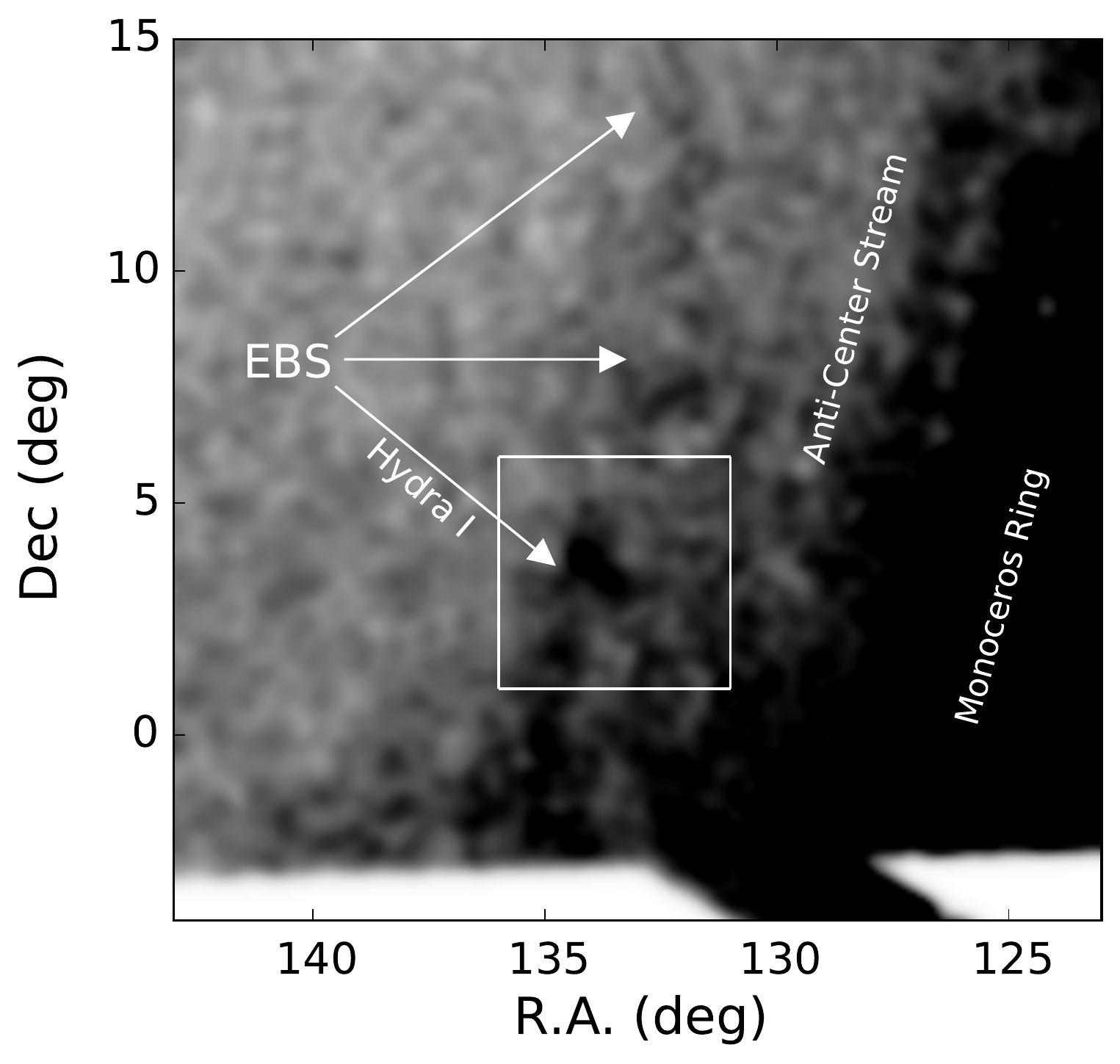}
\includegraphics[width=.48\textwidth,trim={0cm 0cm 0cm 0cm},clip]{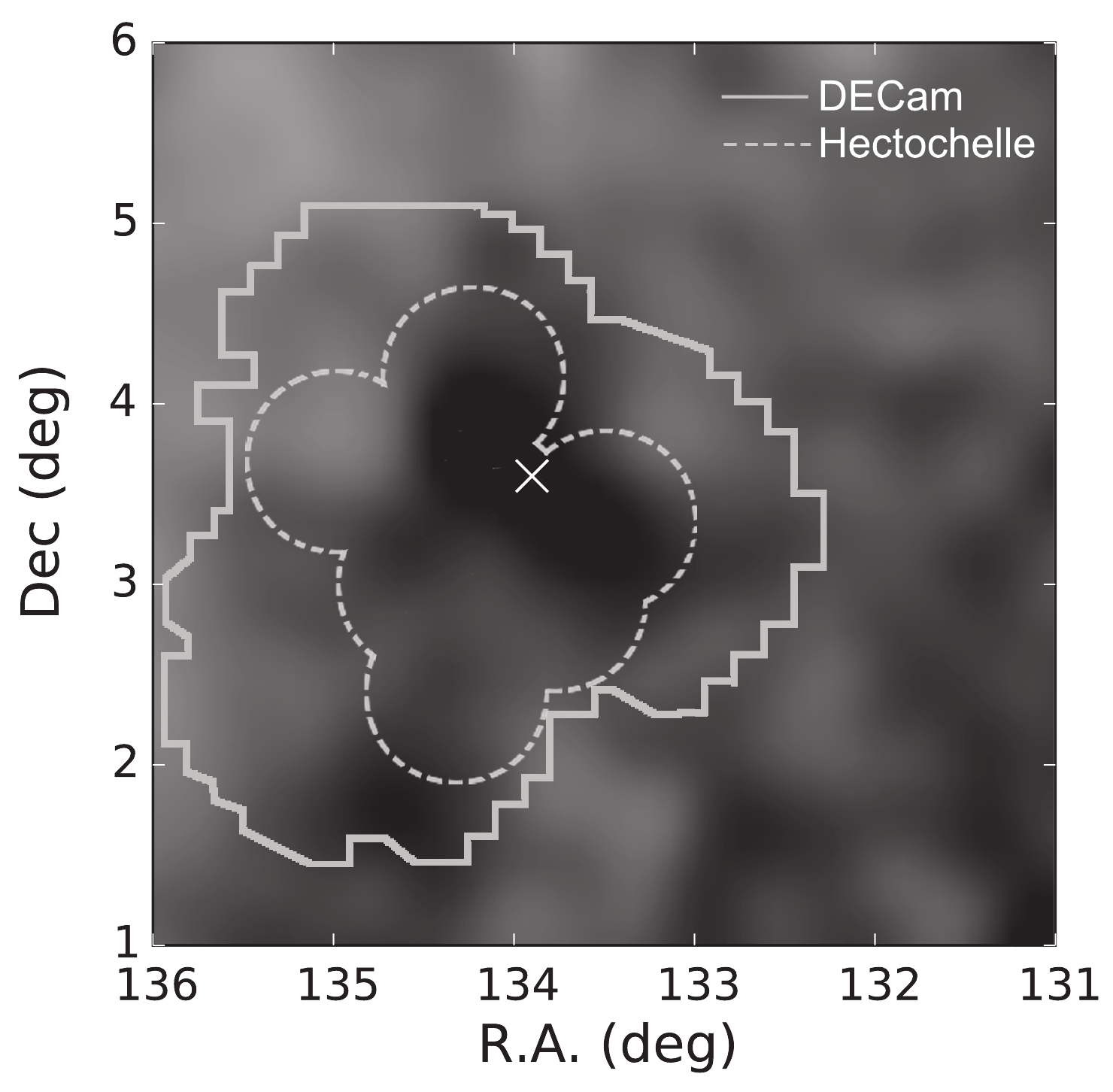}
\caption{Filtered and smoothed SDSS surface density map of the EBS stream and Galactic anti-center region ({\it left}; data from \citealt{grillmair2011}) and the Hydra~I overdensity ({\it right}).  The Anti-Center Stream and Monoceros Ring are labeled for reference.  The right panel is a zoom-in view of Hydra~I (white box in the left panel) shown with the observing footprints of the DECam imaging (solid grey line) and MMT/Hectochelle spectroscopy (dashed grey line).
 }
\label{fig:spatial-ebs}
\end{figure*}

The stellar tidal streams and substructure of the Milky Way (MW) provides an important observational window into the assembly of the Galaxy in a Universe dominated by dark energy and dark matter.  In the standard $\Lambda$CDM model, numerical simulations show that massive galaxies like the Milky Way grow hierarchically, that is, from the continuous accretion and tidal destruction of smaller systems such as dwarf galaxies and globular clusters (e.g., \citealt{bullock2005,johnston2008,cooper2010}).   Observations of the Galaxy's halo support this picture: in addition to the $\sim 40$ known dwarf galaxy satellites of the Milky Way (\citealt{mcconnachie2012,bechtol2015,koposov2015a, drlicawagner2015}), a total of 21 stellar streams have been discovered to-date in wide-field optical imaging surveys such as the Sloan Digital Sky Survey (SDSS; e.g., \citealt{belokurov2006}), PAndAS \citep{martin2014}, Pan-STARRS \citep{bernard2014}, ATLAS \citep{koposov2015}, and DES \citep{balbinot2015}.   

Despite the growing number of stream discoveries, only four stellar streams have known progenitors: the globular clusters Pal~5, NGC~5466, NGC~288 and the Sagittarius dwarf spheroidal. Developing a detailed picture of the origin of the Galaxy's halo requires understanding the relative contribution of tidally-destroyed globular clusters and dwarf galaxies. Searching for and quantifying the ages, metallicities, and kinematics of likely stream progenitors is therefore an important step towards a more complete picture of the hierarchical assembly of the MW's halo in a cosmological context.  

The Eastern Banded Structure (EBS) stream (Figure~\ref{fig:spatial-ebs}) is an ideal target for such a study, as it is relatively nearby ($d\sim 10$ kpc; Grillmair 2011) and therefore may be close enough for proper motion measurements.  \citet{grillmair2006} discovered the EBS in SDSS imaging data as a relatively broad stellar stream near the Monoceros Ring stellar overdensity.  \citet{grillmair2011} identified an extended double-lobed, $\sim 2$ sq. degree feature along the stream -- called Hydra~I -- which has a suface density larger than any other overdensity along the stream (white boxed region in Figure~\ref{fig:spatial-ebs}).  Based on its morphology and prominence, \citet{grillmair2011} suggested that Hydra~I may be the progenitor of the EBS stream.  Using SDSS/SEGUE spectroscopic observations that are spatially coincident with Hydra I, \citet{schlaufman2009} demonstrated the presence of a kinematically cold population of stars, with $V_{\rm helio} \sim 85$ km s$^{-1}$ and the colors and magnitudes expected for an old, metal-poor population at the distance of the EBS.

The origin and nature of the EBS/Hydra~I features are complicated by their proximity to (and possible association with) a number of Galactic anti-center structures, including the Monoceros Ring \citep{li2012,slater2014,xu2015}, the Anti-Center Stream \citep{grillmair2008,carlin2010}, and the Triangulum Andromeda substructures (\citealt{sheffield2014} and references therein).  The origin and relationship between these features remains intensely debated (see \citealt{xu2015} for a recent review).  For example, it is currently unclear whether the Monoceros Ring is tidal debris from an accreted dwarf galaxy \citep{penarrubia2005} or has its origin in the Galactic disk -- as either a normal warp/flare \citep{momany2006} or as the result of kicked-up stars from disk oscillations occurring from an encounter with a massive satellite \citep{kazantzidis2008, pricewhelan2015, xu2015}.  We consider our results in the context of these Galactic anti-center structures in our discussion (Section~\ref{subsec:other}).  

Hydra~I presents a unique opportunity to study a Milky Way satellite undergoing active disruption.  In this work  we determine Hydra~I's stellar population, chemical, and kinematic properties with the aim of exploring whether the object is a disrupting dwarf galaxy or star cluster.  Previous studies of Hydra~I region were limited by a small number candidate member stars (with large velocity uncertainties), prohibiting detailed tests of Hydra~I's nature.  In this work, we combine higher precision photometric and spectroscopic data of Hydra~I -- from the Dark Energy Camera (DECam; \citealt{flaugher2015}) and MMT/Hectochelle spectrograph \citep{szentgyorgyi2011} -- with with archival SDSS data to study Hydra~I using a large sample of candidate stars.  

This paper is organized as follows.  In Section~\ref{sec:data} we describe our observational data sets.  In Section~\ref{sec:sample} we discuss the selection of candidate member stars, and in Section~\ref{sec:results} we present the results of our analysis and the derived properties of Hydra~I.  We discuss possible scenarios concerning the nature of Hydra~I in Section~\ref{sec:discussion}, and present our conclusions in Section~\ref{sec:conclusions}.

\section{Photometric and Spectroscopic Data}
\label{sec:data}

\begin{figure*}
 \epsscale{1.1}
\includegraphics[width=0.52\textwidth,trim={0cm 0cm 0cm 0cm},clip]{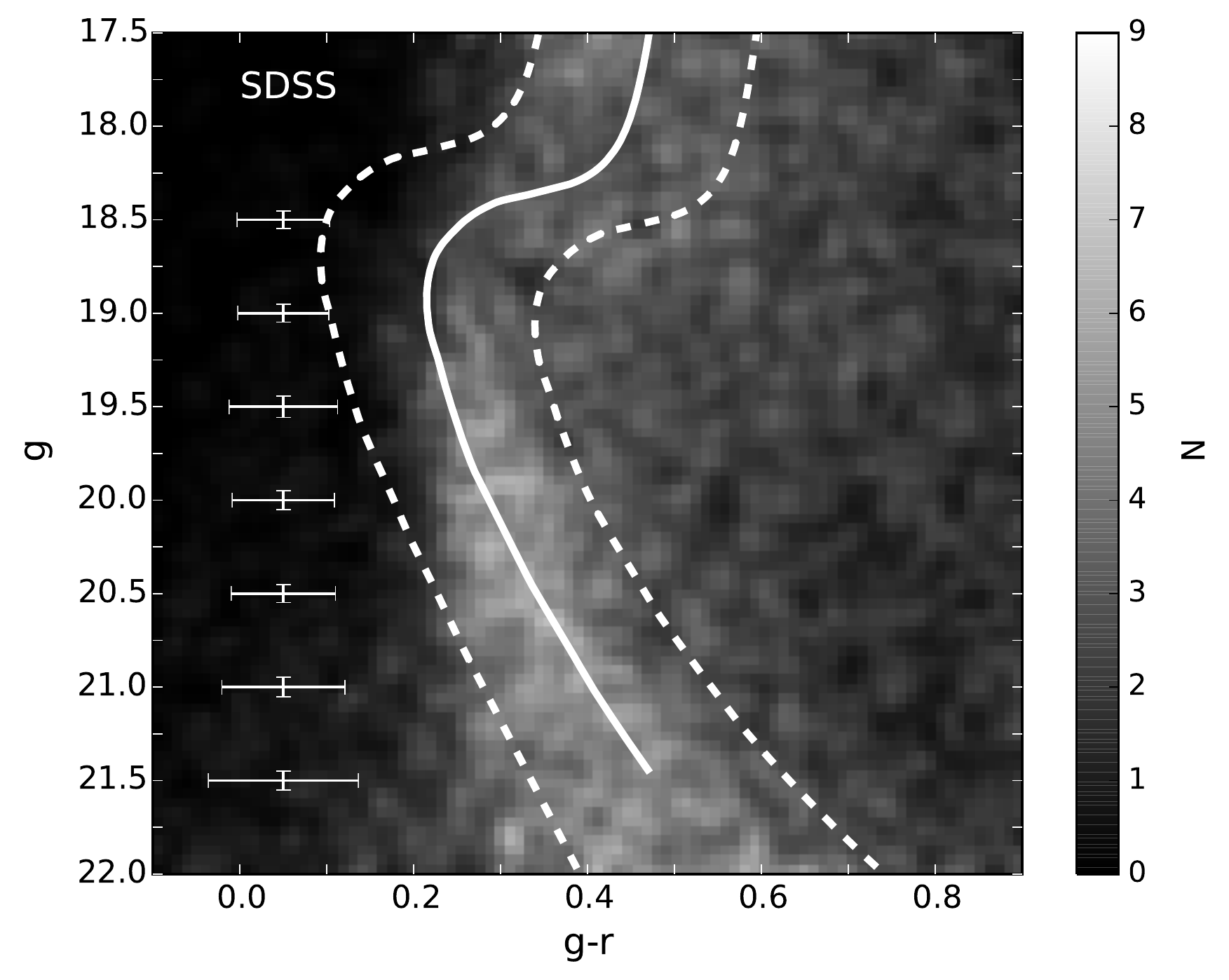}
\includegraphics[width=0.52\textwidth,trim={0cm 0cm 0cm 0cm},clip]{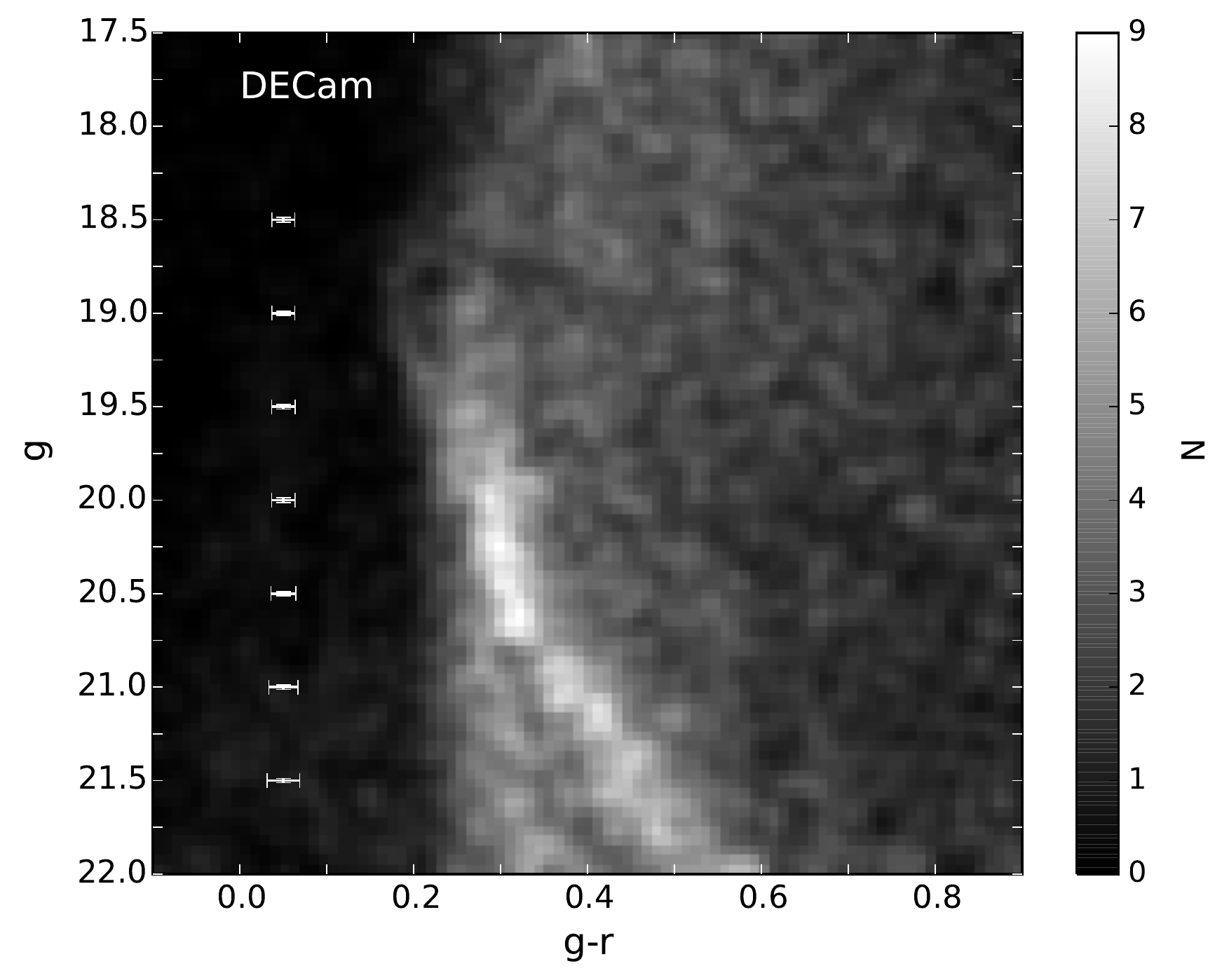}
\caption{SDSS (left) and DECam (right) color-magnitude-number density (Hess) diagrams for stars within a 9 sq. deg region centered on Hydra~I.  The diagrams have been slightly smoothed with a Gaussian filter to enhance low density features.  The dashed white lines denote the CMD filter used to select candidates for spectroscopic follow-up with MMT/Hectochelle.  The isochrone (solid white line, \citealt{dotter2008}) has an [Fe/H] = -1.8, age = 13 Gyr and shifted to a distance of 9.7 kpc \citep{grillmair2011}.  The median color and magnitude uncertainties are shown as white points.  The higher precision DECam photometry clearly shows the detection of a blue MSTO at $g-r \sim 0.18$ not apparent in the SDSS data.}
\label{fig:targeting}
\end{figure*} 

\subsection{DECam Photometry}
\label{sec:decam}

Observations of Hydra~I were obtained on 20-23 March 2014 using the DECam imager installed on the CTIO Blanco 4m telescope.  The DECam imager consists of sixty-two $2048\times4096$ imaging CCDs (chip gaps of $\sim 50\arcsec$) arranged in a circular layout in the focal plane.  On the Blanco telescope, this layout provides imaging over a $2.1$ degree diameter, resulting in a $3$ square degree imaging area with $0.26\arcsec$ pixels.  

We imaged Hydra~I using the $ugri$ filters, but to facilitate comparisons with other studies of Hydra~I/EBS and the Galactic anti-center region, we only present the $gr$ imaging in this work.  We used a series of $6\times180$ sec exposures in each of the $g$ and $r$ filters and dithered between exposures to fill in the chip gaps.  Figure~\ref{fig:spatial-ebs} shows the resulting footprint of the DECam imaging ($\sim 9$ deg$^2$) on the Hydra~I field.  All imaging was obtained under photometric or near-photometric conditions.  The mean FWHM of the image point-spread functions (PSFs) for the exposures ranged from $0.9\arcsec$ to $1.2\arcsec$.

The data were reduced using the NOAO Community Pipeline (CP; \citealt{valdes2014}) and downloaded from the NOAO Science Archive.  The CP performs both a standard instrumental calibration of the raw images (bias, overscan, cross-talk corrections; fringing corrections; dome flat and sky flat field corrections) and corrects for geometric distortions by reprojecting the images onto a common spatial scale.  

We performed photometry on each DECam exposure by separately analyzing each of the 62 reduced, reprojected chips using the DAOPHOTII/ALLSTAR software suite \citep{stetson1987,stetson1994}.  We derive a catalog of point sources using cuts on the DAOPHOT CHI and SHARP parameters which vary as a function of magnitude depth.  To ensure a robust characterization of sources, we require that object must be point-like in each of the $gri$ images.  

The Hydra~I/EBS region is located in the Sloan Digital Sky Survey (SDSS) footprint, and so we derive calibrated magnitudes by matching directly to SDSS Data Release 8 \citep[SDSS DR8;][]{aihara2011}.  We derived a photometric solution for each individual DECam exposure separately, after correcting the instrumental magnitudes for atmospheric extinction.  We derived zero points and linear color terms using the maximum likelihood approach described by \citet{boettcher2013}, using only point sources which have high S/N in both the DECam and SDSS imaging and SDSS colors of $g - r < 1$.  We impose the latter constraint in order to improve the photometric calibration over the color region of interest for this study.  Lastly, all calibrated apparent magnitudes presented in this study were corrected for Galactic extinction using the reddening maps of \citet{schlegel1998} and the reddening coefficients of \citet{schlafly2011}.  The mean color excess in the Hydra~I region imaged by DECam is E(B-V) $\simeq0.04$ mag and varies by $\pm 0.02$ mag across the field of view.  

The final Hydra~I DECam point source catalog was constructed by calculating the weighted mean magnitude of each point source from the individual measurements on each frame.  We used the uncertainties in the calibrated magnitudes  (in flux space) as the weights and used an iterative sigma-clipping algorithm to reject any measurements which fell $> 4\sigma$ from the mean.  The uncertainties in the weighted mean magnitudes were calculated from the inverse square of the flux errors.  The final catalog contains $\sim 17,000$ sources between $17 < g < 22$ mag over a $\sim 9$ deg$^{2}$ area.  We present the entire photometric data set in Table~\ref{tbl:decam}, which is available in machine readable format from the publisher.

The number of point sources as a function of magnitude depth drops dramatically at $g \simeq 23.5$, $r \simeq 22.8$ mag.  We therefore estimate that our imaging is complete at $\sim 1$ mag brighter than these limits.  A completeness limit at $g \sim 22-22.5$ is sufficient for the purposes of this study.  As discussed in Section~\ref{sec:sample}, the main-sequence turn off (MSTO) in the color-magnitude diagram (CMD) of Hydra~I is at $g \simeq 18.5$ ($r \simeq 18.3$) mag, significantly brighter than our estimated incompleteness depths.

\subsection{Hectochelle Spectroscopy}
\label{sec:spectroscopy}

We used MMT/Hectochelle to target stars throughout the Hydra~I overdensity.  Following the previous analysis by \citet{grillmair2011}, we targeted stars from SDSS within 0.1 mag of a 13 Gyr, [Fe/H] = -1.8 isochrone \citep{dotter2008} at a distance of 9.7 kpc (however, see our conclusions in \S~\ref{sec:stellar-pops} regarding the age and metallicity of Hydra~I).  Figure~\ref{fig:targeting} shows the selection filter on the SDSS color-magnitude-number density (Hess) diagram of stars within the $\sim 9$ deg$^2$ footprint of the DECam imaging.  We chose a CMD filter with a relatively blue edge in order to explore the possibility of younger (bluer) stellar populations in Hydra~I.  Because the expected red giant branch (RGB) population of Hydra~I  overlaps with MW thick disk stars in color and magnitude (see Figure~\ref{fig:targeting}), objects fainter than $g=18$ mag were given priority for our Hectochelle observations in order to minimize contamination.  

Spectroscopic observations were obtained over 7 nights using the Hectochelle spectrograph on the 6.5m MMT.  Hectochelle uses 240 fibers across a $1$ degree diameter field of view to take single order spectra at a resolution of $R\sim 38,000$.  We used the RV31 order selecting filter to isolate the 150~\AA~around the Mg b triplet (5167-5184~\AA).  The field of view of Hectochelle is well suited to observing Hydra~I.  We obtained seven pointings around Hydra~I, primarily in relatively bright or grey observing time, with total exposure times ranging from 4800 to 19,8000 sec per pointing.  We targeted the areas on the primary overdensities and along the stream slightly southeast of the two main lobes.  The footprint of the Hectochelle observations are shown in the right panel of Figure~\ref{fig:spatial-ebs}.  We obtained spectra for a total of 796 stars with magnitudes in the range $17 < g < 21.3$. 

The multifiber spectra were reduced in a uniform manner. For each field, the separate exposures were debiased and flat fielded, and then compared before extraction to allow identification and elimination of cosmic rays through interpolation.  Spectra were then extracted, combined and wavelength calibrated using contemporaneous exposures of a ThAr lamp (the spectra were not-linearized in dispersion). Each fiber has a  distinct wavelength dependence in throughput, which can be estimated using exposures of a continuum source or the twilight sky.  A background sky estimate was made by combining data from fibers placed randomly in the focal plane, but excluding those which accidentally fell on a source. This sky spectrum was finely interpolated onto the object spectrum's wavelength grid and then subtracted.

\begin{figure*}
\includegraphics[width=0.5\textwidth,trim={0cm 0cm 0cm 0cm},clip]{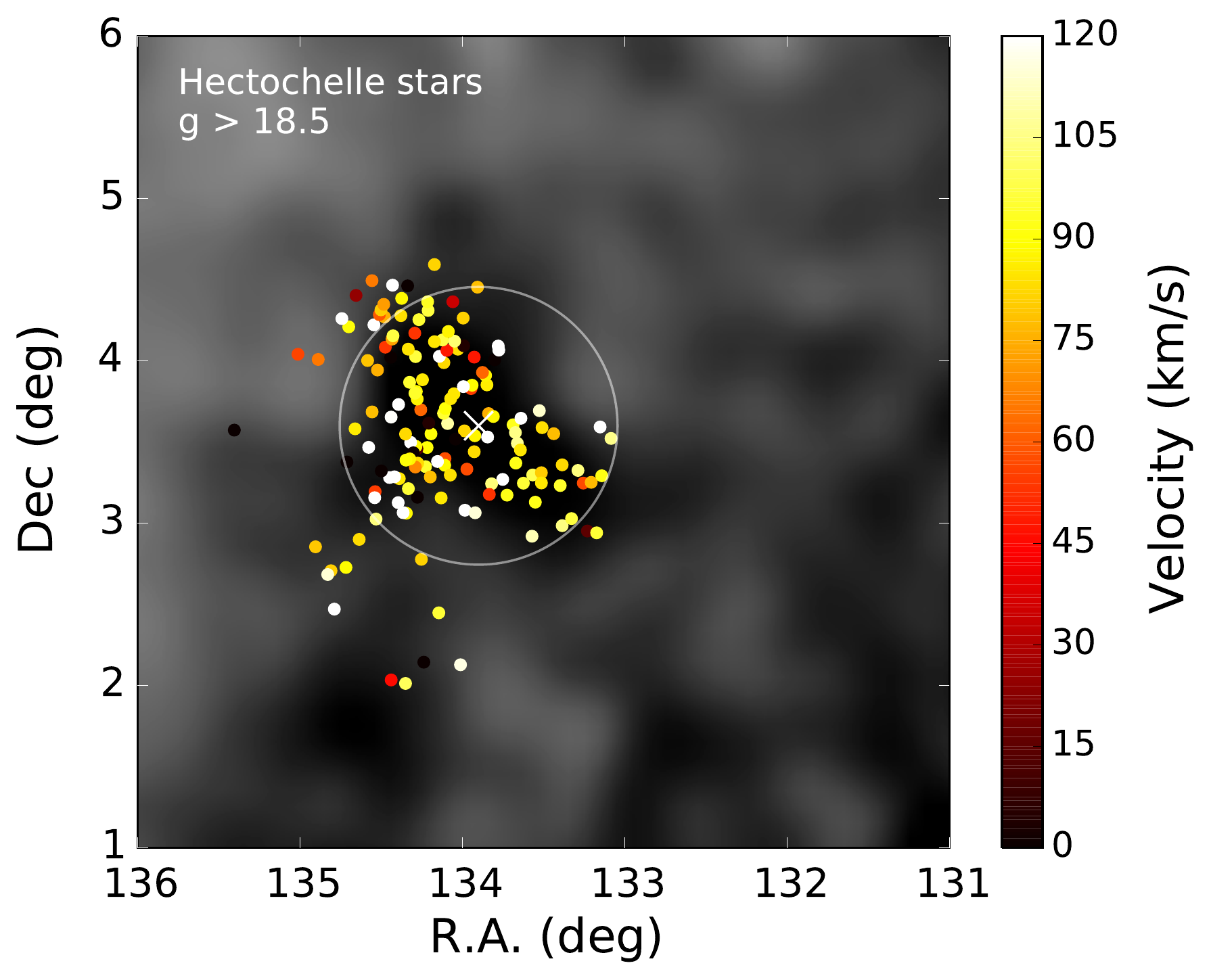}
\includegraphics[width=0.5\textwidth,trim={0cm 0cm 0cm 0cm},clip]{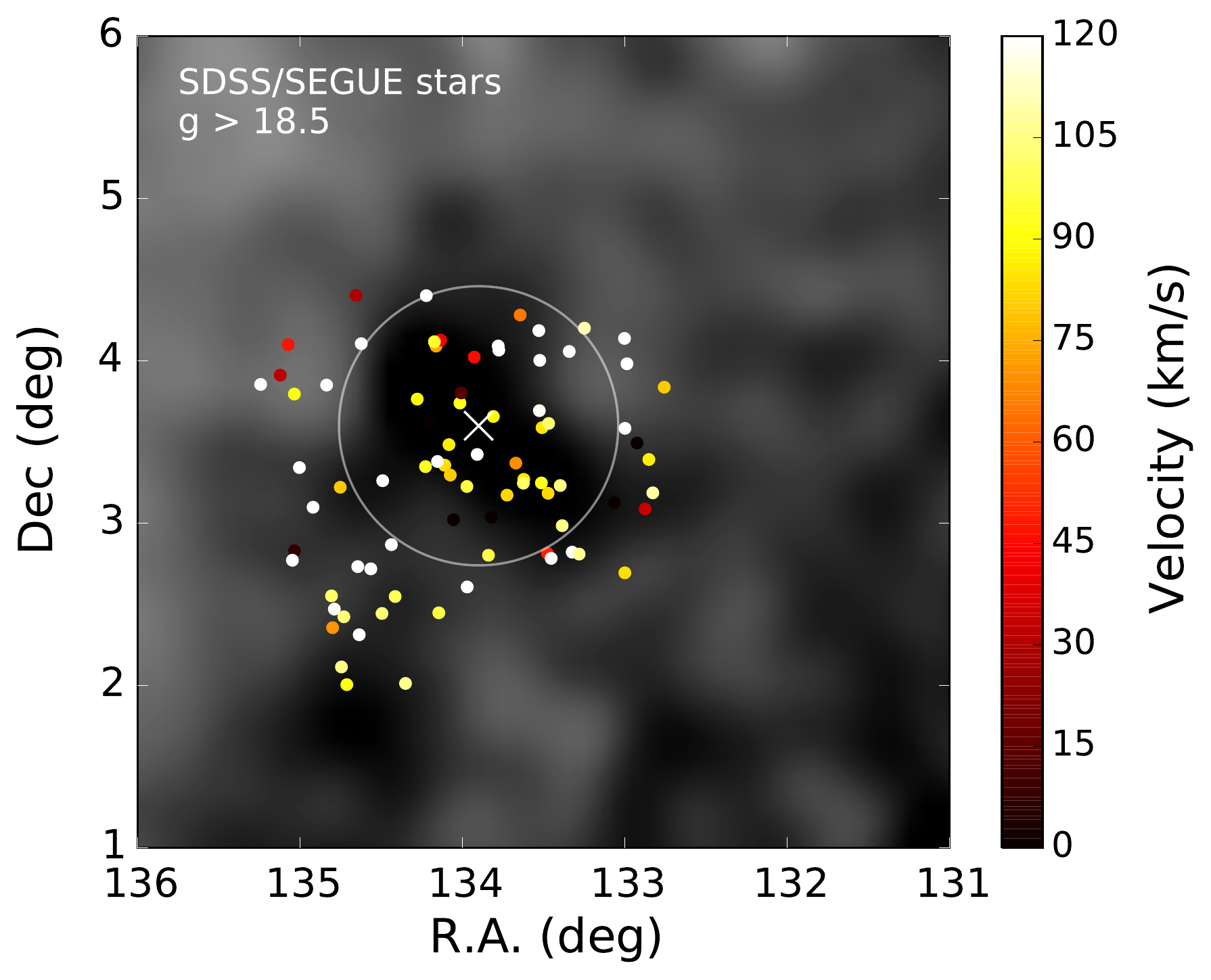}
\caption{Spatial positions of the Hectochelle/MMT sample (left panel) of targeted stars passing spectroscopic quality control cuts (see Section~\ref{sec:spectroscopy}) and the SDSS/SEGUE sample (right panel).  Positions are shown on the smoothed surface density map of the Hydra~I region (white boxed region in Figure~\ref{fig:spatial-ebs}).  The white circle ($r = 0.86$ deg) shows the $2.3$ sq. degree area centered on Hydra I corresponding to the DECam CMD shown in Figures~\ref{fig:decam-cmds} and~\ref{fig:sdss-cmds}.  Only stars fainter than $g = 18.5$ are shown for each sample.  Stars are colored by their measured radial velocities.  The white X shows the spatial center of Hydra I assumed in this study.}
\label{fig:spatial-hydra}
\end{figure*}

We model the Hectochelle spectra using the procedure introduced by \citet[hereafter W15]{walker2015} to model sky-subtracted Hectochelle spectra.   The spectral model is given as:

\begin{equation}
  M(\lambda)=P_l(\lambda)T\biggl (\lambda\biggl [1+\frac{Q_m(\lambda)+v_{\rm los}}{c}\biggr ]\biggr ),
  \label{eq:model}
\end{equation}

\noindent where $c$ is the speed of light and $P_l(\lambda)$ specifies the stellar continuum using an order-$l$ polynomial of the form

\begin{eqnarray}
  P_l(\lambda)\equiv p_0+p_1\biggl [\frac{\lambda-\lambda_0}{\lambda_s}\biggr ]+p_2\biggl [\frac{(\lambda-\lambda_0)}{\lambda_s}\biggr ]^2\nonumber\\
+\ldots+p_l\biggl [\frac{(\lambda-\lambda_0)}{\lambda_s}\biggr ]^l.
  \label{eq:polynomial1}
\end{eqnarray}

\noindent The factor of $T(\lambda[1+(Q_m(\lambda)+v_{\rm los})/c])$ is a continuum-normalized template spectrum that is redshifted according to the line-of-sight velocity ($v_{\rm los}$) and an order-$m$ polynomial of the form

\begin{eqnarray}
  Q_m(\lambda)\equiv q_0+q_1\biggl [\frac{\lambda-\lambda_0}{\lambda_s}\biggr ]+q_2\biggl [\frac{(\lambda-\lambda_0)}{\lambda_s}\biggr ]^2\nonumber\\
  +...+q_m\biggl [\frac{(\lambda-\lambda_0)}{\lambda_s}\biggr ]^m
  \label{eq:polynomial2}
\end{eqnarray}

\noindent that compensates for systematic differences between wavelength functions of target and template spectra (see W15 for details).  Following W15, we choose $m=2$, providing sufficient flexibility to fit the observed continuum shape and to apply low-order corrections to the wavelength solution.  We adopt scale parameters $\lambda_0=5220$ \AA\  and $\lambda_s=60$ \AA, such that $-1 \leq (\lambda-\lambda_0)/\lambda_s \leq +1$ over the entire range (5160 \AA\ - 5280 \AA\ ) considered in the fits. 

Again following W15, we generate template spectra $T$ using the synthetic library that was used for SSPP parameter estimation \citep{lee2008,lee2008b}.  The SSPP library contains rest-frame, continuum-normalized, stellar spectra computed over a grid of atmospheric parameters spanning $4000\leq T_{\mathrm{eff}}/\mathrm{K}\leq 10000$ in effective temperature (with spacing $\Delta T_{\mathrm{eff}}/\mathrm{K}=250$), $0\leq\log_{10}[g/(\mathrm{cm/s^2})]\leq 5$ in surface gravity ($\Delta\log_{10}[g/(\mathrm{cm/s^2})]=0.25$ dex) and $-5\leq \mathrm{[Fe/H]}\leq +1$ in metallicity ($\Delta\mathrm{[Fe/H]}=0.25$ dex).  The library spectra are calculated for a range in [$\alpha$/Fe]; however, this ratio depends on metallicity.  The library has $\alphafe=+0.4$ for spectra with $\feh < -1$, and then $\alphafe$ decreases linearly as metallicity increases from $-1\leq \feh< 0$, with $\alphafe=0$ for $\feh\geq 0$.  The synthetic library spectra are calculated over the range $3000-10000$ \AA\ at resolution $0.01$ \AA\ /pixel, which we degrade to $0.05$ \AA\ per pixel---twice as fine as our Hectochelle spectra.

Our spectral model, $M(\lambda)$, is specified fully by a vector of 13 free parameters.  In order to fit the model and obtain estimates for each parameter, we follow W15's Bayesian approach and adopt the same priors specified in their Table 2.  We use the software package MultiNest \citep{feroz2008, feroz2009} to sample the posterior probability distribution functions (PDFs).  As described in W15, the sampling of the PDFs allows one to evaluate the ``Gaussianity'' of the posterior PDFs by calculating the first four moments (mean, variance, skew, kurtosis) of the 1D marginalized PDFs of each parameter.  W15 note that in the high S/N regime, the posterior PDFs become increasingly Gaussian and therefore allow one to define a measure of quality control.  That is, the Gaussianity measure provides confidence that the calculated variance on the velocity is a good measure of the 1$\sigma$ standard deviation ($68\%$ confidence interval).  In practice, we keep only those spectra for which the model fits yield Gaussian shaped posterior PDFs as defined in W15 (see their Section 4.1 and Figure~4).  

\begin{figure}
\epsscale{1.2}
\plotone{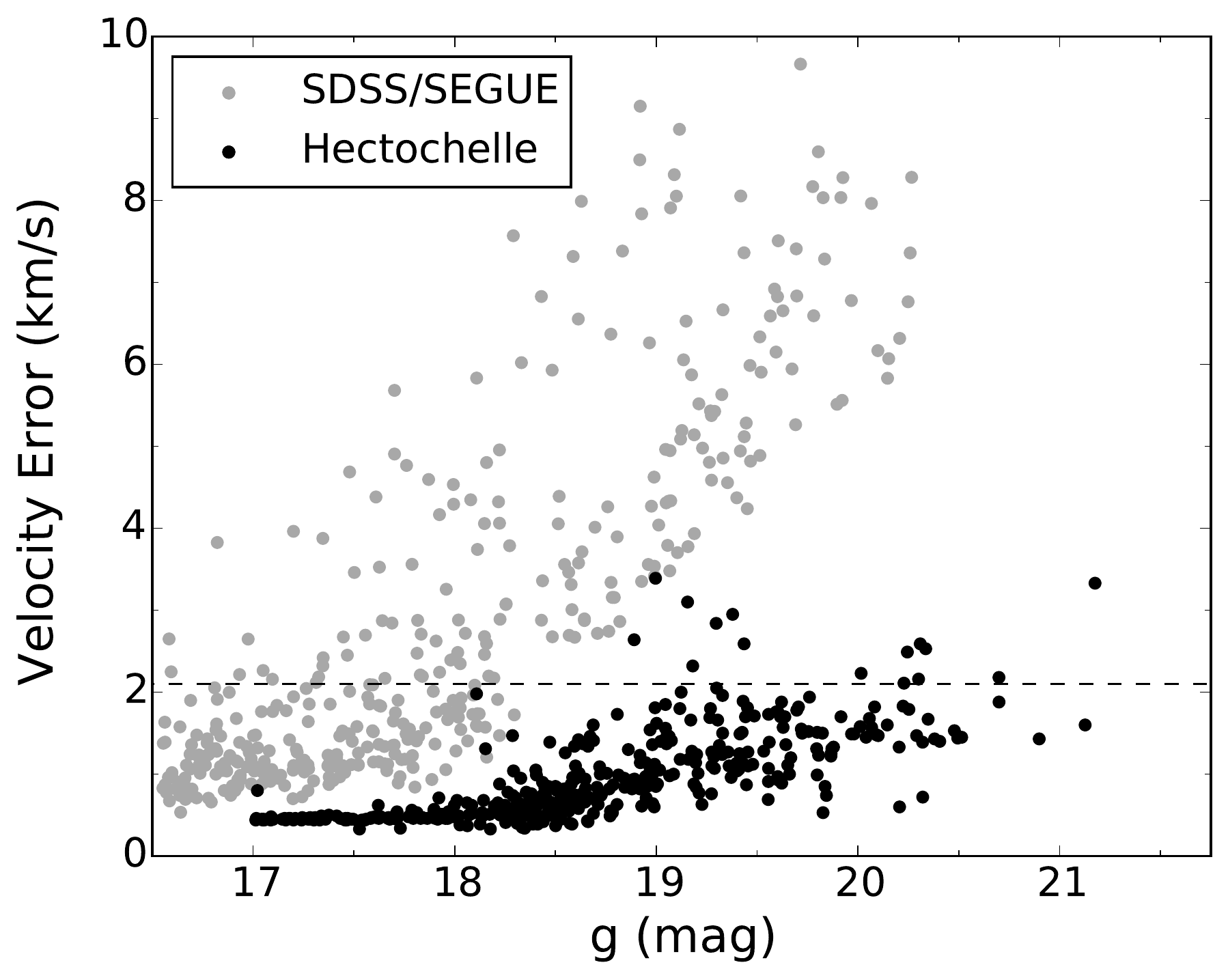}
\caption{Radial velocity error vs $g$ magnitude for the Hectochelle observations ({\it black points}) and archival SDSS/SEGUE data ({\it grey points}). Cross correlation failures resulting in large velocity errors are not shown. The dashed line indicates the velocity error cut (2.1~km s$^{-1}$) adopted in selecting the final, high-quality sample (see Section~\ref{sec:spectroscopy}).}
\label{fig:velocity-error}
\end{figure}

While our fits to the Hectochelle spectra return simultaneous estimates of stellar-atmospheric parameters $T_{\rm eff}$, $\log g$, and [Fe/H], we find that these quantities exhibit systematic as well as random errors that are larger than those found, using the same technique on spectra from the same instrument, by W15.  The reason for this difference is that our current spectra were acquired in relatively bright conditions and our sky-subtracted spectra suffer significantly more residual contamination by scattered sunlight.  Therefore, in this work we shall use only the velocities obtained from our Hectochelle spectra; for stellar-atmospheric parameters we will rely on the SDSS/SEGUE catalog.

We examined the data quality and measurement reliability using both repeat observations and the pipeline data quality measures described above. To ensure a reliable dataset, we adopted cuts on the radial velocity error (v$_{\rm err} <$ 2.1 km s$^{-1}$, or 3 times the median velocity error of $0.7$ km s$^{-1}$) and removed spectra with large residual contamination from scattered sunlight.   After applying these cuts, 411 stars remained in the Hectochelle sample.  These data are presented in Table~\ref{tbl:hecto} and include the DECam magnitudes (corrected for Galactic extinction).  The spatial positions of stars fainter than $g = 18.5$ are shown with respect to the smoothed spatial map of Hydra I in Figure~\ref{fig:spatial-hydra} (left panel).  The radial velocity error of the Hectochelle data as a function of magnitude is shown in Figure~\ref{fig:velocity-error}.   We obtain velocity errors less than $\sim2$~km s$^{-1}$ to a limiting magnitude of $g\sim~20.5$, approximately two magnitudes below the main sequence turnoff (MSTO) of Hydra~I.

\subsection{SDSS/SEGUE Spectroscopy}
\label{sec:sdss_spectroscopy}

We complemented our Hectochelle data with archival spectroscopy from SDSS, adopting the spectroscopic parameters derived from the SEGUE Stellar Parameter Pipeline \citep[SSPP; ][]{lee2008, lee2011}.  Unlike the MMT/Hectochelle target selection, the original SDSS survey spectroscopic selections were designed to meet a number of scientific goals.  We analyze the same SEGUE pointing as Schlaufman et al. (2009, 2011; B-8/PCI-9/PCII-21), but because the SDSS data are more heterogeneously selected, we applied the same isochrone cut in color and magnitude used to select the Hectochelle targets -- i.e., selecting only stars that might reasonably be Hydra~I members.  These data are presented in Table~\ref{tbl:sdss} and include
the DECam magnitudes (corrected for Galactic extinction).  Although SEGUE spectroscopic stars are a biased sampling of stars within this color-magnitude selection, they should still yield a unbiased measurement of its kinematic properties and a lower limit on its spread in [Fe/H] and [$\alpha$/Fe].  The right panel of Figure~\ref{fig:spatial-hydra} shows the spatial distribution of the SDSS/SEGUE stars fainter than  $g=18.5$.

\begin{figure*}[!ht]
\begin{center}
\includegraphics[width=.65\textwidth,trim={0cm 0cm 0cm 0cm},clip]{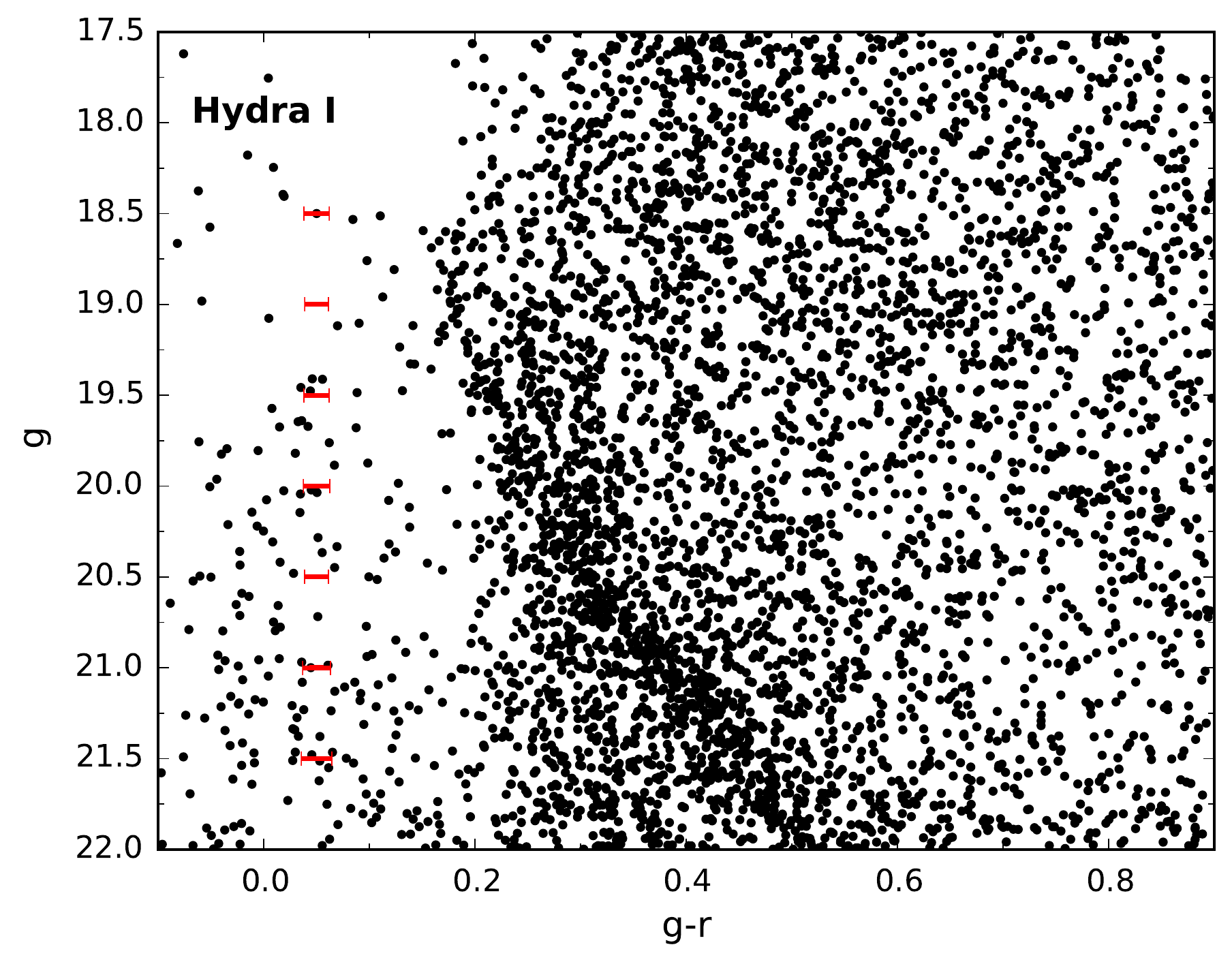}
\includegraphics[width=.65\textwidth,trim={0cm 0cm 0cm 0cm},clip]{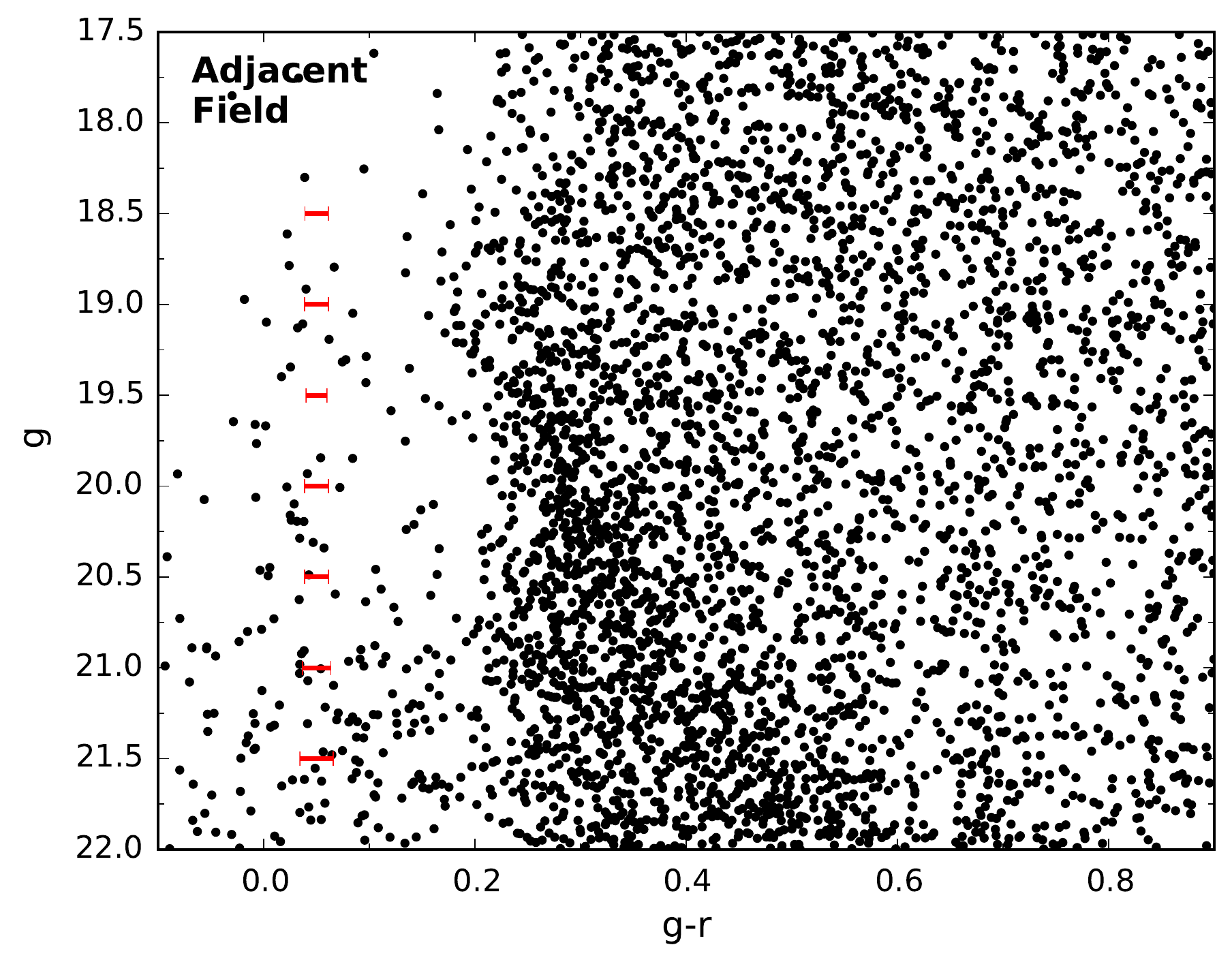}
\end{center} 
\caption{DECam color-magnitude diagram for stars within the $2.3$ sq. deg region ($r < 0.86$ deg) centered on Hydra I ({\it top}).  The CMD of a equal-area region adjacent to Hydra~I is shown for comparison ({\it bottom}).  The area of the Hydra~I CMD corresponds to the white circle shown in Fig~\ref{fig:spatial-hydra}.  The Hydra~I CMD shows a prominent MS with a relatively blue MSTO at $g \sim 18.9$, $g-r \sim 0.18$ mag.  The Hydra~I MS is less apparent in the adjacent field CMD, although the field does contain a significant number of stars similar to those at the MSTO in Hydra~I.
}
\label{fig:decam-with-background}
\end{figure*}

\begin{figure*}[!htb]
\includegraphics[width=.5\textwidth,trim={0cm 0cm 0cm 0cm},clip]{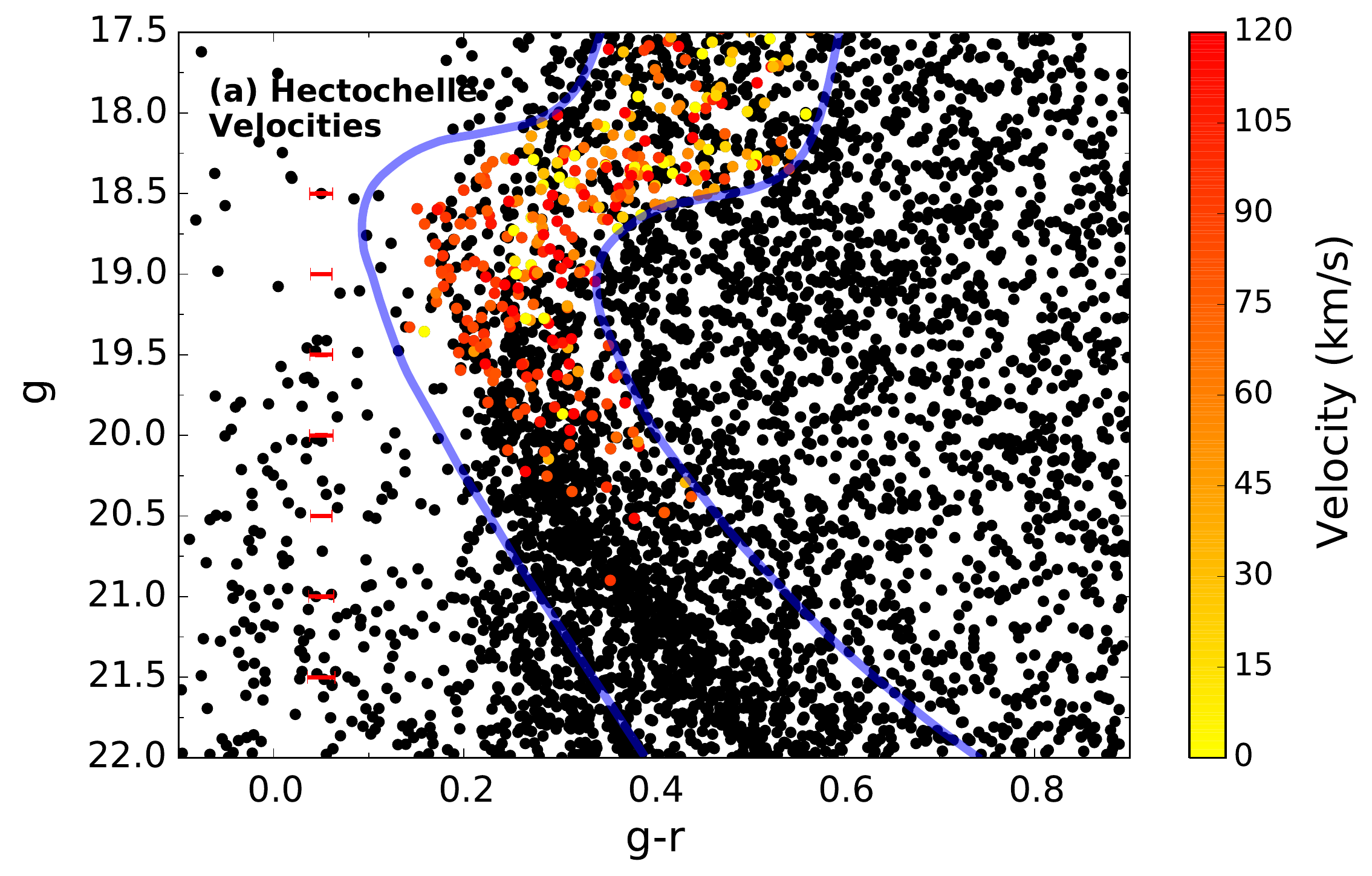}
\includegraphics[width=.5\textwidth,trim={0cm 0cm 0cm 0cm},clip]{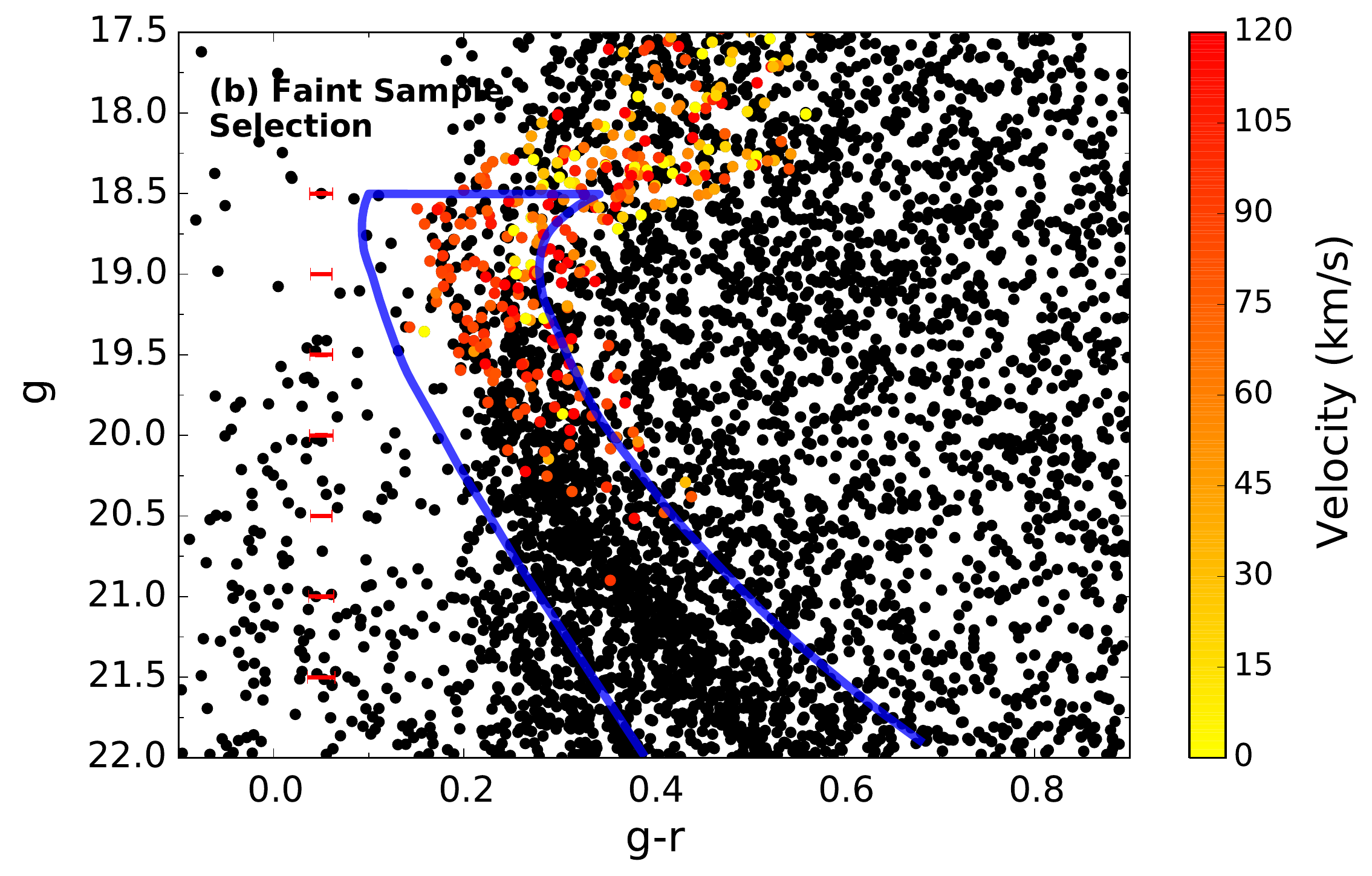}
\includegraphics[width=.5\textwidth,trim={0cm 0cm 0cm 0cm},clip]{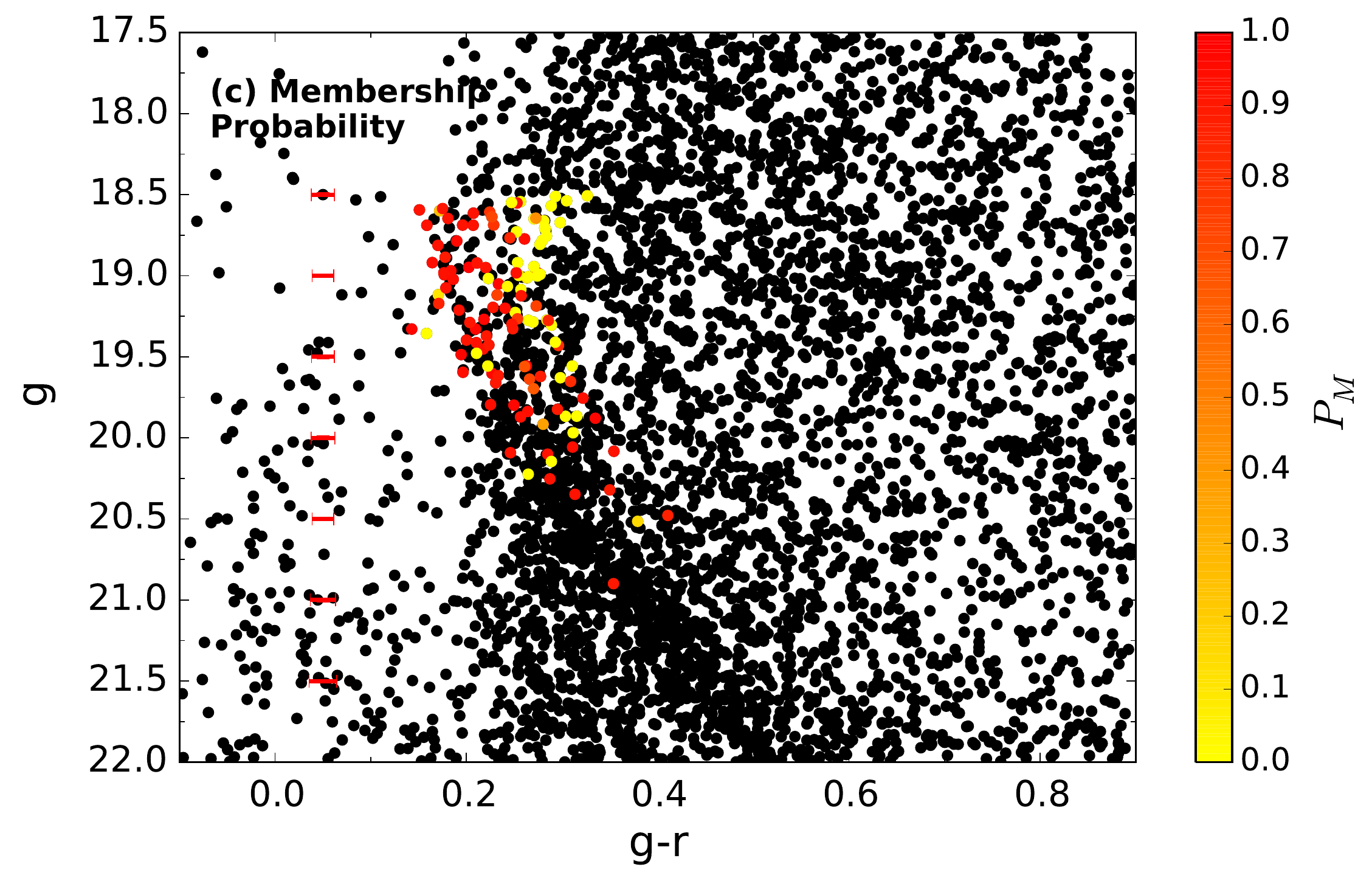}
\hspace{0.075in}\includegraphics[width=0.42\textwidth,trim={0 0 0 0},clip]{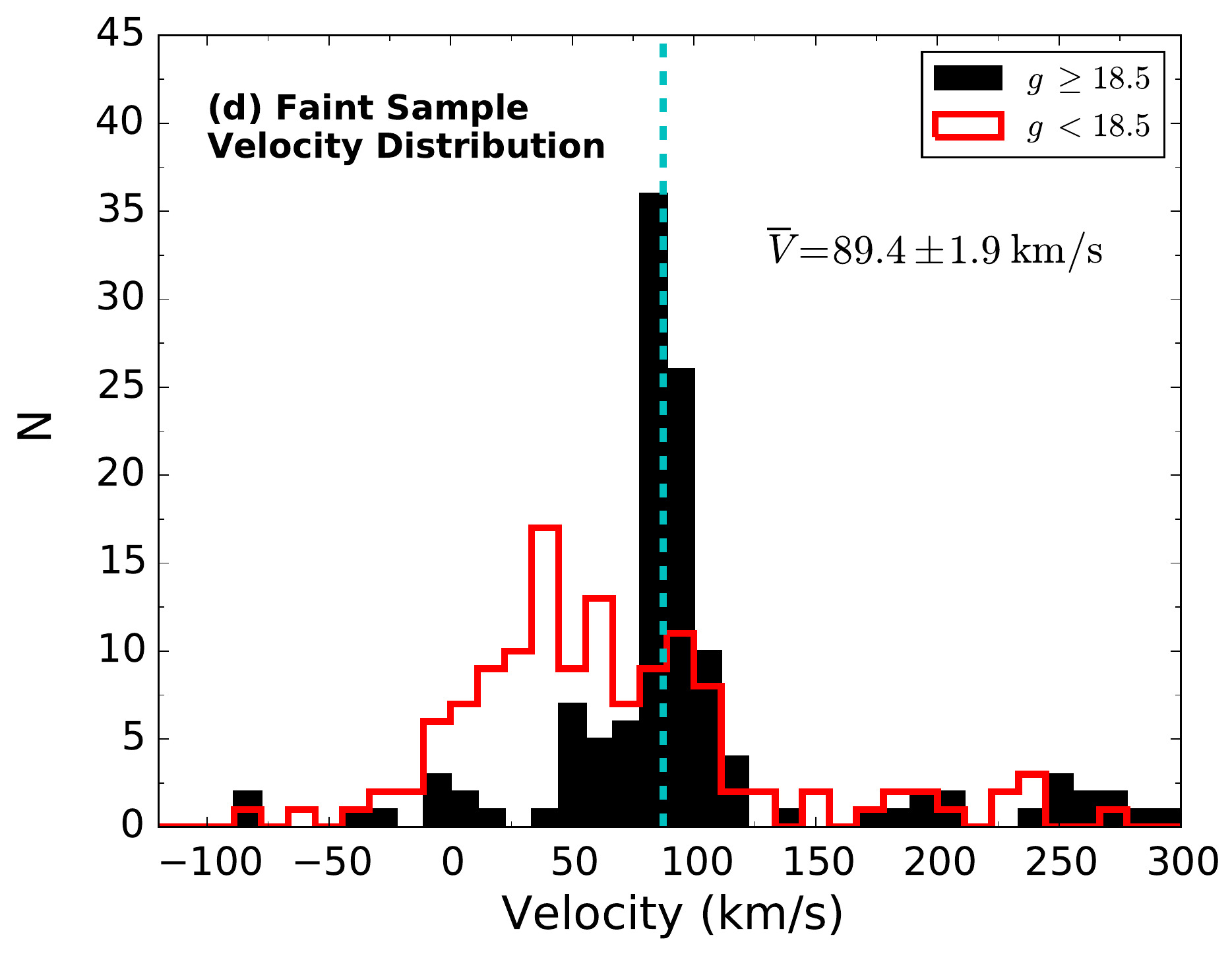}
\caption{DECam CMD of stars within a $2.3$ sq. deg region ($r < 0.86$ deg) centered on Hydra I (white circle in Fig~\ref{fig:spatial-hydra}).  Panel (a) shows all stars (passing the quality control criteria) targeted by Hectochelle/MMT in central region, where the points are colored by their measured velocity.  The blue lines show the SDSS CMD filter used to select observing targets (see also Figure~\ref{fig:targeting}).  Panel (b) shows the selection of the ``faint sample'' used for kinematic membership probability analysis (Sections~\ref{sec:phot_sample} and~\ref{sec:memberships}).  Panel (c) shows the derived membership probabilities $P_M$ for the faint sample as described in Section~\ref{sec:memberships}.  Panel (d) shows the heliocentric velocity distribution of the bright ({\it red histogram}) and faint ({\it black histogram}) Hectochelle samples.
}
\label{fig:decam-cmds}
\end{figure*} 

\section{Photometric and Spectroscopic Selection of Hydra I Stars}
\label{sec:sample}

The DECam photometry reveals a prominent MSTO feature in the Hess diagram (Figure~\ref{fig:targeting}) at $g \simeq 18.9$, $g-r \simeq 0.18$ mag which is not apparent in the SDSS CMD.  Because the DECam imaging covers a relatively large area (compared to the spatial extent of Hydra~I), we consider only  a small region of $2.3$ deg$^2$ (radius of $0.86$ deg) centered on Hydra~I in this study.  This central region is shown on the smoothed spatial map of Hydra~I in Figure~\ref{fig:spatial-hydra}.  Figure~\ref{fig:decam-with-background} shows the DECam CMDs of the central $2.3$ sq. deg region and an equal-area region adjacent to Hydra~I for comparison.  The comparison region was selected from the non-Hydra~I DECam imaging area located to the west of Hydra~I.  Only hints of the Hydra~I MSTO and MS features which are present in the central pointing can be seen in the adjacent pointing, demonstrating that Hydra~I is spatially concentrated.  However, because the adjacent comparison region is located relatively close to Hydra~I, this region may not be truly sampling the background.  We can therefore not rule out the possibility that the MSTO stars in the comparison region are simply EBS/Hydra~I member stars located slightly outside the primary spatial overdensities.

\subsection{Minimizing Contamination in the Spectroscopic Samples}
\label{sec:phot_sample}

\begin{figure*}[!htb]
\includegraphics[width=.5\textwidth,trim={0cm 0cm 0cm 0cm},clip]{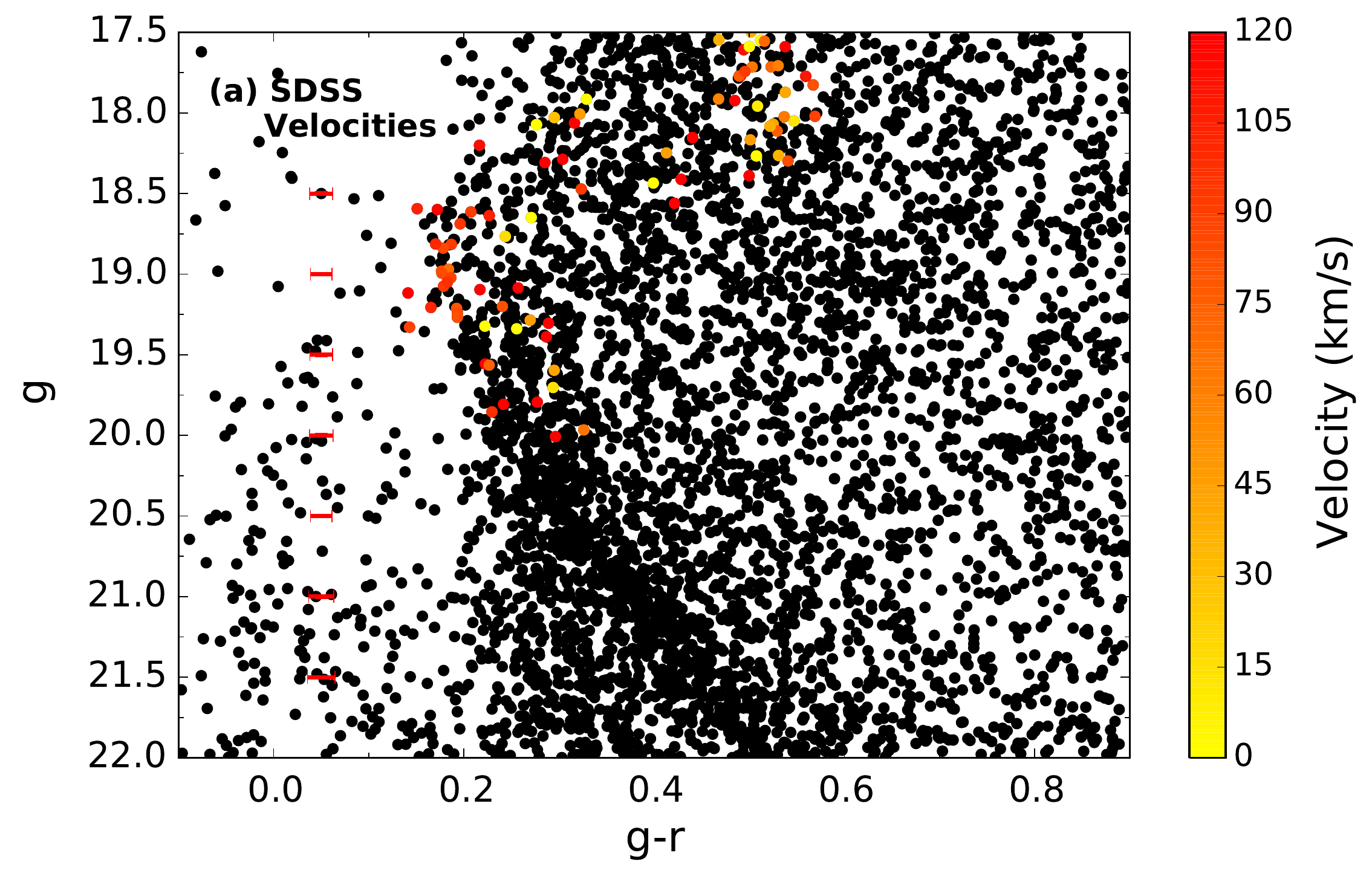}
\includegraphics[width=.5\textwidth,trim={0cm 0cm 0cm 0cm},clip]{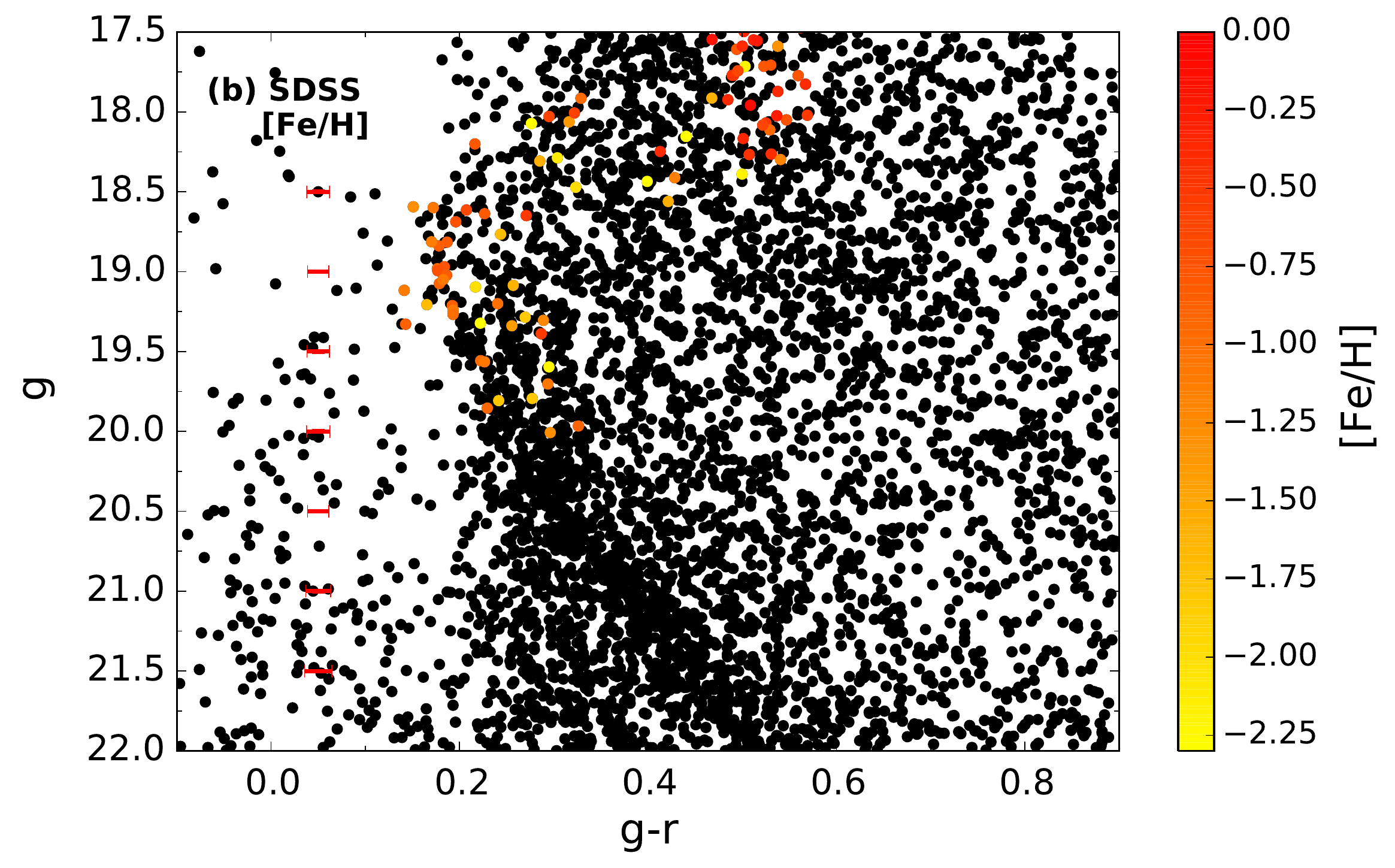}
\includegraphics[width=.5\textwidth,trim={0cm 0cm 0cm 0cm},clip]{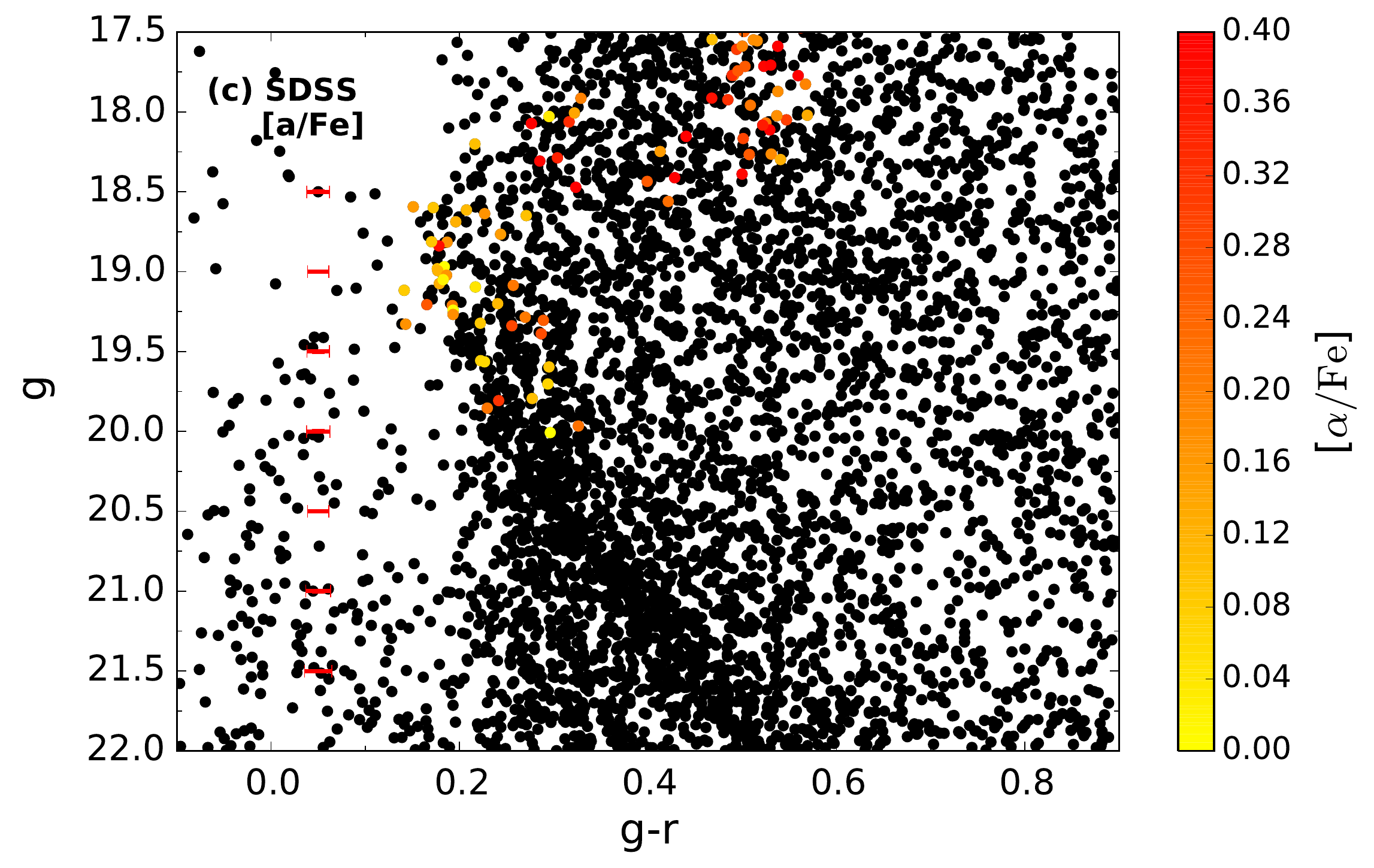}
\includegraphics[width=.5\textwidth,trim={0cm 0cm 0cm 0cm},clip]{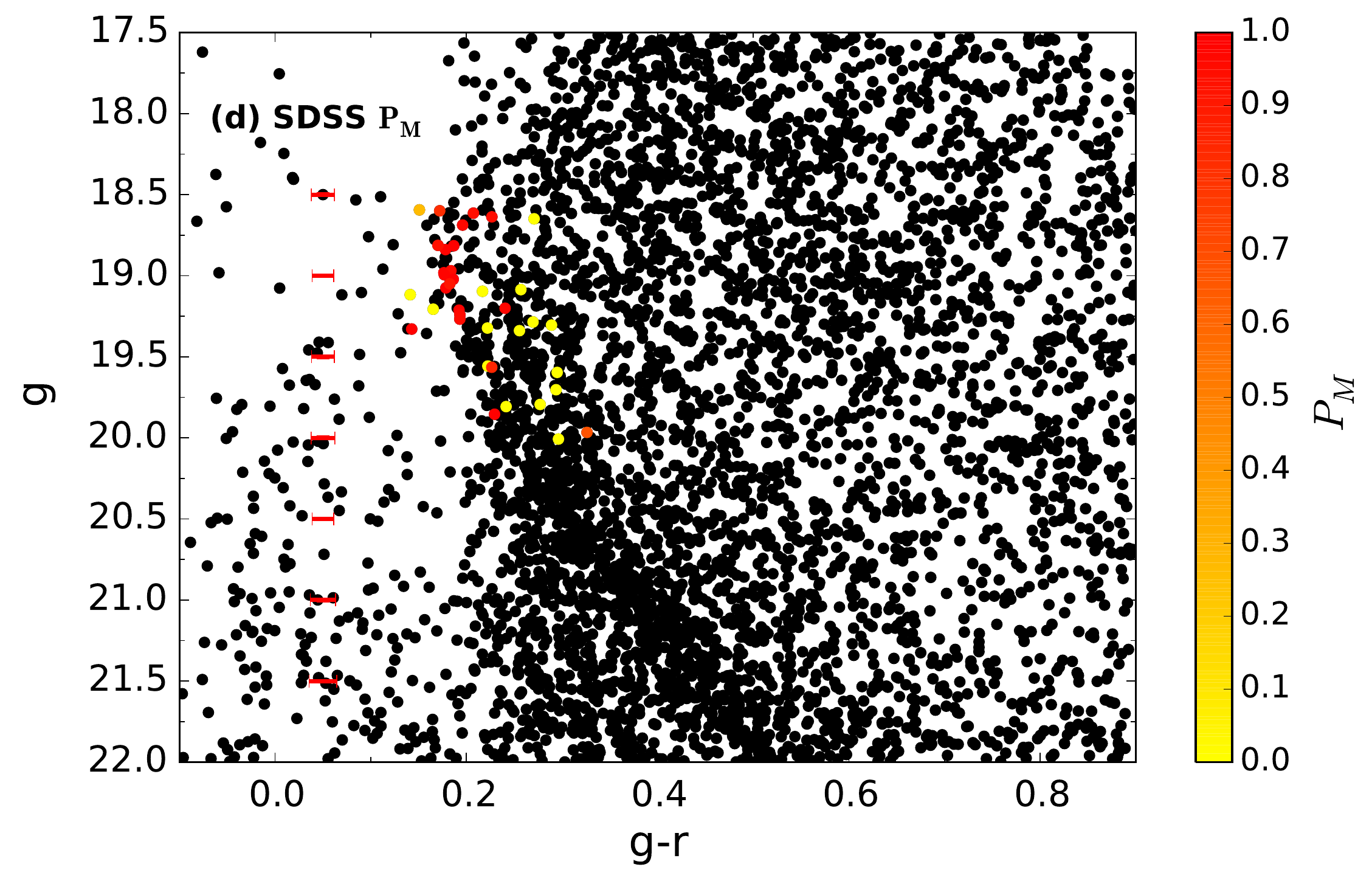}
\caption{Identical to Fig~\ref{fig:decam-cmds} but for the SDSS/SEGUE stars.  The measured velocities, iron abundances, and alpha abundances from the SEGUE SSPP pipeline (see Section~\ref{sec:sdss_spectroscopy}) are shown in panels (a), (b), and (c),  respectively.  The derived membership probabilities $P_M$ (see Section~\ref{sec:memberships}) are shown in panel (d).  We discuss the mean values and spread in [Fe/H] and [$\alpha$/Fe] for the high probability member stars in Section~\ref{sec:feh}.
}
\label{fig:sdss-cmds}
\end{figure*} 

The Hectochelle targeting employed a relatively wide, SDSS-based CMD filter, and given that we have clearly detected the MSTO of Hydra~I, we can remove likely contaminants using their location on the higher precision DECam CMD.  Stars with Hectochelle/MMT observations are shown in Figure~\ref{fig:decam-cmds} and are colored by their measured radial velocities.  The SDSS/SEGUE sample is shown on the same CMD in Figure~\ref{fig:sdss-cmds}, where stars are colored by their velocities (\ref{fig:sdss-cmds}a), iron abundances (\ref{fig:sdss-cmds}b), and alpha abundances (\ref{fig:sdss-cmds}c), respectively.  

The radial velocities of stars along the MS are quite similar ($\sim 90$ km s$^{-1}$; see Figs~\ref{fig:decam-cmds}a and~\ref{fig:sdss-cmds}a), whereas thick disk stars at brighter magnitudes show a larger velocity spread and  have higher iron abundances ($-0.5 < {\rm [Fe/H]} < 0$).  At fainter magnitudes ($g > 18.5$), the SDSS sample shows that both the iron abundance and velocity of MSTO stars are well correlated.  In contrast, stars at slightly redder colors  ($g-r \gtrsim 0.24$) show a wider spread in velocity and may be on average more metal-poor than MSTO stars.

We define a ``faint sample'' of Hectochelle stars using a narrower CMD filter shown by the blue dashed lines in Figure~\ref{fig:decam-cmds}b.  This filter has a red edge which is 0.06 mag bluer than our SDSS selection filter, minimizing contamination from redder stars ($g-r \gtrsim 0.24$).  The filter also removes thick disk stars at brighter magnitudes ($g < 18.5$), slightly brighter than the MSTO.   For our faint sample, we only include stars within the central $2.3$ sq. degrees centered on Hydra~I, resulting in a total of 122 stars.  Figure~\ref{fig:decam-cmds}d shows the velocity histogram of the Hectochelle faint sample.  For comparison we show stars at brighter magnitudes ($g < 18.5$) which would have passed the narrow CMD filter.  The contamination is readily apparent in the velocity histogram of bright stars, which only shows a small excess of stars at the measured systemic velocity of Hydra~I (${V_{rad}} \sim 89$ km s$^{-1}$).  By contrast, the faint sample shows that Hydra~I is a significant overdensity of stars in velocity space, consistent with the results of \citet{schlaufman2009}.

The SDSS faint sample is selected without the use of a narrower, bluer CMD filter due to the larger color measurement uncertainties.  Although this may slightly increase contamination, we use the SSPP [Fe/H] measurements as an additional parameter in our membership probability calculations (see Section~\ref{sec:memberships}).  We keep only stars fainter than $g=18.5$ and within the $2.3$ sq. degree region shown in Figure~\ref{fig:spatial-hydra}.  The SDSS faint sample contains a total of 42 stars.

\subsection{Spectroscopic Membership Probabilities}
\label{sec:memberships}

We determine membership probabilities using the Expectation Maximization (EM) method described in \cite{walker2009a}.  The EM method uses all available data to describe both the member and contamination distributions.  We assume a Gaussian velocity distribution, allowing the algorithm to return membership probabilities for each star along with the mean and variance of the member distribution.  The spatial probability distribution is described as a monotonically decreasing function of the stars' distance from an assumed center between the two primary lobes \citep[RA=$133.9^\circ$, Dec=$3.6^\circ$; see Fig.~\ref{fig:spatial-hydra};][]{grillmair2011}.  The EM algorithm (and the derived membership probabilities) implicitly assumes that there are no gradients in velocity, metallicity, velocity dispersion, or metallicity dispersion.  We use a Besancon model \citep{robin2003} to describe the Galactic contamination in our EM algorithm.  For the faint Hectochelle sample, we use only the spatial and velocity information to determine membership probabilities.  For the SDSS faint sample we also use the iron abundance as an additional parameter in determining memberships.  

\begin{figure}
\epsscale{1.25}
\plotone{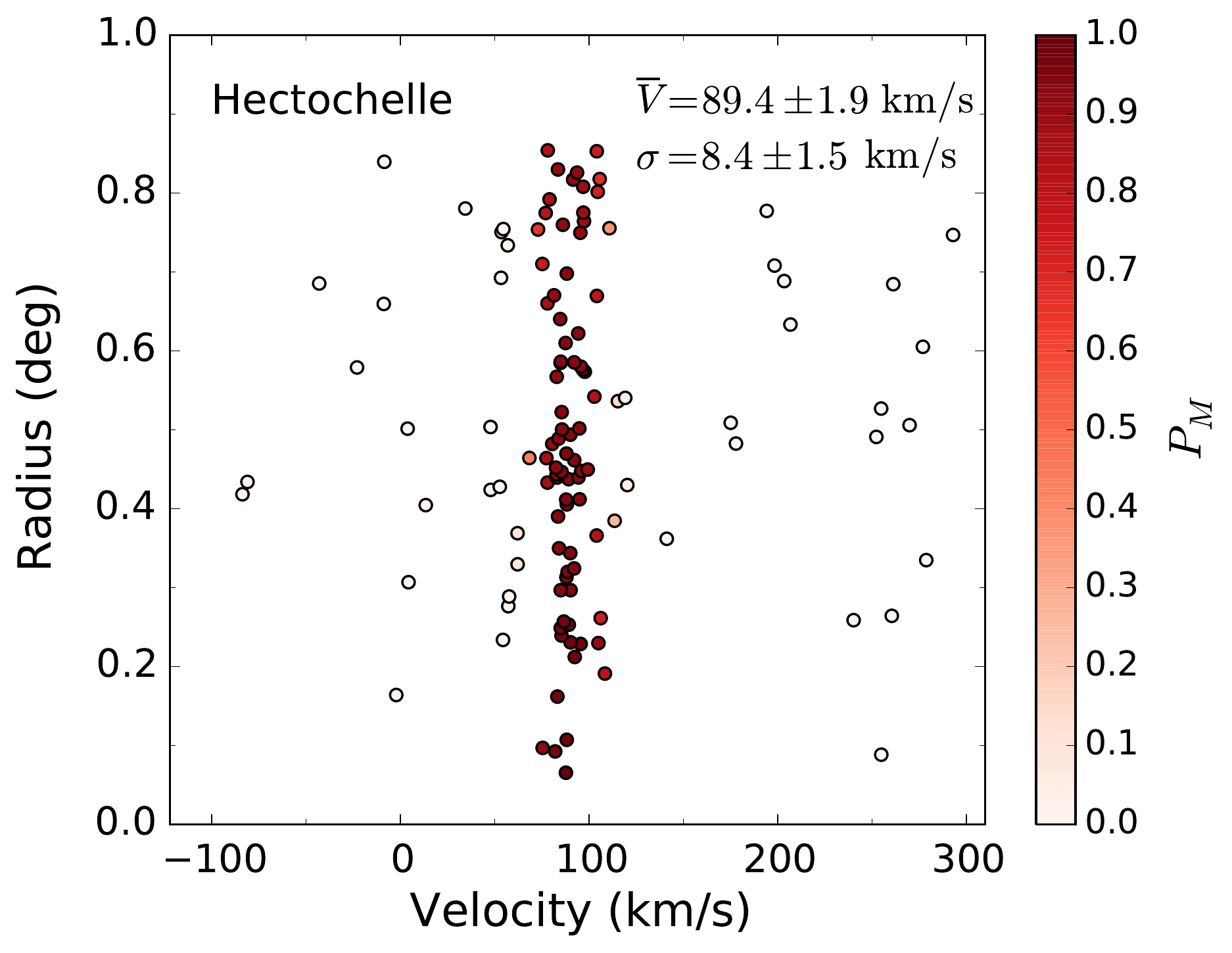}
\caption{Velocity versus radius for stars in the Hectochelle faint sample. Membership probabilities ($P_{M}$) were determined using velocity and spatial information with the EM method (Section~\ref{sec:memberships}).  The mean velocity and velocity dispersion are listed.} 
\label{fig:radius-velocity}
\end{figure}

\begin{figure*}
\begin{center}
\includegraphics[trim={0cm 0cm 0cm 0cm},clip,scale=0.46]{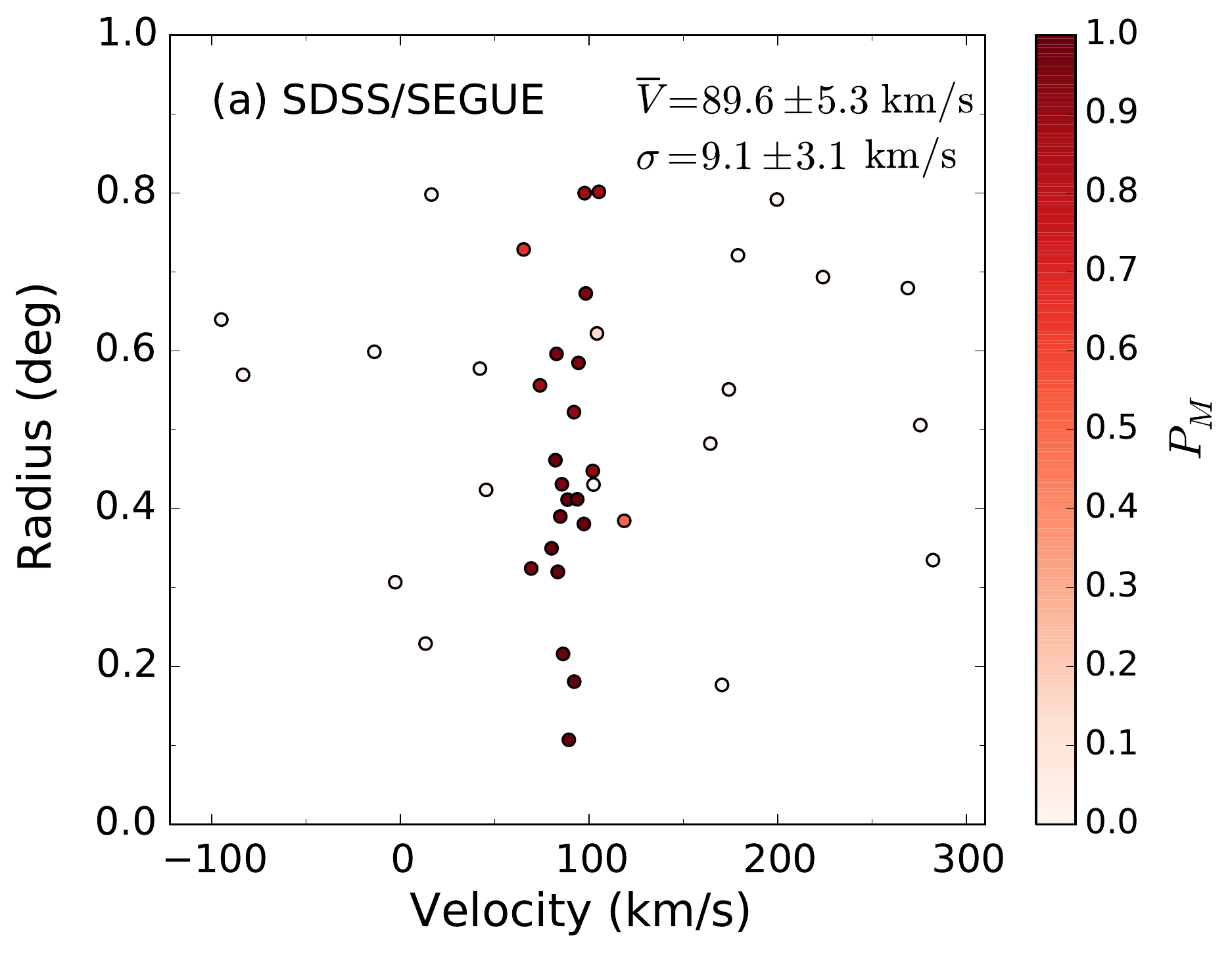}
\includegraphics[trim={0cm 0cm 0cm 0cm},clip,scale=0.46]{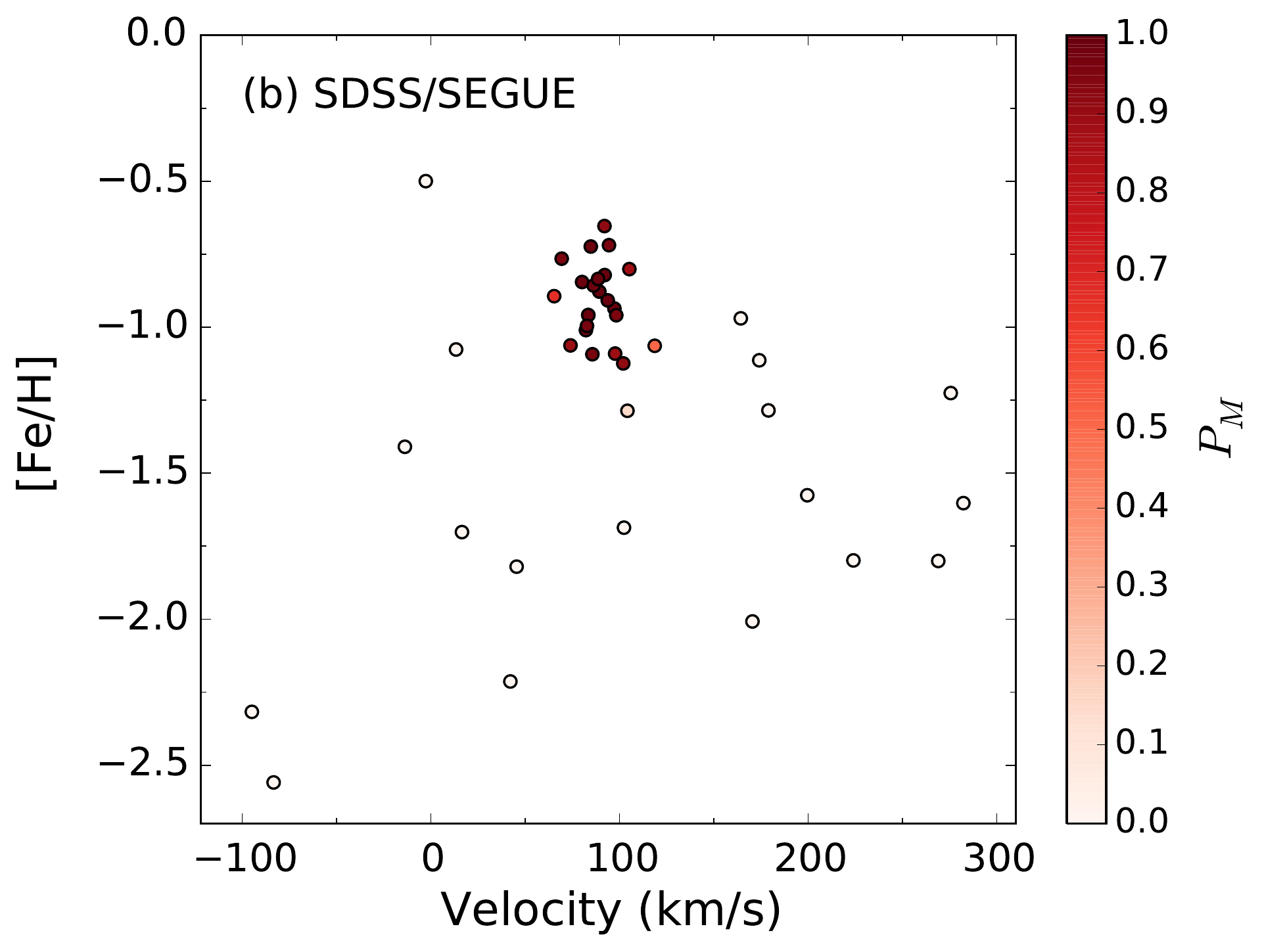}
\end{center}
\caption{Velocity versus radius (left panel) and [Fe/H] versus radius (right panel) for stars in the SDSS/SEGUE sample.  Membership probabilities ($P_{M}$) were determined using velocity, [Fe/H], and spatial information with the EM method (Section~\ref{sec:memberships}). The mean velocity and velocity dispersions are listed.  The mean and dispersion in the iron abundance are discussed in Section~\ref{sec:feh}.}
\label{fig:radius-velocity-sdss}
\end{figure*}

\begin{figure*}
\begin{center}
\includegraphics[width=0.49\textwidth,trim={0cm 0cm 0cm 0cm},clip]{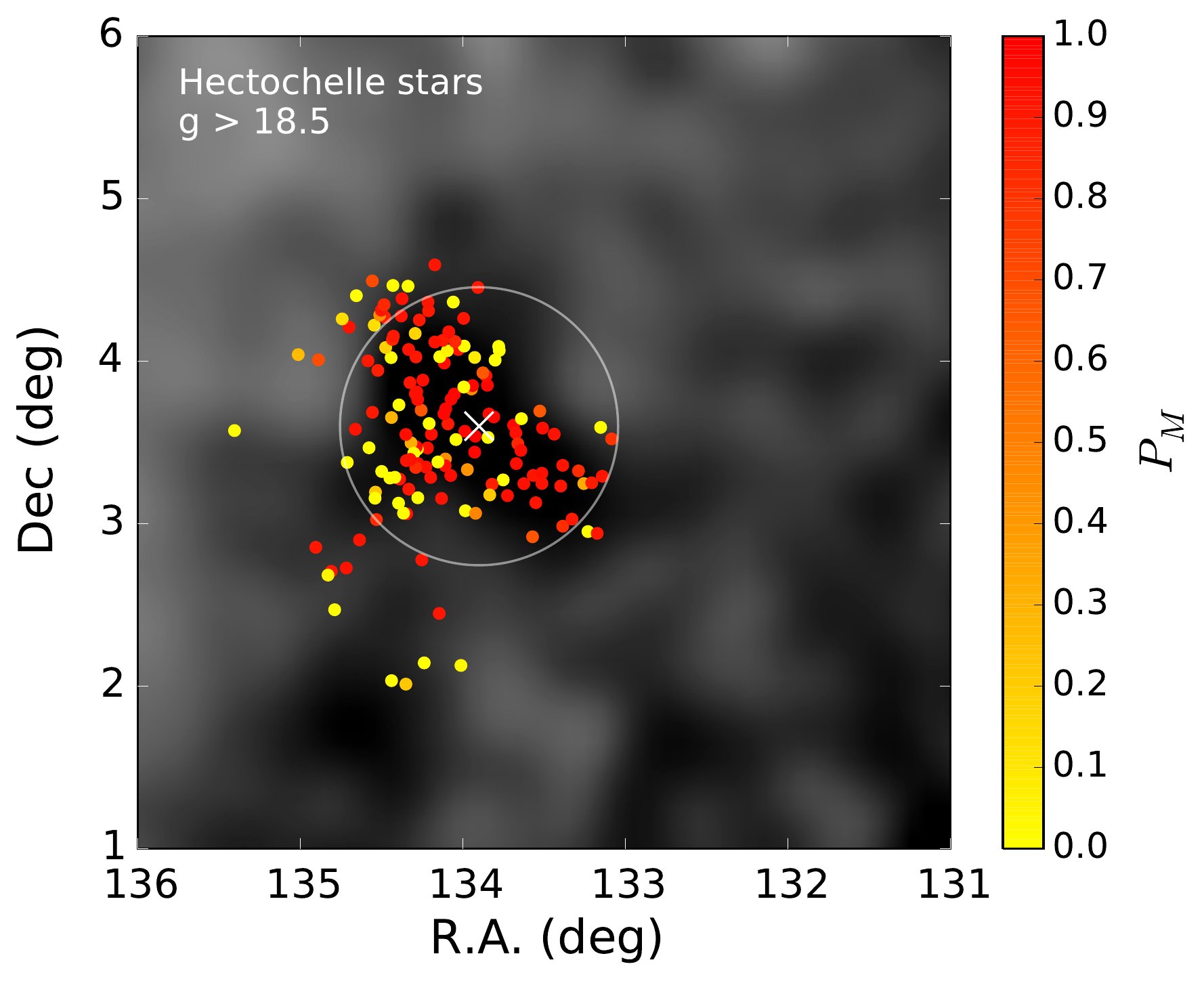}
\includegraphics[width=0.49\textwidth,trim={0cm 0cm 0cm 0cm},clip,scale=0.5]{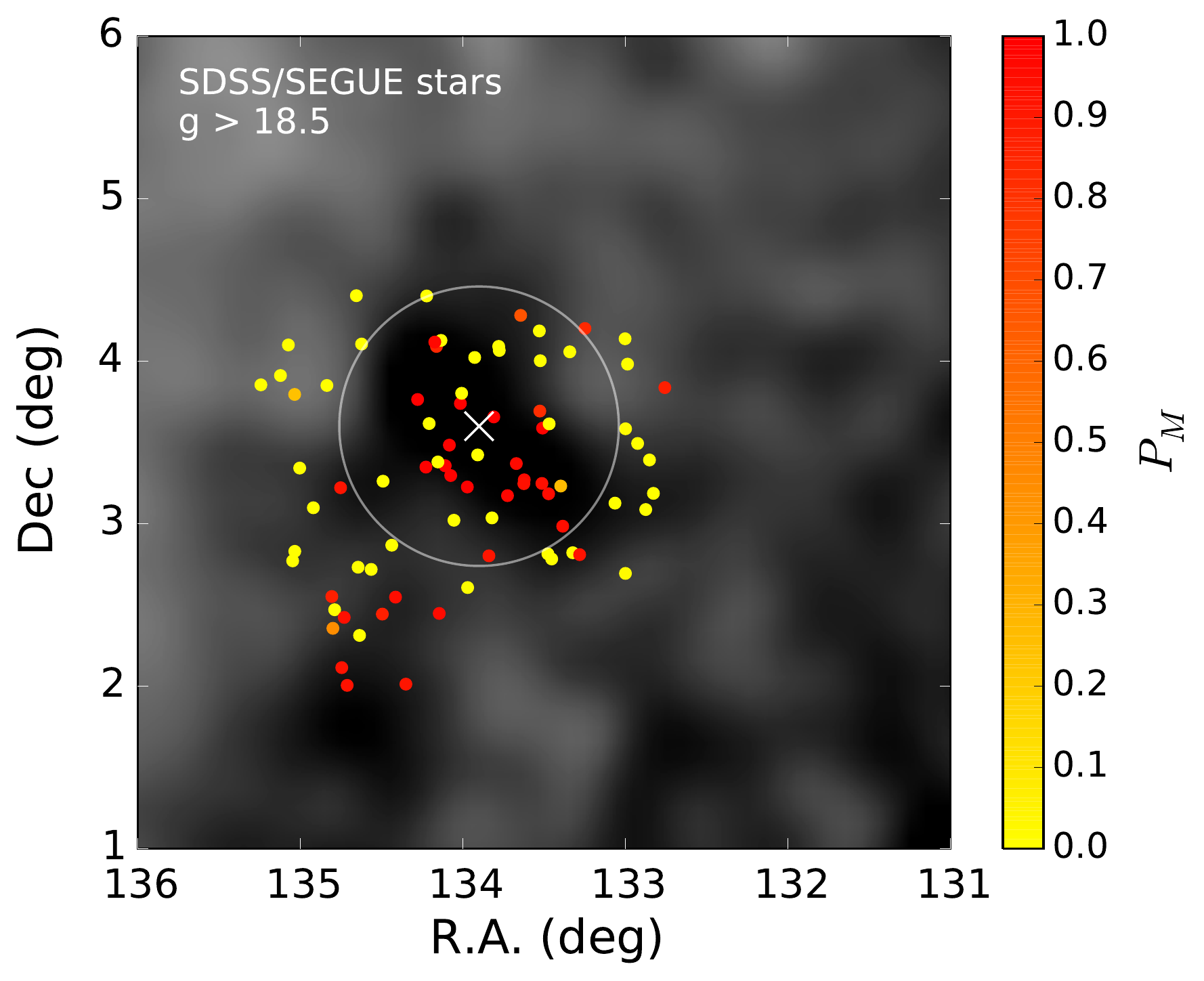}
\end{center}
\caption{Spatial distribution of the spectroscopic samples (identical to Figure~\ref{fig:spatial-hydra}) with stars colored by their EM membership probabilities ($P_{M}$).  Membership probabilities were determined as described in Section~\ref{sec:memberships} using velocity and spatial information; the SDSS sample uses [Fe/H] as an additional criteria for membership.  The SDSS data show a clump of high $P_M$ stars to the southeast of Hydra~I overdensity.  These stars fall along the extent of the EBS stream (see Figure~\ref{fig:spatial-ebs}; also \citealt{grillmair2011}).}
\label{fig:spatial-pmems}
\end{figure*}

The result of the EM algorithm applied to the Hectochelle faint and SDSS samples are shown in Figures~\ref{fig:radius-velocity} and~\ref{fig:radius-velocity-sdss}, respectively.  Each point is colored by its membership probability ($P_{M}$).  The EM algorithm infers a systemic heliocentric velocity of $\overline{V} = 89.4 \pm 1.9$ km s$^{-1}$ and dispersion $\sigma = 8.4 \pm 1.5$ km s$^{-1}$ for the Hectochelle faint sample, and $\overline{V_{SDSS}} = 89.6 \pm 5.3$ km s$^{-1}$ and $\sigma_{SDSS} = 9.1 \pm 3.1$ km s$^{-1}$ for the SDSS faint sample.  Errors on the mean and dispersion were determined via bootstrapping.  \citet{schlaufman2009} found a similar mean velocity of $85$ km s$^{-1}$ from their analysis of the SEGUE pointing coincident with Hydra~I and a higher velocity dispersion of $14.9$ km s$^{-1}$.  

The spatial positions of the Hectochelle and SDSS samples, colored by membership probabilities, are shown on the smoothed surface density map of Hydra~I in Figure~\ref{fig:spatial-pmems}.  The high probability members are well correlated with the photometric overdensities, although contamination from likely non-members is clearly present.  As also noted by \citet{grillmair2011}, the SDSS data show a clump of high $P_M$ stars to the southeast of Hydra~I.  Examining the larger EBS region in Figure~\ref{fig:spatial-ebs} shows that these stars fall along the southeast extent of the EBS stream.

\section{Results}
\label{sec:results}

\subsection{Limits on Rotation and Proper Motion}
\label{sec:pm-rotation}

The presence or lack of rotation in the Hydra~I system is an important clue to its nature and possible ongoing destruction.  To measure its rotation, we use a method similar to the one described by \cite{lane2009, lane2010a, lane2010b} and \cite{bellazzini2012} for stars in the Hectochelle faint sample and SDSS samples with membership probabilities $P_{M}>$~0.5.  In this method, the sample is divided in two by a line that passes through the center of Hydra~I, and a difference in the mean velocity of each subsample is taken.  This velocity difference is calculated for dividing lines with position angles (PA) ranging from 0$<$PA$<$360 deg.  The analysis of both samples show no evidence for rotation at a statistically significant level.  We find rotation amplitudes of $\sim 3-4$ km s$^{-1}$ with bootstrap-derived uncertainties of $\sim 2$ km s$^{-1}$.  We can therefore set a $3\sigma$ upper limit on the rotation of $\sim 6$ km s$^{-1}$.  

We use SDSS DR8 proper motions \citep[USNO B; ][]{munn2004,munn2008} to measure the ensemble proper motion of Hydra~I stars in the Hectochelle sample.  The typical, per star, random measurement uncertainty of this catalog is $\sim 4$ mas/yr \citep{munn2004}.  We find average proper motions of $\overline{\mu_{\alpha} \cos \delta} = -0.04\pm 0.57$~mas/yr ($2.4 \pm 34$ km s$^{-1}$) and $\overline{\mu_{\delta}} = -0.04\pm0.45$~mas/yr ($2.4 \pm 27$ km s$^{-1}$) from Gaussian fits weighted by the EM membership probabilities.  Using the uncertainties in proper motions, we set a $3\sigma$ upper limit to the proper motion of $\mu \sim 2.2$ mas/yr (132 km s$^{-1}$).  Although the measured proper motions are consistent with zero, they are very different than the proper motion of Hydra~I predicted by \citet{grillmair2011}: $\overline{\mu_{\alpha} \cos \delta} = -0.15\pm 0.57$~mas/yr, $\overline{\mu_{\delta}} = -2.67\pm0.45$~mas/yr for a prograde orbit.  A combination of increased numbers of stars and higher precision proper motions -- perhaps from the Gaia mission -- would be necessary for a refined proper motion measurement.

\subsection{[Fe/H] and [$\alpha$/Fe]}
\label{sec:feh}

The mean iron- and alpha-abundances, and possible spreads, are important diagnostics for distinguishing between a globular cluster or dwarf galaxy progenitor for the EBS \citep{willman2012,kirby2013}.  As noted in Section~\ref{sec:spectroscopy}, we use the SDSS sample to measure the iron and alpha-element abundances of Hydra~I stars, adopting the [Fe/H] and [$\alpha$/Fe] values from SSPP and membership probabilities determined using the EM algorithm. 

\begin{figure}
\epsscale{1.2}
\plotone{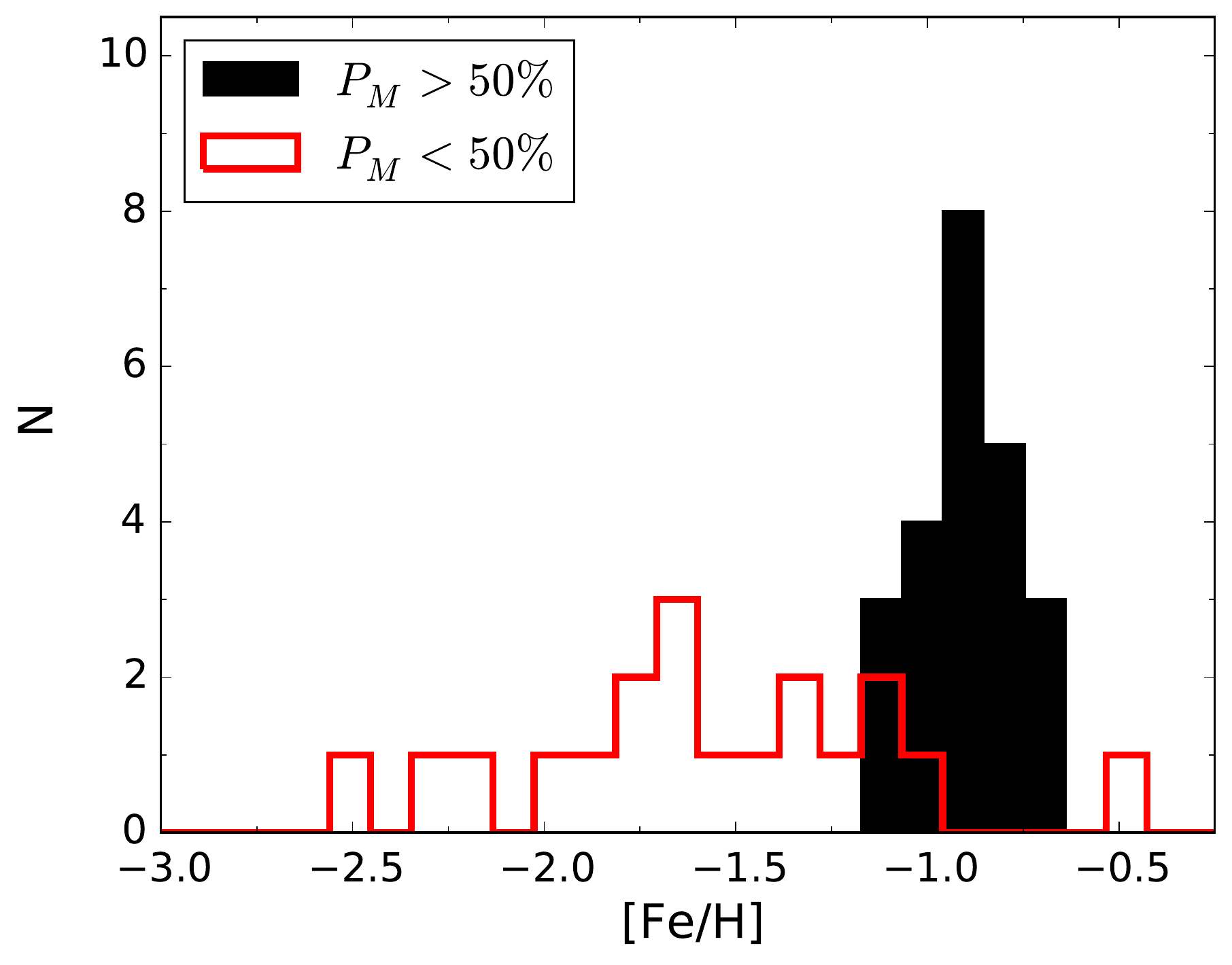}
\caption{Iron abundance histograms of the SDSS data for the probable member stars ($P_M > 0.5$; black histogram) and probable non-member stars $P_M < 0.5$; red histogram).  Using all stars, and weighting by membership probabilities, we find a mean [Fe/H] = $-0.91 \pm 0.03$ dex and $\sigma_{\rm [Fe/H]} = 0.13  \pm 0.02$ dex when adopting the formal SSPP [Fe/H] uncertainties.  Assuming more conservative errors on [Fe/H] measurements (see Section~\ref{sec:feh}) we find $\sigma_{\rm [Fe/H]} = 0.09\ \pm \ 0.03$ dex.}
\label{fig:sdss-feh}
\end{figure}

\begin{figure}
\epsscale{1.2}
\plotone{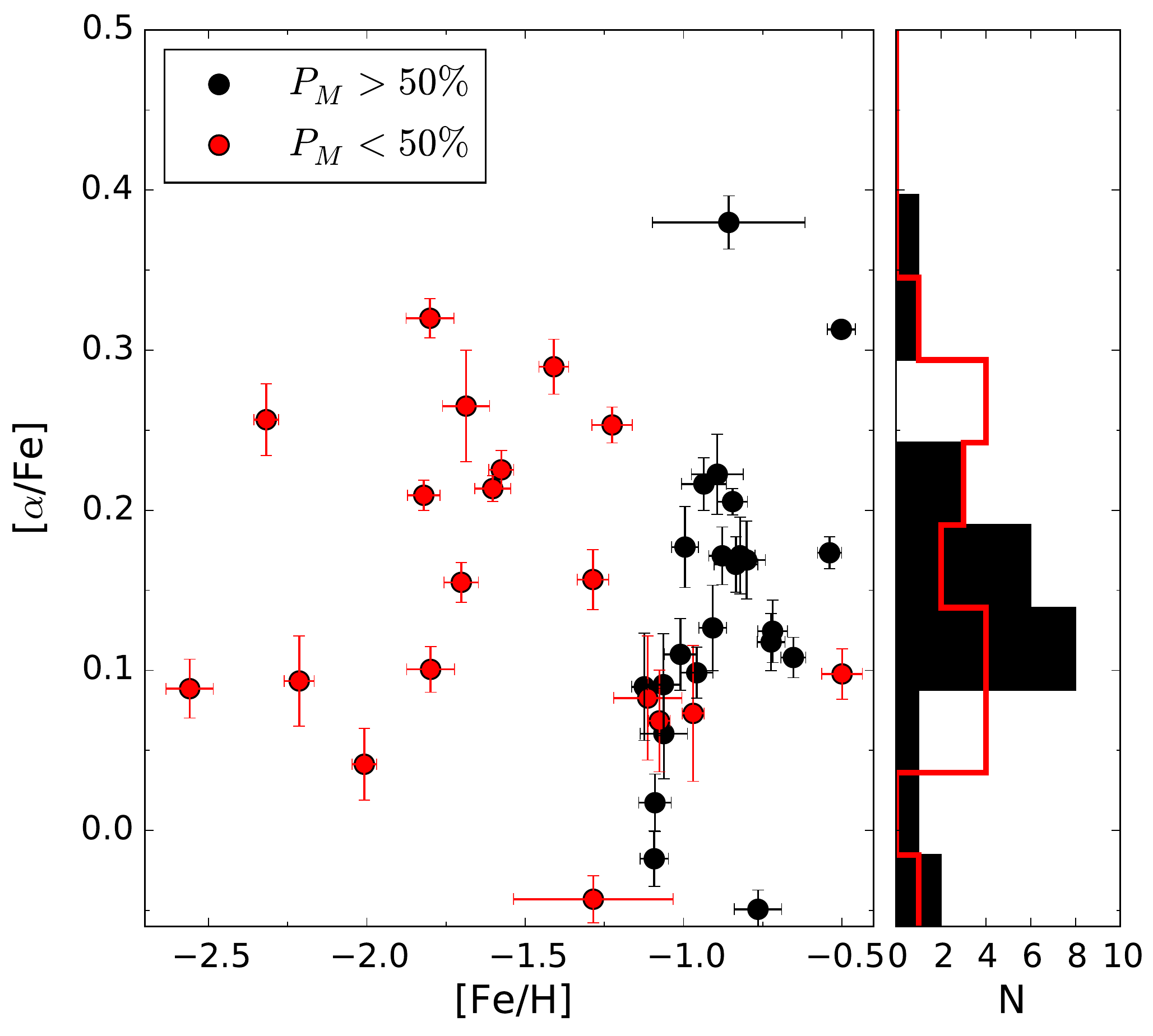}
\caption{[$\alpha$/Fe] vs [Fe/H] for the SDSS sample. Black points indicate likely members of Hydra~I ($P_M > 0.5$) and red points indicate likely non-members ($P_M < 0.5$).  The error bars show the reported internal uncertainties in the measurements as reported in the SSPP catalog.  Using all stars, and weighting by membership probabilities, we find a mean [$\alpha$/Fe]$ = 0.14 \pm 0.02$ and $\sigma_{ [\alpha/{\rm [Fe]}]} = 0.09 \pm 0.02$ when adopting the formal SSPP [$\alpha$/Fe] uncertainties.  Assuming more conservative errors on the [$\alpha$/Fe] measurements (see Section~\ref{sec:feh}), we find $\sigma_{[\alpha/{\rm Fe]}} = 0.08 \pm 0.02$.}
\label{fig:sdss-feh-afe}
\end{figure}

Figure~\ref{fig:sdss-feh} shows the [Fe/H] histograms for the SDSS sample.  Stars with $P_M > 50\%$ show a strong peak in iron abundance space, while lower probability stars show a wider spread in [Fe/H].  The alpha-element [$\alpha$/Fe] abundance measurements from SSPP \citep{lee2011} provide an additional diagnostic on the star-by-star membership, and possible origin, of Hydra~I.  Figure~\ref{fig:sdss-feh-afe} shows [$\alpha$/Fe] versus [Fe/H] for the SDSS sample.  The  [$\alpha$/Fe] distribution of high probability member stars peaks near $\sim0.1$ dex, whereas the low membership probability stars show a broad, non-peaked distribution of [$\alpha$/Fe] values.

We determine the mean and dispersions of the iron and alpha-element abundances using the full sample of SDSS stars, adopting the EM membership probabilities as weights.  We find mean abundances (with standard errors) of $\langle{\rm [Fe/H]} \rangle = -0.91 \pm 0.03$ dex and $\langle{\rm [\alpha/Fe]} \rangle = 0.14 \pm 0.02$.  We measure statistically significant dispersions in both abundances:  $\sigma_{\rm [Fe/H]} = 0.13 \pm 0.02$ dex and $\sigma_{\rm [\alpha/Fe]} = 0.09 \pm 0.02$ dex, respectively.  We discuss the implications of these measurements in Sections~\ref{subsec:dwarf} and~\ref{subsec:other}.

How sensitive are the measured spreads in the abundances to the individual measurement uncertainties?  To assess this, we recalculate the abundance spreads using larger measurement uncertainties of $0.1$ dex when the reported random errors on individual measurements are $<0.1$ dex.  Systematic comparisons to high resolution spectra and repeat SEGUE measurements have shown that the systematic errors on individual measurements are closer to $\sim 0.1$ dex, whereas the random errors on high S/N spectra are typically $\sim 0.05$ \citep{allende2008}.  With these conservative measurement uncertainties, we find $\sigma_{\rm [Fe/H]} = 0.09 \pm 0.03$ dex and $\sigma_{\rm [\alpha/Fe]} = 0.08 \pm 0.02$ dex.  The use of larger uncertainties still yield statistically significant dispersions, and the values themselves remain relatively unchanged at a spread of $\sim 0.1$ dex in the iron and alpha-elements.  We note that adopting errors larger than $\sim 0.12$ dex effectively erases evidence of an abundance spread.

\
\begin{figure*}
\begin{center}
\hspace{-0.8in}
\includegraphics[width=0.525\textwidth,trim={0cm 0cm 0cm 0cm},clip]{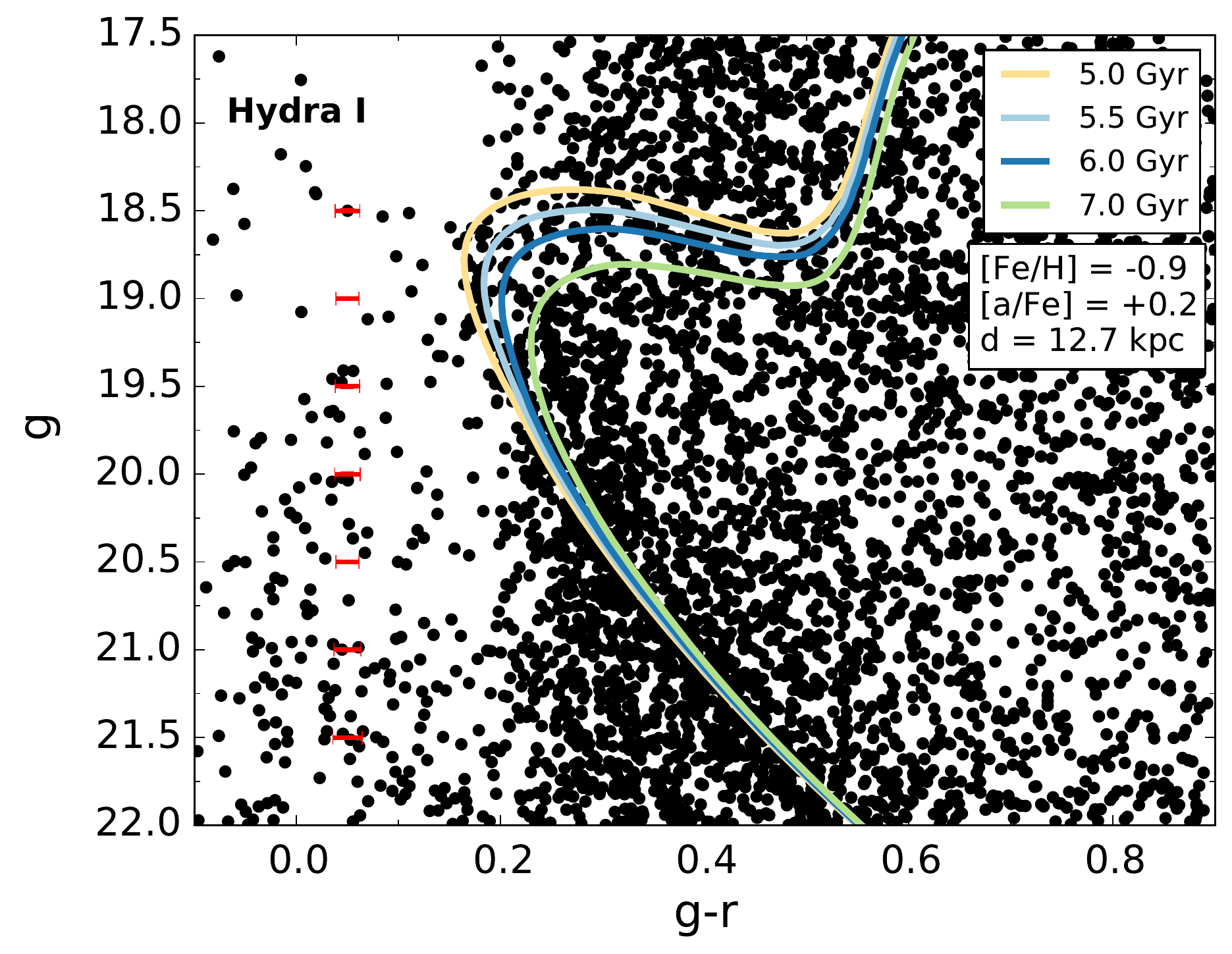}
\includegraphics[width=0.63\textwidth,trim={0cm 0cm 0cm 0cm},clip]{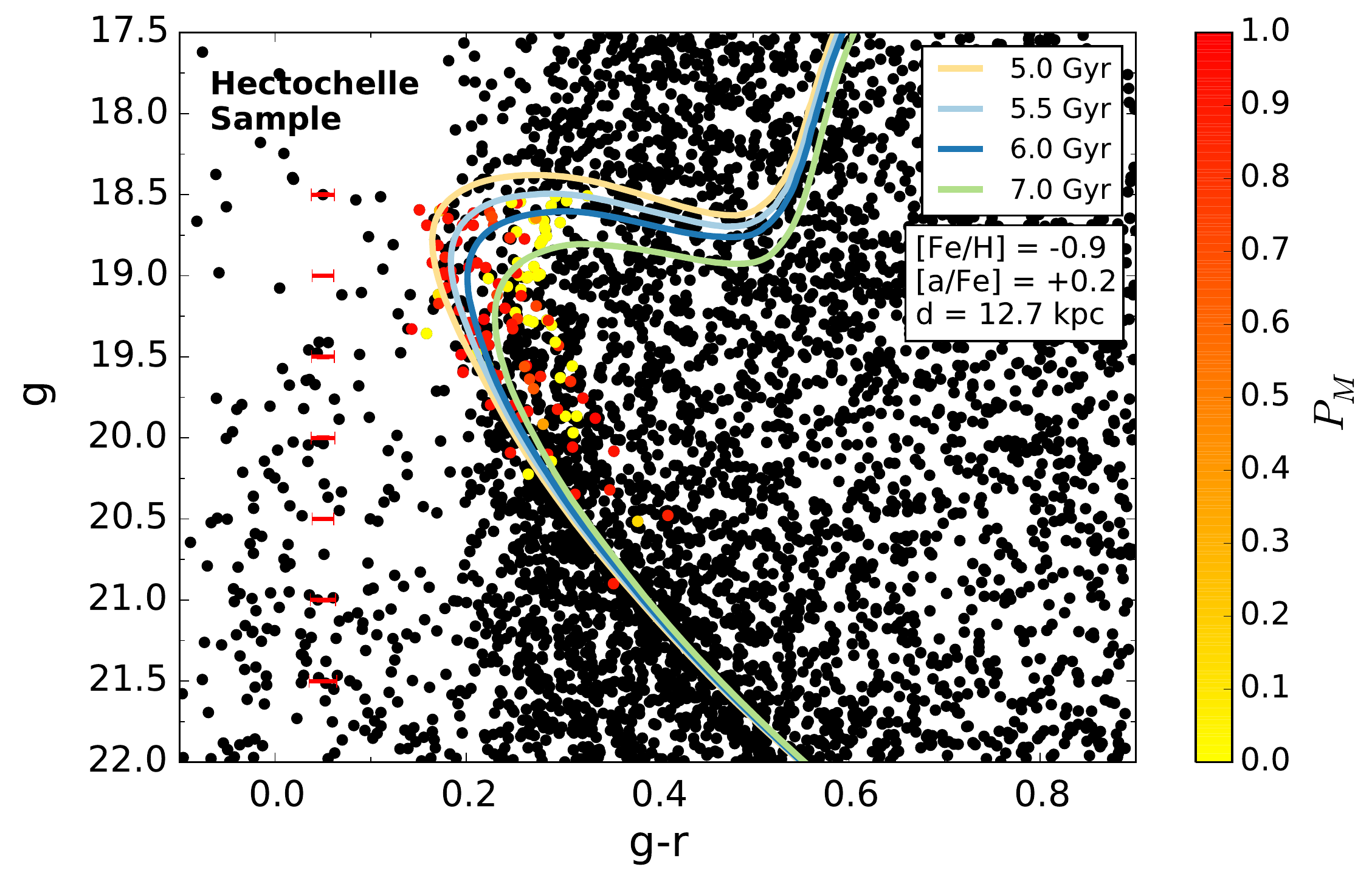}
\includegraphics[width=0.63\textwidth,trim={0cm 0cm 0cm 0cm},clip]{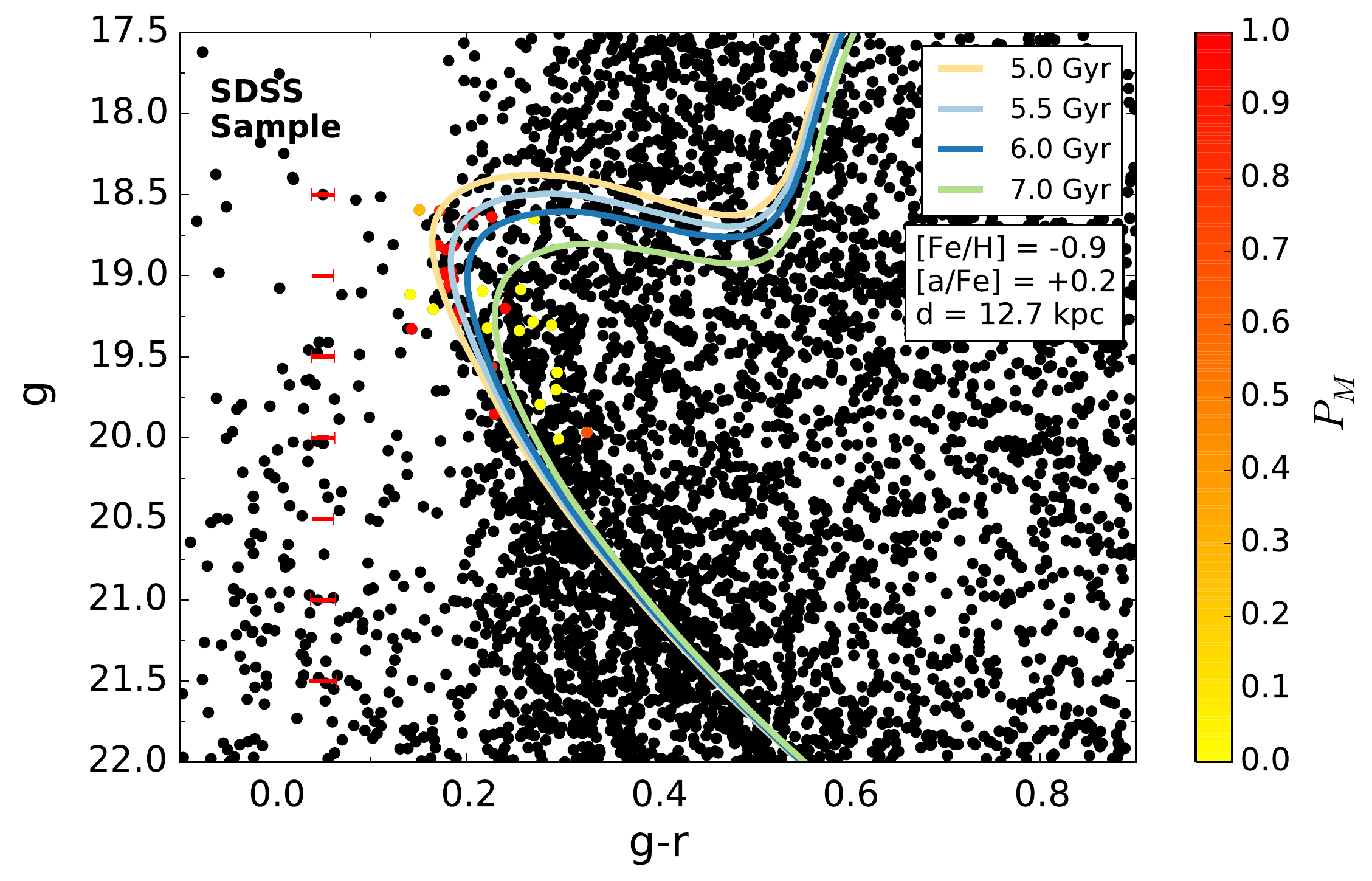}
\caption{Comparison of the Hydra~I CMD to theoretical isochrones.  The top panel reproduces the CMD of the central region shown in Figure~\ref{fig:decam-with-background}.  The middle and bottom panels show the Hectochelle faint and SDSS samples, respectively, colored by the derived membership probabilities $P_M$.  For the SDSS sample, membership probabilities are derived using the [Fe/H] measurements as a second parameter (see Section~\ref{sec:memberships}).  The median uncertainty in the color is shown as a function of magnitude.   Each panel shows four isochrones for ages 5, 5.5, 6, and 7 Gyr for fixed values of [Fe/H] = -0.9, [$\alpha$/Fe]=+0.2, at a distance of $d=12.7$ kpc \citep{dotter2008}.  Hydra~I contains a significant population of bluer stars consistent with ages of $\sim5-6$ Gyr (see Section~\ref{sec:stellar-pops}).
}
\label{fig:cmd-age}
\end{center}
\end{figure*}

\subsection{Stellar Populations and Age Estimates}
\label{sec:stellar-pops}

Given the measurement of a mean iron abundance of [Fe/H] = -0.9 dex from the SDSS sample, we explore the age range of Hydra~I's stellar populations using the observed CMD of spectroscopic members and theoretical isochrones.  Although we used a color-magnitude cut to select spectroscopic targets (see Figures~\ref{fig:targeting} and~\ref{fig:decam-cmds}), we are sensitive to MSTO stellar populations in Hydra~I as young as 4 Gyr (for [Fe/H] = -0.9 dex).

Figure~\ref{fig:cmd-age} shows the CMD of the Hectochelle and SDSS samples compared to isochrones with ages 5-6 Gyr, [Fe/H] = -0.9 dex, and [$\alpha$/Fe] = +0.2 dex \citep{dotter2008}.  We infer a distance of $d = 12.7 \pm 0.3$ kpc based on the prominent location of the MSTO in the CMD.  We find that fitting the blue MSTO population of Hydra~I stars ($g-r \lesssim 0.2$ mag) requires a stellar population with an age of 5-6 Gyr, assuming a mean iron abundance of [Fe/H] = -0.9 dex.  Changes of $\pm 0.1-0.2$ dex in the iron and alpha abundances result in older/younger ages by $\sim 0.5$ Gyr, which is below the age uncertainty of $\sim 1$ Gyr implied by the photometric uncertainties in the DECam $g-r$ colors at the MSTO.  The age of the $\sim 5-6$ Gyr stellar populations correspond to a formation at redshift $z \simeq 1-1.2$ for the standard $\Lambda$CDM cosmological model (\citealt{plank2013}).  We discuss the implications of intermediate age stellar populations in Hydra~I in Sections~\ref{subsec:dwarf} and~\ref{subsec:other}.  

The bluest stars in Hydra~I show good agreement with the $\sim5-6$ Gyr isochrones and are clearly inconsistent with a single old stellar population (age $\gtrsim 7$ Gyr).  The Hectochelle sample shows a number of high probability stars at colors redder than the blue MSTO ($g-r \gtrsim 0.2$), consistent with the expected colors of old, metal-poor MSTO stars.  However, a number of factors limit our ability to draw conclusions about the presence  of older stellar populations, including small numbers of candidate stars in both the Hectochelle and SDSS sample, as well as the limited/incomplete spectroscopic targeting of the reddest stars.  We explored the possibility of a old metal-poor MSTO using a background subtracted Hess diagram of the region but found no statistically significant evidence for such a feature.  Larger spectroscopic samples targeting the expect colors of older stars are necessary to further investigate the presence of complex stellar populations in Hydra~I.  We discuss the implications of the stellar populations analysis in Section~\ref{subsec:dwarf}.

\subsection{Estimated Hydra~I Luminosity}
\label{subsec:luminosity}

We estimate a total of $\sim$400 stars brighter than $g = 22$ in Hydra~I from a comparison of star counts in the overdensity to the adjacent field (Figure~\ref{fig:decam-with-background}).  As noted in Section~\ref{sec:decam}, we expect no issues with completeness at this magnitude limit.  We use the number count estimate, in combination with a luminosity function (LF) from \citet{dotter2008}, to place limits on the total luminosity of Hydra~I.  The LF of Hydra~I was created with a Salpeter IMF, an age of 5.5 Gyr, [Fe/H] = -0.9 dex, and [$\alpha$/Fe]= 0.2 dex. We calculated the total luminosity of Hydra~I by normalizing the LF with our value of 400 stars brighter than $g=22$. The normalized LF was then integrated to find the total luminosity of Hydra~I.  We determine that the object has a total luminosity of $\sim 1000 \pm 200 ~L_{\odot}$, or ${M_g} \simeq-2.4 \pm 0.3$ and ${M_r} \simeq -2.6 \pm 0.3$.  Our uncertainties are determined assuming Poisson statistics describe the number of star counts.  Using the SDSS-Johnson filter transformations of \citet{jordi2006}, we find ${M_V} \simeq -2.5\pm 0.3$ mag. 

\subsection{Constraints on Dynamical Mass and Tidal Radius}
\label{subsec:mass-dynamical}

The measured velocity dispersion of Hydra~I ($\sigma = 8.4 \pm 1.5$), in combination with a half-light or other characteristic radius, can yield an estimate of the dynamical mass of the system \citep{walker2009b,wolf2010}.  The spatial morphology of Hydra~I indicates that the object is likely undergoing significant tidal disturbance, so any measure of the dynamical mass is limited by the extent to which the present day stellar motions trace Hydra~I's underlying gravitational potential.  We discuss this assumption further in Section~\ref{subsec:dwarf}.  

Within the inner $r=0.86$ deg (190 pc) region of interest, we find a velocity dispersion $\sigma = 8.4$ km s$^{-1}$.  If we assume this radius correspond to the half light radius of Hydra~I, we can use Equation~2 from \citet{wolf2010} to estimate the dynamical mass within the half light radius, $M_{1/2}$.  We derive a dynamical mass estimate of $M_{1/2} = 9 \times 10^6$ $M_{\odot}$.  If this mass estimate is within two orders of magnitude of the true dynamical mass, then Hydra~I must have a significant dark matter component given the luminosity of $\sim$1000 L$_{\odot}$ derived in \ref{subsec:luminosity}. 

Given our stellar and dynamical mass estimates: Is the observation that Hydra~I appears to be experiencing stellar mass loss consistent with the expected tidal radius of the object at its distance from the Galactic center?  The tidal radius $r_{tidal}$ of an object can be approximated using Equation~2 from \citet{bellazzini2004}:

\begin{equation}
\label{eq:tidal-radius}
r_{tidal}  = \frac{2}{3}\left[ \frac{m_{\rm Hydra}}{2M_{\rm MW}(d_{\rm Hydra})} \right]^{1/3} d_{\rm Hydra},
\end{equation}

\noindent where $m_{\rm Hydra}$ is the mass of Hydra~I, $d_{\rm Hydra}$ is the galactocentric distance of Hydra~I ($\sim$19~kpc), and $M_{MW}(d_{\rm Hydra})$ is the Milky Way mass within the distance to Hydra~I.  We calculate the Milky Way mass $M_{MW}(d_{\rm Hydra})$ assuming an isothermal sphere with circular velocity of $V_{circ}$ = 220 $\pm$ 40 km s$^{-1}$ \citep{bellazzini2004}.  

We calculate the tidal radius ($r_{tidal}$) using both dynamical and stellar mass estimates.  For a dynamical mass of $9 \times 10^6$ $M_{\odot}$, we find tidal radius of $1.7$ deg ($\sim 370$ pc).  However,  assuming a stellar mass to light ratio $M_{*}/L = 1$ and $L= 1000~L_{\odot}$, we find a tidal radius of $r_{tidal} \sim$0.1 deg ($\sim 22$ pc).  

The instantaneous tidal radii of both the ``only stars" and the high dynamical mass scenarios appear to be consistent with our observations of Hydra~I.  The small tidal radius, expected if Hydra~I only contains stellar mass, is easily in good agreement with the spatial extent  ($\sim$ 1 deg radius) and asymmetric morphology of Hydra~I.  The larger tidal radius of $\sim 1-2$ deg, expected from the estimated dynamical mass, is consistent with mass loss at Hydra~I's edges (possibly resulting in the EBS stream).  It could also be consistent  with significant tidal disturbance within Hydra I's inner $\sim1-2$ degrees, given that the instantaneous tidal radius may exceed its orbit averaged tidal radius and that we may be overestimating the dynamical mass.  

If, however, the system is unbound, we can estimate the dissolution time for the system based on the crossing time -- the timescale for an unbound star to reach the observed extent of the object.  For simplicity, we assume a characteristic size of $1$ degree and that stars are moving at their measured velocity dispersion ($\sim 8$ km s$^{-1}$).  We find that the system would have a dissolution time of only $\sim 30 \times 10^6$ years.  Given such a short dissolution timescale, it would be surprising to observe this stellar overdensity at the present day if Hydra~I were truly unbound.  Although this provides general support for Hydra~I as a bound object, additional kinematic data and numerical modeling would be necessary to draw more detailed conclusions.

\section{Discussion}
\label{sec:discussion}


\begin{deluxetable}{lr}
\tablecolumns{2}
\tablecaption{{Properties of Hydra~I}}
\tablehead{
   \colhead{Parameter} & \colhead{Value} \\
   }
\startdata
    Right Ascension &  133.9 deg\\
    Declination & 3.6 deg\\
    Galactic Longitude & 224.7 deg\\
    Galactic Latitude & 29.1 deg\\
    Distance & $12.7 \pm 0.3$ kpc \\
    $L$ & $\sim 1000 \pm 200$ $L_\odot$ \\
    $M_V $ & $\sim -2.5 \pm 0.3$ mag \\
    Age\footnote{Youngest stellar populations only (see Section~\ref{sec:stellar-pops}).  } & $\sim5-6$ Gyr  \\
    $V_{\rm helio}$ \footnote{\label{fn:hect}Derived from the faint Hectochelle sample.}   & $89.4 \pm 1.4$ km  s$^{-1}$\\
    $V_{\rm GSR}$ \textsuperscript{\ref{fn:hect}} & $-55.5 \pm 1.4$ km s$^{-1}$ \\
    $\sigma_V$ \textsuperscript{\ref{fn:hect}} & $8.4 \pm 1.5$  km s$^{-1}$ \\
    Rotation Amplitude\textsuperscript{\ref{fn:hect}} & $\lesssim 6$ km s$^{-1}$ \\
    $\langle{\rm [Fe/H]}\rangle$ \footnote{\label{fn:sdss}Derived from the SDSS sample.}  & $-0.91 \pm 0.03$  dex \\ 	
    $\sigma_{\rm [Fe/H]}$\textsuperscript{\ref{fn:sdss}}        &  $0.13 \pm 0.02$ dex\\
    $\langle{\rm [\alpha/Fe]}\rangle$\textsuperscript{\ref{fn:sdss}}  & $0.14 \pm 0.02$  dex\\
    $\sigma_{\rm [\alpha/Fe]}$\textsuperscript{\ref{fn:sdss}}      &  $0.09 \pm 0.03$  dex
\enddata
\label{tab:summary}
\end{deluxetable}

We summarize the observed and derived properties from our study of Hydra~I in Table~\ref{tab:summary}.  In the following discussion, we consider the consistency of these properties with different scenarios for the possible progenitor of the EBS stream.

\subsection{Is Hydra~I a Disrupting Dwarf Galaxy?}
\label{subsec:dwarf}

The hypothesis that Hydra~I is a disrupting dwarf galaxy has been considered previously, albeit indirectly, by \citet{schlaufman2011} in the discussion of their cold halo objects (ECHOs) sample.  They argued that the relatively high velocity dispersions of the ECHOs sample ($\langle \sigma_{\rm vel}\rangle \sim 20$ km s$^{-1}$) are difficult to explain with globular cluster progenitors,  which (on average) show velocity dispersions of $\sim$7 km s$^{-1}$ \citep{kimmig2015}.  However, see discussion below (Section~\ref{subsec:glob}) for a counterpoint to this broad dynamical argument.

If Hydra~I is a dwarf galaxy, then the mass-metallicity and luminosity-metallicity relations for Local Group dwarf galaxies provide a means to estimate the stellar mass and luminosity of the stream progenitor \citep[][]{kirby2013}.  For a mean iron abundance of [Fe/H]$=-1$ dex, a dwarf galaxy stream progenitor would have been relatively massive -- approximately $M_*\sim 10^9 M_\odot$ -- similar to a present-day Fornax dSph.   Given our stellar estimate of $M_* \sim 1000~M_\odot$ for Hydra~I (see Sections~\ref{subsec:luminosity}), this implies that a dwarf galaxy stream progenitor would have lost $\gtrsim 99.99\%$ of its stellar mass.  If Hydra~I has undergone such substantial stellar mass loss, it is unlikely that our analysis in Section~\ref{subsec:mass-dynamical} yields the true, gravitationally bound mass of the remnant.  \citet{smith2013} find that the velocity dispersion of a dwarf galaxy remnant is only a reasonable measure of the bound mass when the object has retained $\gtrsim 10\%$ of the initial mass.  When significant mass loss  ($\gtrsim 98\%$) has taken place and only a small bound object remains --  the remnant mass can be overestimated by as many as three orders of magnitude. We therefore don't consider the large dynamical ``mass" of Hydra~I to provide support for a dwarf galaxy hypothesis. 

The iron abundance spread $\sigma_{\rm [Fe/H]}$ is a potentially strong discriminator between dwarf galaxies and GCs \citep{willman2012}.  Comparing a sample of Milky Way dwarfs and GCs,  \citet{willman2012} find $\sigma_{\rm [Fe/H]} \sim 0.3-0.7$ dex for dwarfs, while GCs less luminous than $M_V = -10$ have almost no measurable spread above the measurement uncertainties.  Our estimate of $\sigma_{\rm [Fe/H]} \simeq 0.1 \pm 0.03$ dex is statistically significant  and consistent with a dwarf galaxy progenitor hypothesis.  Although the brightest (i.e., classical; $M_V < -8$) dwarfs generally have larger iron abundance spreads ($\sim 0.2-0.3$ dex) compared to Hydra~I, a relatively small value $\sigma_{\rm [Fe/H]}$ is consistent with the apparent decrease in $\sigma_{\rm [Fe/H]}$ with increasing galaxy luminosity noted by \citet{willman2012}.  

All Milky Way dwarf galaxies with comprehensive studies of their stellar populations have shown evidence for an old, metal-poor stellar population \citep[e.g.,][]{kirby2013,brown2014, weisz2014}.  The addition of this population to the observed metallicity distribution would significantly increase its dispersion in [Fe/H], making the value more consistent with other luminous dwarf galaxies. If Hydra I is indeed a dwarf galaxy, our non-detection of this population could be partially due to our selection function (see Section~\ref{sec:stellar-pops}). Another possibility is that Hydra I might have once possessed such a population, but due to its larger spatial extent \citep[e.g.,][]{battaglia2006,bate2015}, it has preferentially been tidally stripped compared to the more metal-rich subpopulation.  We emphasize that our analysis does not rule out the possibility that Hydra~I contains an old, metal-poor population.  Additional spectroscopic targeting of candidate stars in the expected region of color-magnitude space and/or additional wide-field imaging would be necessary to find evidence for an older population.

The presence of intermediate age stellar population in Hydra~I (ages $\sim 5-6$ Gyr; Section~\ref{sec:stellar-pops}) is consistent with a dwarf galaxy progenitor hypothesis.  Although MW dwarf spheroidals, on average, formed the majority of their stars prior to $z\sim2$ ($\sim 10$ Gyr ago), there is significant variation from galaxy to galaxy.  Moreover, more massive dwarfs form a larger fraction of their stellar mass at later times \citep{weisz2014}.  For example, $\sim 50\%$ of the stellar mass in Fornax formed in the last $\sim 8$ Gyr and $\sim 40\%$ was formed in the last $\sim 6$ Gyr \citep{weisz2014}.  If Hydra~I is a disrupting dwarf galaxy and was relatively massive at infall (as implied by the measured [Fe/H]), then the presence of an intermediate age stellar population is consistent with our expectations from the SFHs of Local Group dwarfs.

\subsection{Is Hydra~I a Disrupting Star Cluster?}
\label{subsec:glob}

The presence of tidal streams associated with Milky Way GCs \citep[e.g., Pal~5;][]{odenkirchen2001} raises the possibility that Hydra~I could be a GC in the final throes of destruction.  \citet{grillmair2011} discussed this possibility and concluded that,  because EBS/Hydra~I may be on a highly eccentric orbit that brings it into the inner Galactic potential, dynamical heating from frequent encounters with massive structures (such as giant molecular clouds) or dark matter subhalos could be significant.  Thus the large observed velocity dispersion for Hydra~I could be due to dynamical heating rather than being evidence of a dwarf galaxy progenitor.

\begin{figure}
\epsscale{1.2}
\plotone{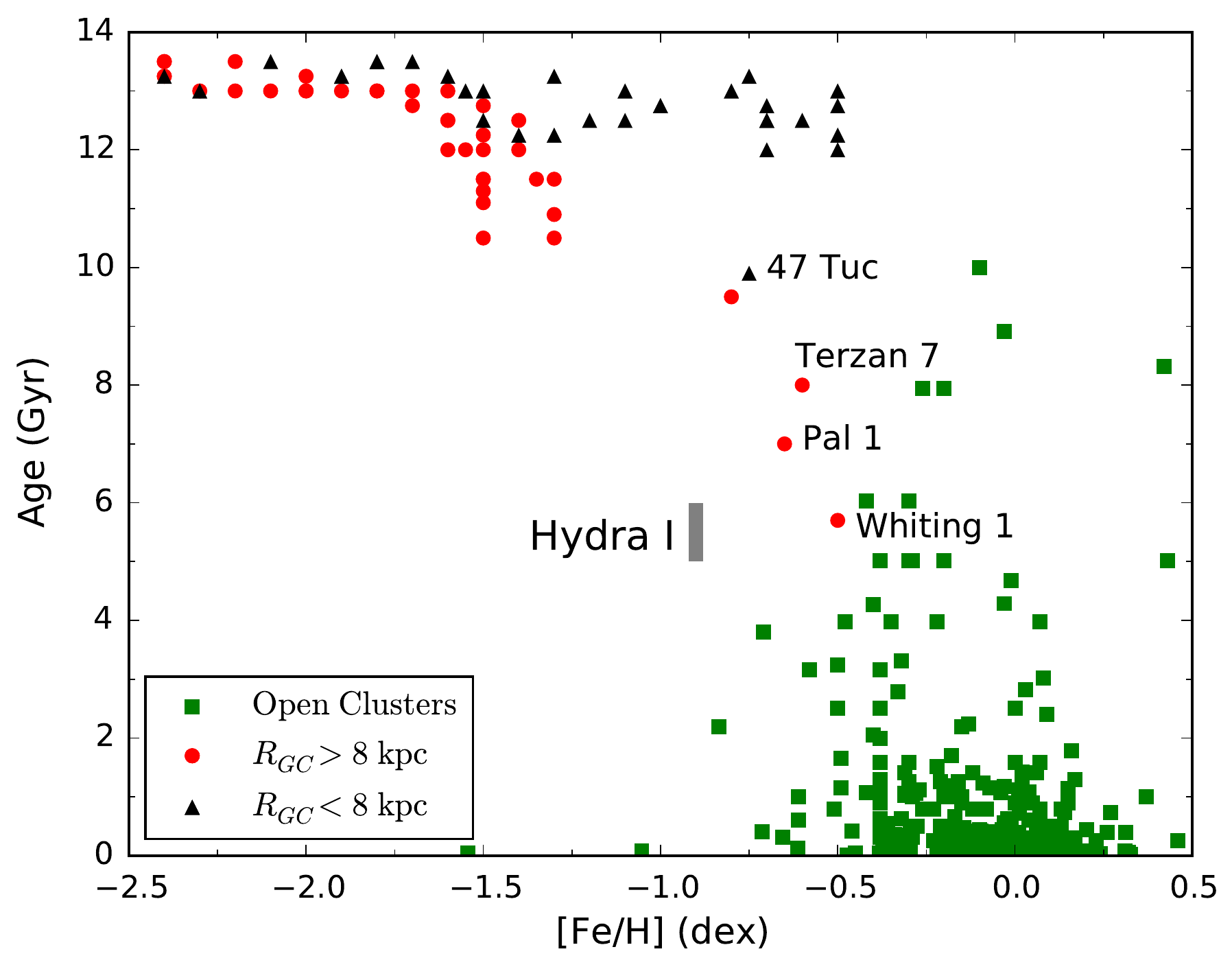}
\caption{Age vs [Fe/H] for Galactic GCs ({\it red circles, black triangles}; \citealt{dotter2010,dotter2011}) and open clusters ({\it green squares}; \citealt{dias2014}).  Ages and [Fe/H] values for younger globular clusters 47 Tuc, Whiting 1, and Pal 1 from \cite{hansen2013}, \cite{valcheva2015}, and \cite{rosenberg1998,rosenberg1998b}, respectively, are also included and labeled.  As in \citet{dotter2011}, individual GCs are identified by their galactocentric distance $R_{GC}> 8$ kpc or $R_{GC} \leq 8$ kpc.  The age range and mean metallicity of the stellar populations of Hydra~I are shown as the grey boxed region.  Very few known MW GCs or open clusters have ages and iron abundances similar to Hydra~I.}
\label{fig:amr}
\end{figure}

To further explore the GC progenitor hypothesis, we consider the location of Hydra~I in the age-metallicity relation (AMR) for MW globular and open clusters in Figure~\ref{fig:amr}.  The AMR of GCs has been well studied both in the Milky Way \citep{forbes2010, dotter2011} and M31 \citep{caldwell2011}, as well as the LMC and SMC \citep{pagel1998}.  For the MW, GCs within 8~kpc of the Galactic center show a narrow range of old ages ($\sim 13$ Gyr) but span a wide metallicity range.  Outside a galactocentric distance of 8~kpc, however, GCs exhibit an AMR more like that of the LMC and SMC, with a trend showing younger ages at higher metallicities.  

If Hydra~I is a disrupting GC, then we expect it to contain (to first order) a single stellar population with the age of $\sim 5-6$ Gyr as inferred from our stellar populations analysis (Section~\ref{sec:stellar-pops}).  The MW has few known, well studied GCs in the intermediate age ($\sim 6-8$ Gyr) range.  These include Pal 1 \citep{rosenberg1998,rosenberg1998b}, Pal 12, Terzan 7 \citep{dotter2010}, Whiting 1 \citep{valcheva2015}, and 47 Tuc \citep{hansen2013}.    With the exception of Whiting 1, all of these GCs have galactocentric distances $\lesssim 18$ kpc \citep[][]{harris1996}, similar to that of Hydra~I ($d_{\rm MW} = 19$ kpc). The youngest GCs of this subset (Pal~1 and Whiting~1), however, have higher iron abundances than Hydra~I.  

Outside of the MW, some evidence exists for massive star clusters with intermediate ages and metallicities.  However, there are significant measurement uncertainties associated with these observations.  For example, M31 contains a large population of massive [Fe/H]$\sim -1$ GCs, but age determinations from integrated light measurements (e.g., Lick indices) at this metallicity are highly sensitive to the presence of blue horizontal branch stars, resulting in artificially young and thus highly uncertain ages \citep[see discussion in][]{caldwell2011}.  The Large Magellanic Cloud (LMC) has a well known ``age-gap'' in its star cluster population \citep[e.g., ][]{harris2009}: there are no massive clusters with ages from $\sim5-12$ Gyr, and only one relatively low mass, intermediate metallicity cluster ($M\sim10^4~M_\odot$; [Fe/H]$\sim -1$) with an age $\sim8-9$ Gyr  \citep{mackey2006}.  In contrast, the Small Magellanic Cloud may contain a few $M\sim 10^4-10^5~M_\odot$ clusters consistent with the age and metallicity of Hydra~I (see \citealt{dias2010} for a recent review), although the wide range of age estimates for individual clusters highlights the large uncertainty in their observed ages. 

Given Hydra~I's outlying age and metallicity, relative to other known MW clusters, and the lack of strong evidence for analog clusters in M31 and the Magellanic Clouds, we conclude that the star cluster hypothesis is less likely than the dwarf galaxy hypothesis.  If Hydra~I is indeed a GC, the observed age and metallicity imply that it would have likely formed in dwarf galaxy (perhaps smaller than the LMC or SMC) which was subsequently accreted by the Milky Way.  However, the lack of known GCs with ages and metallicities consistent with Hydra~I places limits on the plausibility of such a scenario.

\subsection{Possible Association of Hydra~I With a Larger Galactic Structure}
\label{subsec:other}

Lastly, we consider the possibility that Hydra~I and EBS may be part of a larger Galactic structure --  in particular the Monoceros Ring stellar overdensity.  The distance and metallicity of Hydra~I are consistent with photometric and spectroscopic studies of the larger Monoceros Ring region \citep{conn2012,meisner2012}.  \citet{slater2014} recently presented a Pan-STARRS1 stellar density map for the Monoceros Ring, constructed using MSTO stars from a $\sim 9$ Gyr population with [Fe/H] = -1 (colors in the range $0.2 < (g-r)_0 < 0.3$).  They argue that the EBS and Anticenter streams are extended stellar arcs that are physically associated with the Monoceros Ring.  Conversely, \citet{grillmair2011} argued that the EBS stream is kinematically distinct from the Monoceros Ring, given the high eccentricity orbital solution needed to explain EBS's observed spatial curvature.  The Monoceros Ring is believed to be on a nearly circular orbit \citep{crane2003, martin2006}.  Although our study cannot shed additional light on whether either Hydra~I or EBS might be associated with the Monoceros Ring, a more refined orbital solution for the EBS stream -- using GAIA proper motions and a large sample of radial velocity measurements -- is a promising avenue for investigating this hypothesis.

Given the proximity of Hydra~I to the Monoceros Ring, it is possible we have isolated a younger population of stars recently speculated to be part of this structure.  \citet{bello2015} noted that along the line of sight to the Monoceros Ring (at $l \sim 180^{\rm o}, 195^{\rm o}; b \sim 25^{\rm o}$), there exists a population of stars bluer than the $\sim 9$ Gyr MSTO of Monoceros Ring stars.  They suggested that this could be a $\sim 4$ Gyr population (with an iron abundance of [Fe/H]$\sim -0.9$ dex) associated with the Monoceros Ring.  \citet{bello2015} were unable to clearly detect a MSTO feature in the SDSS data but estimated a distance of $\sim 10$ kpc for the population.  The systemic velocities of the blue population -- as well as Hydra~I -- are consistent with model predictions from \citet{penarrubia2005} for the origin of the Monoceros Ring as an accreted dwarf galaxy.  Although suggestive, additional precision photometry over a much larger field will be necessary to further explore any spatial connections between the \citet{bello2015} detections and the EBS/Hydra~I.

The average [Fe/H] and [$\alpha$/Fe] observed for Hydra~I limits the possibility that it may just represent an overdensity of stars originating in the MW's halo or thick disk.  \citet{schlaufman2011} studied the SDSS/SEGUE observations of a set of cold halo substructures (ECHOs), one of which was Hydra~I.  They compared the mean iron and alpha abundances of ECHOs to samples of thin disk, think disk, and halo stars compiled by \citet{venn2004}.  They show, and we confirm, that their ECHO spatially associated with Hydra~I is more iron-rich and less alpha-enhanced than either the smooth halo or the thick disk in that direction ([Fe/H]$_{\rm halo}$ $\sim -1.4$; [$\alpha$/Fe]$_{\rm halo}$ $\sim 0.25$]).

\section{Conclusions}
\label{sec:conclusions}

In this paper we present a wide-field DECam imaging and  MMT/Hectochelle + archival SDSS  spectroscopic study of Hydra~I -- the double-lobed, $\sim 2$ deg$^2$ halo overdensity previously suggested to be the progenitor of the EBS stream -- and quantify its present day chemo-dynamical properties.  The results inferred using both Hectochelle and SDSS/SEGUE data are consistent.  Hydra~I contains an intermediate age stellar population of $\sim$ 5-6 Gyr with a mean iron abundance of [Fe/H] = $-0.91 \pm 0.02$ dex and alpha abundance of [$\alpha$/Fe] = $0.15 \pm 0.03$ dex.  We find spreads in both the iron and alpha abundances of $\sigma_{\rm [Fe/H]} \simeq 0.10 \pm 0.03$ dex and $\sigma_{[\alpha/{\rm Fe}]} \simeq 0.09 \pm 0.03$ dex, respectively.  We confirm previous observations that Hydra~I is a kinematically cold structure, with a velocity dispersion of $8.4 \pm 1.5$ km s$^{-1}$ at a systemic heliocentric velocity of $V = 89.4 \pm 1.4$ km s$^{-1}$ ($V_{\rm GSR} = -55.5 \pm 1.4$ km s$^{-1}$).  

Hydra~I's stellar population appears initially inconsistent with earlier studies reporting that an old- and metal-poor stellar population yielded the strongest overdensity in matched-filtering maps of the Hydra~I region  \citep{grillmair2006,grillmair2011}.  Old and metal-poor populations can have MSTO populations as blue as those of intermediate age and [Fe/H] rich populations.  Moreover, many approaches to searching for and mapping streams in the MW's halo assume that streams will be composed of old and metal-poor populations.  Assuming a metal-poor population can thus lead to erroneous conclusions about a substructure's age and distance.   Complimentary spectroscopy to measure the metal content of substructures provides a critical constraint on both their ages and distances, and ultimately provides a more accurate picture of the accretion events that formed the Galactic halo.

Hydra~I's intermediate age ($\sim5-6$~Gyr) and [Fe/H] ($\sim$-1~dex) make it an outlier among known MW globular cluster and open clusters.  Conversely, Hydra~I's stellar populations, along with its spread in [Fe/H], is consistent with those of a Fornax-like dwarf galaxy progenitor.  If Hydra~I's progenitor was a Fornax-like dwarf galaxy, then it must have lost $>99.99\%$ of its stars to bear the low luminosity it displays today.  Hydra~I may thus be the first example of a system truly in its final throes of disruption, making it an excellent benchmark of tidal disruption models.

Although evidence strongly points towards a dwarf galaxy interpretation of Hydra~I, we cannot rule out the hypothesis that Hydra~I originated as a star cluster or that it is physically associated with the Monoceros Ring.  Higher S/N spectroscopic follow up of high probability member stars is a promising avenue for a more certain test of the dwarf galaxy hypothesis.  With high resolution spectra, the apparent [Fe/H] spread presented in this work could be measured with higher fidelity.  Combined with Gaia proper motions and deeper wide-field imaging, Hydra~I modeling will further test whether it could be kinematically associated with the Monoceros Ring and will constrain the impact of secular and environmental factors on its observable properties.

\section{Acknowledgements}

BW, JRH, and BK were supported by an NSF Faculty Early Career Development (CAREER) award (AST-1151462).  MGW is supported by National Science Foundation grants AST-1313045, AST-1412999.  JS acknowledges support from NSF grant AST-1514763.  JRH would like to acknowledge useful conversations with Keith Hawkins, Ting Li, and Jennifer Marshall during the course of this work.  JRH would also like to thank Kathy Vivas both for useful conversations and observing support during the DECam observing at CTIO.  This work was supported in part by National Science Foundation Grant No. PHYS-1066293 and the hospitality of the Aspen Center for Physics. 

This project used data obtained with the Dark Energy Camera, which was constructed by the Dark Energy Survey collaboration. Funding for the DES Projects has been provided by the DOE and NSF(USA), MISE(Spain), STFC(UK), HEFCE(UK). NCSA(UIUC), KICP(U. Chicago), CCAPP(Ohio State), MIFPA(Texas A\&M), CNPQ, FAPERJ, FINEP (Brazil), MINECO(Spain), DFG(Germany) and the collaborating institutions in the Dark Energy Survey, which are Argonne Lab, UC Santa Cruz, University of Cambridge, CIEMAT-Madrid, University of Chicago, University College London, DES-Brazil Consortium, University of Edinburgh, ETH Zurich, Fermilab, University of Illinois, ICE (IEEC-CSIC), IFAE Barcelona, Lawrence Berkeley Lab, LMU Munchen and the associated Excellence Cluster Universe, University of Michigan, NOAO, University of Nottingham, Ohio State University, University of Pennsylvania, University of Portsmouth, SLAC National Lab, Stanford University, University of Sussex, and Texas A\&M University.

\bibliographystyle{apj}


\begin{thebibliography}{}
\expandafter\ifx\csname natexlab\endcsname\relax\def\natexlab#1{#1}\fi

\bibitem[{{Aihara} {et~al.}(2011){Aihara}, {Allende Prieto}, {An}, {Anderson},
  {Aubourg}, {Balbinot}, {Beers}, {Berlind}, {Bickerton}, {Bizyaev}, {Blanton},
  {Bochanski}, {Bolton}, {Bovy}, {Brandt}, {Brinkmann}, {Brown}, {Brownstein},
  {Busca}, {Campbell}, {Carr}, {Chen}, {Chiappini}, {Comparat}, {Connolly},
  {Cortes}, {Croft}, {Cuesta}, {da Costa}, {Davenport}, {Dawson}, {Dhital},
  {Ealet}, {Ebelke}, {Edmondson}, {Eisenstein}, {Escoffier}, {Esposito},
  {Evans}, {Fan}, {Femen{\'{\i}}a Castell{\'a}}, {Font-Ribera}, {Frinchaboy},
  {Ge}, {Gillespie}, {Gilmore}, {Gonz{\'a}lez Hern{\'a}ndez}, {Gott}, {Gould},
  {Grebel}, {Gunn}, {Hamilton}, {Harding}, {Harris}, {Hawley}, {Hearty}, {Ho},
  {Hogg}, {Holtzman}, {Honscheid}, {Inada}, {Ivans}, {Jiang}, {Johnson},
  {Jordan}, {Jordan}, {Kazin}, {Kirkby}, {Klaene}, {Knapp}, {Kneib},
  {Kochanek}, {Koesterke}, {Kollmeier}, {Kron}, {Lampeitl}, {Lang}, {Le Goff},
  {Lee}, {Lin}, {Long}, {Loomis}, {Lucatello}, {Lundgren}, {Lupton}, {Ma},
  {MacDonald}, {Mahadevan}, {Maia}, {Makler}, {Malanushenko}, {Malanushenko},
  {Mandelbaum}, {Maraston}, {Margala}, {Masters}, {McBride}, {McGehee},
  {McGreer}, {M{\'e}nard}, {Miralda-Escud{\'e}}, {Morrison}, {Mullally},
  {Muna}, {Munn}, {Murayama}, {Myers}, {Naugle}, {Neto}, {Nguyen}, {Nichol},
  {O'Connell}, {Ogando}, {Olmstead}, {Oravetz}, {Padmanabhan},
  {Palanque-Delabrouille}, {Pan}, {Pandey}, {P{\^a}ris}, {Percival},
  {Petitjean}, {Pfaffenberger}, {Pforr}, {Phleps}, {Pichon}, {Pieri}, {Prada},
  {Price-Whelan}, {Raddick}, {Ramos}, {Reyl{\'e}}, {Rich}, {Richards}, {Rix},
  {Robin}, {Rocha-Pinto}, {Rockosi}, {Roe}, {Rollinde}, {Ross}, {Ross},
  {Rossetto}, {S{\'a}nchez}, {Sayres}, {Schlegel}, {Schlesinger}, {Schmidt},
  {Schneider}, {Sheldon}, {Shu}, {Simmerer}, {Simmons}, {Sivarani}, {Snedden},
  {Sobeck}, {Steinmetz}, {Strauss}, {Szalay}, {Tanaka}, {Thakar}, {Thomas},
  {Tinker}, {Tofflemire}, {Tojeiro}, {Tremonti}, {Vandenberg}, {Vargas
  Maga{\~n}a}, {Verde}, {Vogt}, {Wake}, {Wang}, {Weaver}, {Weinberg}, {White},
  {White}, {Yanny}, {Yasuda}, {Yeche}, \& {Zehavi}}]{aihara2011}
{Aihara}, H., {Allende Prieto}, C., {An}, D., {et~al.} 2011, \apjs, 193, 29

\bibitem[{{Allende Prieto} {et~al.}(2008){Allende Prieto}, {Sivarani}, {Beers},
  {Lee}, {Koesterke}, {Shetrone}, {Sneden}, {Lambert}, {Wilhelm}, {Rockosi},
  {Lai}, {Yanny}, {Ivans}, {Johnson}, {Aoki}, {Bailer-Jones}, \& {Re
  Fiorentin}}]{allende2008}
{Allende Prieto}, C., {Sivarani}, T., {Beers}, T.~C., {et~al.} 2008, \aj, 136,
  2070

\bibitem[{Balbinot {et~al.}(2015)Balbinot, Yanny, Li, Santiago, Marshall,
  Finley, Pieres, Abbott, Abdalla, Allam, Benoit-Levy, Bernstein, Bertin,
  Brooks, Burke, Carnero~Rosell, Carrasco~Kind, Carretero, Cunha, da~Costa,
  DePoy, Desai, Diehl, Doel, Estrada, Flaugher, Frieman, Gerdes, Gruen,
  Gruendl, Honscheid, James, Kuehn, Kuropatkin, Lahav, March, Martini, Miquel,
  Nichol, Ogando, Romer, Sanchez, Schubnell, Sevilla-Noarbe, Smith,
  Soares-Santos, Sobreira, Suchyta, Tarle, Thomas, Tucker, \&
  Walker}]{balbinot2015}
Balbinot, E., Yanny, B., Li, T.~S., {et~al.} 2015, eprint arXiv:1509.04283,
  1509.04283

\bibitem[{{Bate} {et~al.}(2015){Bate}, {McMonigal}, {Lewis}, {Irwin},
  {Gonzalez-Solares}, {Shanks}, \& {Metcalfe}}]{bate2015}
{Bate}, N.~F., {McMonigal}, B., {Lewis}, G.~F., {et~al.} 2015, \mnras, 453, 690

\bibitem[{{Battaglia} {et~al.}(2006){Battaglia}, {Tolstoy}, {Helmi}, {Irwin},
  {Letarte}, {Jablonka}, {Hill}, {Venn}, {Shetrone}, {Arimoto}, {Primas},
  {Kaufer}, {Francois}, {Szeifert}, {Abel}, \& {Sadakane}}]{battaglia2006}
{Battaglia}, G., {Tolstoy}, E., {Helmi}, A., {et~al.} 2006, \aap, 459, 423

\bibitem[{Bechtol {et~al.}(2015)Bechtol, Drlica-Wagner, Balbinot, Pieres,
  Simon, Yanny, Santiago, Wechsler, Frieman, Walker, Williams, Rozo, Rykoff,
  Queiroz, Luque, Benoit-L?vy, Tucker, Sevilla, Gruendl, da~Costa, Neto, Maia,
  Abbott, Allam, Armstrong, Bauer, Bernstein, Bernstein, Bertin, Brooks,
  Buckley-Geer, Burke, Rosell, Castander, Covarrubias, D?Andrea, DePoy, Desai,
  Diehl, Eifler, Estrada, Evrard, Fernandez, Finley, Flaugher, Gaztanaga,
  Gerdes, Girardi, Gladders, Gruen, Gutierrez, Hao, Honscheid, Jain, James,
  Kent, Kron, Kuehn, Kuropatkin, Lahav, Li, Lin, Makler, March, Marshall,
  Martini, Merritt, Miller, Miquel, Mohr, Neilsen, Nichol, Nord, Ogando,
  Peoples, Petravick, Plazas, Romer, Roodman, Sako, Sanchez, Scarpine,
  Schubnell, Smith, Soares-Santos, Sobreira, Suchyta, Swanson, Tarle, Thaler,
  Thomas, Wester, Zuntz, \& Collaboration}]{bechtol2015}
Bechtol, K., Drlica-Wagner, A., Balbinot, E., {et~al.} 2015, The Astrophysical
  Journal, 807, 50

\bibitem[{{Bellazzini}(2004)}]{bellazzini2004}
{Bellazzini}, M. 2004, \mnras, 347, 119

\bibitem[{{Bellazzini} {et~al.}(2012){Bellazzini}, {Bragaglia}, {Carretta},
  {Gratton}, {Lucatello}, {Catanzaro}, \& {Leone}}]{bellazzini2012}
{Bellazzini}, M., {Bragaglia}, A., {Carretta}, E., {et~al.} 2012, \aap, 538,
  A18

\bibitem[{{Belokurov} {et~al.}(2006){Belokurov}, {Zucker}, {Evans}, {Gilmore},
  {Vidrih}, {Bramich}, {Newberg}, {Wyse}, {Irwin}, {Fellhauer}, {Hewett},
  {Walton}, {Wilkinson}, {Cole}, {Yanny}, {Rockosi}, {Beers}, {Bell},
  {Brinkmann}, {Ivezi{\'c}}, \& {Lupton}}]{belokurov2006}
{Belokurov}, V., {Zucker}, D.~B., {Evans}, N.~W., {et~al.} 2006, \apjl, 642,
  L137

\bibitem[{{Bernard} {et~al.}(2014){Bernard}, {Ferguson}, {Schlafly}, {Abbas},
  {Bell}, {Deacon}, {Martin}, {Rix}, {Sesar}, {Slater}, {Pe{\~n}arrubia},
  {Wyse}, {Burgett}, {Chambers}, {Draper}, {Hodapp}, {Kaiser}, {Kudritzki},
  {Magnier}, {Metcalfe}, {Morgan}, {Price}, {Tonry}, {Wainscoat}, \&
  {Waters}}]{bernard2014}
{Bernard}, E.~J., {Ferguson}, A.~M.~N., {Schlafly}, E.~F., {et~al.} 2014,
  \mnras, 443, L84

\bibitem[{{Boettcher} {et~al.}(2013){Boettcher}, {Willman}, {Fadely},
  {Strader}, {Baker}, {Hopkins}, {Tasnim Ananna}, {Cunningham}, {Douglas},
  {Gilbert}, {Preston}, \& {Sturner}}]{boettcher2013}
{Boettcher}, E., {Willman}, B., {Fadely}, R., {et~al.} 2013, \aj, 146, 94

\bibitem[{Brown {et~al.}(2014)Brown, Tumlinson, Geha, Simon, Vargas,
  VandenBerg, Kirby, Kalirai, Avila, Gennaro, Ferguson, Mu{\~n}oz,
  Guhathakurta, \& Renzini}]{brown2014}
Brown, T.~M., Tumlinson, J., Geha, M., {et~al.} 2014, The Astrophysical
  Journal, 796, 91

\bibitem[{{Bullock} \& {Johnston}(2005)}]{bullock2005}
{Bullock}, J.~S., \& {Johnston}, K.~V. 2005, \apj, 635, 931

\bibitem[{{Caldwell} {et~al.}(2011){Caldwell}, {Schiavon}, {Morrison}, {Rose},
  \& {Harding}}]{caldwell2011}
{Caldwell}, N., {Schiavon}, R., {Morrison}, H., {Rose}, J.~A., \& {Harding}, P.
  2011, \aj, 141, 61

\bibitem[{{Carballo-Bello} {et~al.}(2015){Carballo-Bello}, {Mu{\~n}oz},
  {Carlin}, {C{\^o}t{\'e}}, {Geha}, {Simon}, \& {Djorgovski}}]{bello2015}
{Carballo-Bello}, J.~A., {Mu{\~n}oz}, R.~R., {Carlin}, J.~L., {et~al.} 2015,
  \apj, 805, 51

\bibitem[{{Carlin} {et~al.}(2010){Carlin}, {Casetti-Dinescu}, {Grillmair},
  {Majewski}, \& {Girard}}]{carlin2010}
{Carlin}, J.~L., {Casetti-Dinescu}, D.~I., {Grillmair}, C.~J., {Majewski},
  S.~R., \& {Girard}, T.~M. 2010, \apj, 725, 2290

\bibitem[{{Conn} {et~al.}(2012){Conn}, {No{\"e}l}, {Rix}, {Lane}, {Lewis},
  {Irwin}, {Martin}, {Ibata}, {Dolphin}, \& {Chapman}}]{conn2012}
{Conn}, B.~C., {No{\"e}l}, N.~E.~D., {Rix}, H.-W., {et~al.} 2012, \apj, 754,
  101

\bibitem[{{Cooper} {et~al.}(2010){Cooper}, {Cole}, {Frenk}, {White}, {Helly},
  {Benson}, {De Lucia}, {Helmi}, {Jenkins}, {Navarro}, {Springel}, \&
  {Wang}}]{cooper2010}
{Cooper}, A.~P., {Cole}, S., {Frenk}, C.~S., {et~al.} 2010, \mnras, 406, 744

\bibitem[{{Crane} {et~al.}(2003){Crane}, {Majewski}, {Rocha-Pinto},
  {Frinchaboy}, {Skrutskie}, \& {Law}}]{crane2003}
{Crane}, J.~D., {Majewski}, S.~R., {Rocha-Pinto}, H.~J., {et~al.} 2003, \apjl,
  594, L119

\bibitem[{{Dias} {et~al.}(2010){Dias}, {Coelho}, {Barbuy}, {Kerber}, \&
  {Idiart}}]{dias2010}
{Dias}, B., {Coelho}, P., {Barbuy}, B., {Kerber}, L., \& {Idiart}, T. 2010,
  \aap, 520, A85

\bibitem[{{Dias} {et~al.}(2014){Dias}, {Alessi}, {Moitinho}, \&
  {Lepine}}]{dias2014}
{Dias}, W.~S., {Alessi}, B.~S., {Moitinho}, A., \& {Lepine}, J.~R.~D. 2014,
  VizieR Online Data Catalog, 1, 2022

\bibitem[{{Dotter} {et~al.}(2008){Dotter}, {Chaboyer}, {Jevremovi{\'c}},
  {Kostov}, {Baron}, \& {Ferguson}}]{dotter2008}
{Dotter}, A., {Chaboyer}, B., {Jevremovi{\'c}}, D., {et~al.} 2008, \apjs, 178,
  89

\bibitem[{{Dotter} {et~al.}(2011){Dotter}, {Sarajedini}, \&
  {Anderson}}]{dotter2011}
{Dotter}, A., {Sarajedini}, A., \& {Anderson}, J. 2011, \apj, 738, 74

\bibitem[{{Dotter} {et~al.}(2010){Dotter}, {Sarajedini}, {Anderson},
  {Aparicio}, {Bedin}, {Chaboyer}, {Majewski}, {Mar{\'{\i}}n-Franch}, {Milone},
  {Paust}, {Piotto}, {Reid}, {Rosenberg}, \& {Siegel}}]{dotter2010}
{Dotter}, A., {Sarajedini}, A., {Anderson}, J., {et~al.} 2010, \apj, 708, 698

\bibitem[{{Drlica-Wagner} {et~al.}(2015){Drlica-Wagner}, {Bechtol}, {Rykoff},
  {Luque}, {Queiroz}, {Mao}, {Wechsler}, {Simon}, {Santiago}, {Yanny},
  {Balbinot}, {Dodelson}, {Fausti Neto}, {James}, {Li}, {Maia}, {Marshall},
  {Pieres}, {Stringer}, {Walker}, {Abbott}, {Abdalla}, {Allam}, {Benoit-Levy},
  {Bernstein}, {Bertin}, {Brooks}, {Buckley-Geer}, {Burke}, {Carnero Rosell},
  {Carrasco Kind}, {Carretero}, {Crocce}, {da Costa}, {Desai}, {Diehl},
  {Dietrich}, {Doel}, {Eifler}, {Evrard}, {Finley}, {Fosalba}, {Frieman},
  {Gaztanaga}, {Gerdes}, {Gruen}, {Gruendl}, {Gutierrez}, {Honscheid}, {Kuehn},
  {Kuropatkin}, {Lahav}, {Martini}, {Miquel}, {Nord}, {Ogando}, {Plazas},
  {Reil}, {Roodman}, {Sako}, {Sanchez}, {Scarpine}, {Schubnell},
  {Sevilla-Noarbe}, {Smith}, {Soares-Santos}, {Sobreira}, {Suchyta}, {Swanson},
  {Tarle}, {Tucker}, {Vikram}, {Wester}, {Zhang}, \&
  {Zuntz}}]{drlicawagner2015}
{Drlica-Wagner}, A., {Bechtol}, K., {Rykoff}, E.~S., {et~al.} 2015, ArXiv
  e-prints, arXiv:1508.03622

\bibitem[{{Feroz} \& {Hobson}(2008)}]{feroz2008}
{Feroz}, F., \& {Hobson}, M.~P. 2008, \mnras, 384, 449

\bibitem[{{Feroz} {et~al.}(2009){Feroz}, {Hobson}, \& {Bridges}}]{feroz2009}
{Feroz}, F., {Hobson}, M.~P., \& {Bridges}, M. 2009, \mnras, 398, 1601

\bibitem[{{Flaugher} {et~al.}(2015){Flaugher}, {Diehl}, {Honscheid}, {Abbott},
  {Alvarez}, {Angstadt}, {Annis}, {Antonik}, {Ballester}, {Beaufore},
  {Bernstein}, {Bernstein}, {Bigelow}, {Bonati}, {Boprie}, {Brooks},
  {Buckley-Geer}, {Campa}, {Cardiel-Sas}, {Castander}, {Castilla}, {Cease},
  {Cela-Ruiz}, {Chappa}, {Chi}, {Cooper}, {da Costa}, {Dede}, {Derylo},
  {DePoy}, {de Vicente}, {Doel}, {Drlica-Wagner}, {Eiting}, {Elliott}, {Emes},
  {Estrada}, {Fausti Neto}, {Finley}, {Flores}, {Frieman}, {Gerdes},
  {Gladders}, {Gregory}, {Gutierrez}, {Hao}, {Holland}, {Holm}, {Huffman},
  {Jackson}, {James}, {Jonas}, {Karcher}, {Karliner}, {Kent}, {Kessler},
  {Kozlovsky}, {Kron}, {Kubik}, {Kuehn}, {Kuhlmann}, {Kuk}, {Lahav}, {Lathrop},
  {Lee}, {Levi}, {Lewis}, {Li}, {Mandrichenko}, {Marshall}, {Martinez},
  {Merritt}, {Miquel}, {Munoz}, {Neilsen}, {Nichol}, {Nord}, {Ogando}, {Olsen},
  {Palio}, {Patton}, {Peoples}, {Plazas}, {Rauch}, {Reil}, {Rheault}, {Roe},
  {Rogers}, {Roodman}, {Sanchez}, {Scarpine}, {Schindler}, {Schmidt},
  {Schmitt}, {Schubnell}, {Schultz}, {Schurter}, {Scott}, {Serrano}, {Shaw},
  {Smith}, {Soares-Santos}, {Stefanik}, {Stuermer}, {Suchyta}, {Sypniewski},
  {Tarle}, {Thaler}, {Tighe}, {Tran}, {Tucker}, {Walker}, {Wang}, {Watson},
  {Weaverdyck}, {Wester}, {Woods}, \& {Yanny}}]{flaugher2015}
{Flaugher}, B., {Diehl}, H.~T., {Honscheid}, K., {et~al.} 2015, ArXiv e-prints,
  arXiv:1504.02900

\bibitem[{{Forbes} \& {Bridges}(2010)}]{forbes2010}
{Forbes}, D.~A., \& {Bridges}, T. 2010, \mnras, 404, 1203

\bibitem[{{Grillmair}(2006)}]{grillmair2006}
{Grillmair}, C.~J. 2006, \apjl, 651, L29

\bibitem[{{Grillmair}(2011)}]{grillmair2011}
---. 2011, \apj, 738, 98

\bibitem[{{Grillmair} {et~al.}(2008){Grillmair}, {Carlin}, \&
  {Majewski}}]{grillmair2008}
{Grillmair}, C.~J., {Carlin}, J.~L., \& {Majewski}, S.~R. 2008, \apjl, 689,
  L117

\bibitem[{{Hansen} {et~al.}(2013){Hansen}, {Kalirai}, {Anderson}, {Dotter},
  {Richer}, {Rich}, {Shara}, {Fahlman}, {Hurley}, {King}, {Reitzel}, \&
  {Stetson}}]{hansen2013}
{Hansen}, B.~M.~S., {Kalirai}, J.~S., {Anderson}, J., {et~al.} 2013, \nat, 500,
  51

\bibitem[{{Harris} \& {Zaritsky}(2009)}]{harris2009}
{Harris}, J., \& {Zaritsky}, D. 2009, \aj, 138, 1243

\bibitem[{{Harris}(1996)}]{harris1996}
{Harris}, W.~E. 1996, \aj, 112, 1487

\bibitem[{{Johnston} {et~al.}(2008){Johnston}, {Bullock}, {Sharma}, {Font},
  {Robertson}, \& {Leitner}}]{johnston2008}
{Johnston}, K.~V., {Bullock}, J.~S., {Sharma}, S., {et~al.} 2008, \apj, 689,
  936

\bibitem[{{Jordi} {et~al.}(2006){Jordi}, {Grebel}, \& {Ammon}}]{jordi2006}
{Jordi}, K., {Grebel}, E.~K., \& {Ammon}, K. 2006, \aap, 460, 339

\bibitem[{{Kazantzidis} {et~al.}(2008){Kazantzidis}, {Bullock}, {Zentner},
  {Kravtsov}, \& {Moustakas}}]{kazantzidis2008}
{Kazantzidis}, S., {Bullock}, J.~S., {Zentner}, A.~R., {Kravtsov}, A.~V., \&
  {Moustakas}, L.~A. 2008, \apj, 688, 254

\bibitem[{{Kimmig} {et~al.}(2015){Kimmig}, {Seth}, {Ivans}, {Strader},
  {Caldwell}, {Anderton}, \& {Gregersen}}]{kimmig2015}
{Kimmig}, B., {Seth}, A., {Ivans}, I.~I., {et~al.} 2015, \aj, 149, 53

\bibitem[{{Kirby} {et~al.}(2013){Kirby}, {Cohen}, {Guhathakurta}, {Cheng},
  {Bullock}, \& {Gallazzi}}]{kirby2013}
{Kirby}, E.~N., {Cohen}, J.~G., {Guhathakurta}, P., {et~al.} 2013, \apj, 779,
  102

\bibitem[{{Koposov} {et~al.}(2015){Koposov}, {Belokurov}, {Torrealba}, \&
  {Evans}}]{koposov2015a}
{Koposov}, S.~E., {Belokurov}, V., {Torrealba}, G., \& {Evans}, N.~W. 2015,
  \apj, 805, 130

\bibitem[{{Koposov} {et~al.}(2014){Koposov}, {Irwin}, {Belokurov},
  {Gonzalez-Solares}, {Yoldas}, {Lewis}, {Metcalfe}, \& {Shanks}}]{koposov2015}
{Koposov}, S.~E., {Irwin}, M., {Belokurov}, V., {et~al.} 2014, \mnras, 442, L85

\bibitem[{{Lane} {et~al.}(2009){Lane}, {Kiss}, {Lewis}, {Ibata}, {Siebert},
  {Bedding}, \& {Sz{\'e}kely}}]{lane2009}
{Lane}, R.~R., {Kiss}, L.~L., {Lewis}, G.~F., {et~al.} 2009, \mnras, 400, 917

\bibitem[{{Lane} {et~al.}(2010{\natexlab{a}}){Lane}, {Kiss}, {Lewis}, {Ibata},
  {Siebert}, {Bedding}, \& {Sz{\'e}kely}}]{lane2010a}
---. 2010{\natexlab{a}}, \mnras, 401, 2521

\bibitem[{{Lane} {et~al.}(2010{\natexlab{b}}){Lane}, {Kiss}, {Lewis}, {Ibata},
  {Siebert}, {Bedding}, {Sz{\'e}kely}, {Balog}, \& {Szab{\'o}}}]{lane2010b}
---. 2010{\natexlab{b}}, \mnras, 406, 2732

\bibitem[{{Lee} {et~al.}(2008{\natexlab{a}}){Lee}, {Beers}, {Sivarani},
  {Allende Prieto}, {Koesterke}, {Wilhelm}, {Re Fiorentin}, {Bailer-Jones},
  {Norris}, {Rockosi}, {Yanny}, {Newberg}, {Covey}, {Zhang}, \&
  {Luo}}]{lee2008}
{Lee}, Y.~S., {Beers}, T.~C., {Sivarani}, T., {et~al.} 2008{\natexlab{a}}, \aj,
  136, 2022

\bibitem[{{Lee} {et~al.}(2008{\natexlab{b}}){Lee}, {Beers}, {Sivarani},
  {Johnson}, {An}, {Wilhelm}, {Allende Prieto}, {Koesterke}, {Re Fiorentin},
  {Bailer-Jones}, {Norris}, {Yanny}, {Rockosi}, {Newberg}, {Cudworth}, \&
  {Pan}}]{lee2008b}
---. 2008{\natexlab{b}}, \aj, 136, 2050

\bibitem[{{Lee} {et~al.}(2011){Lee}, {Beers}, {Allende Prieto}, {Lai},
  {Rockosi}, {Morrison}, {Johnson}, {An}, {Sivarani}, \& {Yanny}}]{lee2011}
{Lee}, Y.~S., {Beers}, T.~C., {Allende Prieto}, C., {et~al.} 2011, \aj, 141, 90

\bibitem[{{Li} {et~al.}(2012){Li}, {Newberg}, {Carlin}, {Deng}, {Newby},
  {Willett}, {Xu}, \& {Luo}}]{li2012}
{Li}, J., {Newberg}, H.~J., {Carlin}, J.~L., {et~al.} 2012, \apj, 757, 151

\bibitem[{{Mackey} {et~al.}(2006){Mackey}, {Payne}, \& {Gilmore}}]{mackey2006}
{Mackey}, A.~D., {Payne}, M.~J., \& {Gilmore}, G.~F. 2006, \mnras, 369, 921

\bibitem[{{Martin} {et~al.}(2006){Martin}, {Irwin}, {Ibata}, {Conn}, {Lewis},
  {Bellazzini}, {Chapman}, \& {Tanvir}}]{martin2006}
{Martin}, N.~F., {Irwin}, M.~J., {Ibata}, R.~A., {et~al.} 2006, \mnras, 367,
  L69

\bibitem[{{Martin} {et~al.}(2014){Martin}, {Ibata}, {Rich}, {Collins},
  {Fardal}, {Irwin}, {Lewis}, {McConnachie}, {Babul}, {Bate}, {Chapman},
  {Conn}, {Crnojevi{\'c}}, {Ferguson}, {Mackey}, {Navarro}, {Pe{\~n}arrubia},
  {Tanvir}, \& {Valls-Gabaud}}]{martin2014}
{Martin}, N.~F., {Ibata}, R.~A., {Rich}, R.~M., {et~al.} 2014, \apj, 787, 19

\bibitem[{{McConnachie}(2012)}]{mcconnachie2012}
{McConnachie}, A.~W. 2012, \aj, 144, 4

\bibitem[{{Meisner} {et~al.}(2012){Meisner}, {Frebel}, {Juri{\'c}}, \&
  {Finkbeiner}}]{meisner2012}
{Meisner}, A.~M., {Frebel}, A., {Juri{\'c}}, M., \& {Finkbeiner}, D.~P. 2012,
  \apj, 753, 116

\bibitem[{{Momany} {et~al.}(2006){Momany}, {Zaggia}, {Gilmore}, {Piotto},
  {Carraro}, {Bedin}, \& {de Angeli}}]{momany2006}
{Momany}, Y., {Zaggia}, S., {Gilmore}, G., {et~al.} 2006, \aap, 451, 515

\bibitem[{{Munn} {et~al.}(2004){Munn}, {Monet}, {Levine}, {Canzian}, {Pier},
  {Harris}, {Lupton}, {Ivezi{\'c}}, {Hindsley}, {Hennessy}, {Schneider}, \&
  {Brinkmann}}]{munn2004}
{Munn}, J.~A., {Monet}, D.~G., {Levine}, S.~E., {et~al.} 2004, \aj, 127, 3034

\bibitem[{{Munn} {et~al.}(2008){Munn}, {Monet}, {Levine}, {Canzian}, {Pier},
  {Harris}, {Lupton}, {Ivezi{\'c}}, {Hindsley}, {Hennessy}, {Schneider}, \&
  {Brinkmann}}]{munn2008}
---. 2008, \aj, 136, 895

\bibitem[{{Odenkirchen} {et~al.}(2001){Odenkirchen}, {Grebel}, {Rockosi},
  {Dehnen}, {Ibata}, {Rix}, {Stolte}, {Wolf}, {Anderson}, {Bahcall},
  {Brinkmann}, {Csabai}, {Hennessy}, {Hindsley}, {Ivezi{\'c}}, {Lupton},
  {Munn}, {Pier}, {Stoughton}, \& {York}}]{odenkirchen2001}
{Odenkirchen}, M., {Grebel}, E.~K., {Rockosi}, C.~M., {et~al.} 2001, \apjl,
  548, L165

\bibitem[{{Pagel} \& {Tautvaisiene}(1998)}]{pagel1998}
{Pagel}, B.~E.~J., \& {Tautvaisiene}, G. 1998, \mnras, 299, 535

\bibitem[{{Pe{\~n}arrubia} {et~al.}(2005){Pe{\~n}arrubia},
  {Mart{\'{\i}}nez-Delgado}, {Rix}, {G{\'o}mez-Flechoso}, {Munn}, {Newberg},
  {Bell}, {Yanny}, {Zucker}, \& {Grebel}}]{penarrubia2005}
{Pe{\~n}arrubia}, J., {Mart{\'{\i}}nez-Delgado}, D., {Rix}, H.~W., {et~al.}
  2005, \apj, 626, 128

\bibitem[{{Planck Collaboration} {et~al.}(2014){Planck Collaboration}, {Ade},
  {Aghanim}, {Armitage-Caplan}, {Arnaud}, {Ashdown}, {Atrio-Barandela},
  {Aumont}, {Baccigalupi}, {Banday}, \& et~al.}]{plank2013}
{Planck Collaboration}, {Ade}, P.~A.~R., {Aghanim}, N., {et~al.} 2014, \aap,
  571, A16

\bibitem[{{Price-Whelan} {et~al.}(2015){Price-Whelan}, {Johnston}, {Sheffield},
  {Laporte}, \& {Sesar}}]{pricewhelan2015}
{Price-Whelan}, A.~M., {Johnston}, K.~V., {Sheffield}, A.~A., {Laporte},
  C.~F.~P., \& {Sesar}, B. 2015, \mnras, 452, 676

\bibitem[{{Robin} {et~al.}(2003){Robin}, {Reyl{\'e}}, {Derri{\`e}re}, \&
  {Picaud}}]{robin2003}
{Robin}, A.~C., {Reyl{\'e}}, C., {Derri{\`e}re}, S., \& {Picaud}, S. 2003,
  \aap, 409, 523

\bibitem[{{Rosenberg} {et~al.}(1998{\natexlab{a}}){Rosenberg}, {Piotto},
  {Saviane}, {Aparicio}, \& {Gratton}}]{rosenberg1998}
{Rosenberg}, A., {Piotto}, G., {Saviane}, I., {Aparicio}, A., \& {Gratton}, R.
  1998{\natexlab{a}}, \aj, 115, 658

\bibitem[{{Rosenberg} {et~al.}(1998{\natexlab{b}}){Rosenberg}, {Saviane},
  {Piotto}, {Aparicio}, \& {Zaggia}}]{rosenberg1998b}
{Rosenberg}, A., {Saviane}, I., {Piotto}, G., {Aparicio}, A., \& {Zaggia},
  S.~R. 1998{\natexlab{b}}, \aj, 115, 648

\bibitem[{{Schlafly} \& {Finkbeiner}(2011)}]{schlafly2011}
{Schlafly}, E.~F., \& {Finkbeiner}, D.~P. 2011, \apj, 737, 103

\bibitem[{{Schlaufman} {et~al.}(2011){Schlaufman}, {Rockosi}, {Lee}, {Beers},
  \& {Allende Prieto}}]{schlaufman2011}
{Schlaufman}, K.~C., {Rockosi}, C.~M., {Lee}, Y.~S., {Beers}, T.~C., \&
  {Allende Prieto}, C. 2011, \apj, 734, 49

\bibitem[{{Schlaufman} {et~al.}(2009){Schlaufman}, {Rockosi}, {Allende Prieto},
  {Beers}, {Bizyaev}, {Brewington}, {Lee}, {Malanushenko}, {Malanushenko},
  {Oravetz}, {Pan}, {Simmons}, {Snedden}, \& {Yanny}}]{schlaufman2009}
{Schlaufman}, K.~C., {Rockosi}, C.~M., {Allende Prieto}, C., {et~al.} 2009,
  \apj, 703, 2177

\bibitem[{{Schlegel} {et~al.}(1998){Schlegel}, {Finkbeiner}, \&
  {Davis}}]{schlegel1998}
{Schlegel}, D.~J., {Finkbeiner}, D.~P., \& {Davis}, M. 1998, \apj, 500, 525

\bibitem[{{Sheffield} {et~al.}(2014){Sheffield}, {Johnston}, {Majewski},
  {Damke}, {Richardson}, {Beaton}, \& {Rocha-Pinto}}]{sheffield2014}
{Sheffield}, A.~A., {Johnston}, K.~V., {Majewski}, S.~R., {et~al.} 2014, \apj,
  793, 62

\bibitem[{{Slater} {et~al.}(2014){Slater}, {Bell}, {Schlafly}, {Morganson},
  {Martin}, {Rix}, {Pe{\~n}arrubia}, {Bernard}, {Ferguson}, {Martinez-Delgado},
  {Wyse}, {Burgett}, {Chambers}, {Draper}, {Hodapp}, {Kaiser}, {Magnier},
  {Metcalfe}, {Price}, {Tonry}, {Wainscoat}, \& {Waters}}]{slater2014}
{Slater}, C.~T., {Bell}, E.~F., {Schlafly}, E.~F., {et~al.} 2014, \apj, 791, 9

\bibitem[{{Smith} {et~al.}(2013){Smith}, {Fellhauer}, {Candlish}, {Wojtak},
  {Farias}, \& {Bla{\~n}a}}]{smith2013}
{Smith}, R., {Fellhauer}, M., {Candlish}, G.~N., {et~al.} 2013, \mnras, 433,
  2529

\bibitem[{{Stetson}(1987)}]{stetson1987}
{Stetson}, P.~B. 1987, \pasp, 99, 191

\bibitem[{{Stetson}(1994)}]{stetson1994}
---. 1994, \pasp, 106, 250

\bibitem[{Szentgyorgyi {et~al.}(2011)Szentgyorgyi, F.Furesz, Cheimets, Conroy,
  Eng, Fabricant, Fata, Gauron, Geary, McLeod, Zajac, Amato, Bergner, Caldwell,
  Dupree, Goddard, Johnston, Meibom, Mink, Pieri, Roll, Tokarz, Wyatt, Epps,
  Hartmann, \& Meszaros}]{szentgyorgyi2011}
Szentgyorgyi, A., F.Furesz, Cheimets, P., {et~al.} 2011, Publications of the
  Astronomical Society of the Pacific, 123, 1188

\bibitem[{{Valcheva} {et~al.}(2015){Valcheva}, {Ovcharov}, {Lalova},
  {Nedialkov}, {Ivanov}, \& {Carraro}}]{valcheva2015}
{Valcheva}, A.~T., {Ovcharov}, E.~P., {Lalova}, A.~D., {et~al.} 2015, \mnras,
  446, 730

\bibitem[{{Valdes} {et~al.}(2014){Valdes}, {Gruendl}, \& {DES
  Project}}]{valdes2014}
{Valdes}, F., {Gruendl}, R., \& {DES Project}. 2014, in Astronomical Society of
  the Pacific Conference Series, Vol. 485, Astronomical Data Analysis Software
  and Systems XXIII, ed. N.~{Manset} \& P.~{Forshay}, 379

\bibitem[{{Venn} {et~al.}(2004){Venn}, {Irwin}, {Shetrone}, {Tout}, {Hill}, \&
  {Tolstoy}}]{venn2004}
{Venn}, K.~A., {Irwin}, M., {Shetrone}, M.~D., {et~al.} 2004, \aj, 128, 1177

\bibitem[{{Walker} {et~al.}(2009{\natexlab{a}}){Walker}, {Mateo}, {Olszewski},
  {Pe{\~n}arrubia}, {Wyn Evans}, \& {Gilmore}}]{walker2009b}
{Walker}, M.~G., {Mateo}, M., {Olszewski}, E.~W., {et~al.} 2009{\natexlab{a}},
  \apj, 704, 1274

\bibitem[{{Walker} {et~al.}(2009{\natexlab{b}}){Walker}, {Mateo}, {Olszewski},
  {Sen}, \& {Woodroofe}}]{walker2009a}
{Walker}, M.~G., {Mateo}, M., {Olszewski}, E.~W., {Sen}, B., \& {Woodroofe}, M.
  2009{\natexlab{b}}, \aj, 137, 3109

\bibitem[{{Walker} {et~al.}(2015){Walker}, {Olszewski}, \&
  {Mateo}}]{walker2015}
{Walker}, M.~G., {Olszewski}, E.~W., \& {Mateo}, M. 2015, \mnras, 448, 2717

\bibitem[{{Weisz} {et~al.}(2014){Weisz}, {Dolphin}, {Skillman}, {Holtzman},
  {Gilbert}, {Dalcanton}, \& {Williams}}]{weisz2014}
{Weisz}, D.~R., {Dolphin}, A.~E., {Skillman}, E.~D., {et~al.} 2014, \apj, 789,
  147

\bibitem[{{Willman} \& {Strader}(2012)}]{willman2012}
{Willman}, B., \& {Strader}, J. 2012, \aj, 144, 76

\bibitem[{{Wolf} {et~al.}(2010){Wolf}, {Martinez}, {Bullock}, {Kaplinghat},
  {Geha}, {Mu{\~n}oz}, {Simon}, \& {Avedo}}]{wolf2010}
{Wolf}, J., {Martinez}, G.~D., {Bullock}, J.~S., {et~al.} 2010, \mnras, 406,
  1220

\bibitem[{{Xu} {et~al.}(2015){Xu}, {Newberg}, {Carlin}, {Liu}, {Deng}, {Li},
  {Sch{\"o}nrich}, \& {Yanny}}]{xu2015}
{Xu}, Y., {Newberg}, H.~J., {Carlin}, J.~L., {et~al.} 2015, \apj, 801, 105

\end{thebibliography}

\clearpage

\begin{deluxetable}{lllllll}
\tablecolumns{7}
\tablewidth{0pt} 
\tabletypesize{\scriptsize}
\tablecaption{{DECam Photometry}}
\tablehead{
\colhead{R.A.} & \colhead{Dec.} & \colhead{$g_0$} & \colhead{$\sigma_{g_0}$}  & \colhead{$r_0$} & \colhead{$\sigma_{r_0}$}  & \colhead{$E(B-V)$} \\
\colhead{(deg)} & \colhead{(deg)} & \colhead{mag} & \colhead{(mag)} & \colhead{(mag)} & \colhead{(mag)} & \colhead{(mag)}
}
\startdata
132.348949 & 3.153806 & 19.116 & 0.015 & 18.187 & 0.009 & 0.0318   \\
132.350640 & 3.344006 & 18.993 & 0.013 & 18.308 & 0.009 & 0.0339   \\
132.351339 & 3.363441 & 22.279 & 0.042 & 21.675 & 0.033 & 0.0340   \\
132.351990 & 3.397592 & 22.226 & 0.068 & 21.581 & 0.038 & 0.0340   \\
132.352598 & 3.438227 & 19.357 & 0.012 & 18.727 & 0.009 & 0.0338   \\
132.352717 & 3.150096 & 22.082 & 0.037 & 22.058 & 0.050 & 0.0319   \\
132.354884 & 3.307660 & 20.540 & 0.014 & 20.283 & 0.013 & 0.0341   \\
132.355516 & 3.424209 & 20.756 & 0.016 & 20.193 & 0.011 & 0.0339   \\
132.356115 & 3.223843 & 21.874 & 0.027 & 21.533 & 0.025 & 0.0334  \\
132.358492 & 3.395449 & 23.487 & 0.106 & 22.591 & 0.070 & 0.0341 \\
... & & & & & & 
\enddata
\tablecomments{DECam photometric point-source catalog of the EBS/Hydra~I region (see footprint in Figure~\ref{fig:spatial-ebs}).  The photometric data have been corrected for Galactic extinction (see Section~\ref{sec:decam}) using the listed values of $E(B-V)$.  We provide photometry only for stars with $g - r < 1$ due to the photometric calibration method employed in this study.  Point-source selection was performed using the DAOPHOT CHI and SHARP parameters.  Table~\ref{tbl:decam} is published in its entirety in the electronic edition of The Astronomical Journal.   A portion is shown here for guidance regarding its form and content.}
\label{tbl:decam}
\end{deluxetable}

\begin{deluxetable*}{lllllllll}
\tablecolumns{9}
\tablewidth{0pt} 
\tablecaption{{Hectochelle/MMT Spectroscopy}}
\tablehead{
\colhead{R.A.} & \colhead{Dec.} & \colhead{$V_R$} & \colhead{$\sigma_{V_R}$} & \colhead{$g_0$} & \colhead{$\sigma_{g_0}$}  & \colhead{$r_0$} & \colhead{$\sigma_{r_0}$}  & \colhead{$E(B-V)$} \\
\colhead{(deg)} & \colhead{(deg)} & \colhead{(km s$^{-1}$)} & \colhead{(km s$^{-1}$)} & \colhead{(mag)} & \colhead{(mag)} & \colhead{(mag)} & \colhead{(mag)} & \colhead{(mag)} 
}
\startdata
134.208389 & 3.950312 & 96.84 & 0.44 & 17.241 & 0.005 & 16.744 & 0.003 & 0.0499  \\
134.244110 & 3.806066 & 92.04 & 0.53 & 18.404 & 0.003 & 18.182 & 0.002 & 0.0479  \\ 
134.257034 & 4.086879 & 39.57 & 1.10 & 19.455 & 0.005 & 19.146 & 0.004 & 0.0512   \\
134.273636 & 4.103878 & 63.54 & 0.44 & 17.330 & 0.005 & 16.867 & 0.003 & 0.0510   \\
134.260468 & 4.003685 & 97.81 & 0.44 & 17.555 & 0.005 & 17.140 & 0.003 & 0.0505   \\
134.115387 & 3.871296 & 178.97 & 0.35 & 18.335 & 0.004 & 17.902 & 0.003 & 0.0514   \\
134.117737 & 3.833826 & 31.85 & 1.47 & 20.291 & 0.006 & 19.858 & 0.004 & 0.0501   \\
134.071915 & 3.766812 & 85.44 & 0.97 & 19.602 & 0.004 & 19.375 & 0.003 & 0.0486   \\
134.084824 & 3.745721 & 65.47 & 0.34 & 17.731 & 0.003 & 17.330 & 0.003 & 0.0475   \\
134.113266 & 3.739799 & 20.37 & 0.46 & 17.905 & 0.004 & 17.449 & 0.003 & 0.0464  \\
... & & & & & & & & 
\enddata
\tablecomments{Hectochelle/MMT catalog of all 411 stars passing the spectroscopic quality control cuts and located within the SDSS CMD filter shown in Figure~\ref{fig:targeting}a.  Only a subset of these spectra were used in our analysis of Hydra~I (see Section~\ref{sec:phot_sample}).    Heliocentric radial velocities ($V_R$) and $1\sigma$ uncertainties were derived using the method described in Section~\ref{sec:spectroscopy}.  The DECam photometry has been corrected for Galactic extinction (see Section~\ref{sec:decam}) using the listed values of $E(B-V)$.  Table~\ref{tbl:hecto} is published in its entirety in the electronic edition of The Astronomical Journal.   A portion is shown here for guidance regarding its form and content.}
\label{tbl:hecto}
\end{deluxetable*}

\begin{deluxetable*}{lllllllllllll}
\tablecolumns{13}
\tablewidth{0pt} 
\tablecaption{{SDSS/SEGUE Spectroscopy}}
\tablehead{
\colhead{R.A.} & \colhead{Dec.} & \colhead{$V_R$} & \colhead{$\sigma_{V_R}$} & \colhead{[Fe/H]} & \colhead{$\sigma_{[Fe/H]}$} & \colhead{[$\alpha$/Fe]} & \colhead{$\sigma_{[\alpha/Fe]}$} & \colhead{$g_0$} & \colhead{$\sigma_{g_0}$}  & \colhead{$r_0$} & \colhead{$\sigma_{r_0}$}  & \colhead{$E(B-V)$} \\
\colhead{(deg)} & \colhead{(deg)} & \colhead{(km s$^{-1}$)} & \colhead{(km s$^{-1}$)} & \colhead{(dex)} & \colhead{(dex)} & \colhead{(dex)} & \colhead{(dex)} & \colhead{(mag)} & \colhead{(mag)} & \colhead{(mag)} & \colhead{(mag)} & \colhead{(mag)} 
}
\startdata
132.572900 & 3.247481 & 54.42 & 0.70 & -0.35 & 0.051 & 0.18 & 0.014 & 17.217 & 0.006 & 16.636 & 0.006 & 0.0348 \\
132.673340 & 3.591954 & 9.72 & 0.80 & -0.20 & 0.056 & 0.13 & 0.016 & 16.880 & 0.006 & 16.377 & 0.004 & 0.0445  \\
132.678310 & 2.828570 & 269.38 & 8.09 & -0.87 & 0.066 & 0.46 & 0.018 & 19.960 & 0.013 & 19.574 & 0.009 & 0.0304  \\
132.684040 & 3.827853 & 28.81 & 1.49 & -0.47 & 0.023 & 0.31 & 0.012 & 18.131 & 0.011 & 17.584 & 0.006 & 0.0384  \\
132.685030 & 3.294454 & 44.80 & 0.97 & -0.36 & 0.040 & 0.21 & 0.014 & 17.402 & 0.006 & 16.843 & 0.003 & 0.0343  \\
132.691000 & 3.215672 & 214.44 & 2.48 & -1.57 & 0.065 & 0.24 & 0.018 & 18.027 & 0.008 & 17.526 & 0.007 & 0.0340  \\
132.703350 & 2.779862 & 26.07 & 1.24 & -0.58 & 0.031 & 0.43 & 0.013 & 16.762 & 0.010 & 16.250 & 0.007 & 0.0307  \\
132.709140 & 3.524834 & 296.70 & 1.97 & -1.01 & 0.074 & 0.34 & 0.014 & 18.124 & 0.005 & 17.600 & 0.003 & 0.0420  \\
132.734190 & 3.252893 & 53.05 & 0.74 & -0.40 & 0.011 & 0.22 & 0.013 & 16.587 & 0.005 & 16.118 & 0.004 & 0.0339  \\
... & & &  & & & & & & & & &
\enddata
\tablecomments{SDSS/SEGUE catalog of all 292 stars located in the Hydra~I region and within the SDSS CMD filter shown in Figure~\ref{fig:targeting}a.  All spectroscopic parameters were taken from the SSPP pipeline.  Only a subset of these spectra were used in our analysis of Hydra~I (see Section~\ref{sec:phot_sample}).  The DECam photometry has been corrected for Galactic extinction (see Section~\ref{sec:decam}) using the listed values of $E(B-V)$.  Table~\ref{tbl:hecto} is published in its entirety in the electronic edition of The Astronomical Journal.   A portion is shown here for guidance regarding its form and content.}
\label{tbl:sdss}
\end{deluxetable*}

\end{document}